\documentclass[preprint2]{aastex}

\shorttitle{On the TP-AGB contribution to the light of nearby disk galaxies}
\shortauthors{Mart\'{\i}nez-Garc\'ia et al.}

\newcommand{\tn}{\tablenotemark}

\begin{document}

\title{On the Thermally Pulsing Asymptotic Giant Branch Contribution to the Light of Nearby Disk Galaxies}

\author{Eric E. Mart\'inez-Garc\'ia\altaffilmark{1}, 
Gustavo Bruzual\altaffilmark{2},
Rosa A. Gonz\'alez-L\'opezlira\altaffilmark{2}, and
Lino H. Rodr\'iguez-Merino\altaffilmark{3}}

\affil{1 CONACYT-Instituto Nacional de Astrof\'isica, \'Optica y Electr\'onica, Luis E. Erro 1, Tonantzintla, Puebla, C.P. 72840, M\'exico}
\affil{2 Instituto de Radioastronom\'ia y Astrof\'isica, UNAM, Campus Morelia, Michoac\'an, M\'exico, C.P. 58089}
\affil{3 Instituto Nacional de Astrof\'isica, \'Optica y Electr\'onica, Luis E. Erro 1, Tonantzintla, Puebla, C.P. 72840, M\'exico}

\email{ericmartinez@inaoep.mx}

\begin{abstract}

The study of the luminosity contribution from thermally pulsing asymptotic giant branch (TP-AGB) stars to the
stellar populations of galaxies is crucial to determine their physical parameters (e.g., stellar mass and age). We use
a sample of 84 nearby disk galaxies to explore diverse stellar population synthesis models with different luminosity
contributions from TP-AGB stars. We fit the models to optical and near-infrared (NIR) photometry, on a pixel-bypixel
basis. The statistics of the fits show a preference for a low-luminosity contribution (i.e., high mass-to-light
ratio in the NIR) from TP-AGB stars. Nevertheless, for 30\%–40\% of the pixels in our sample a high-luminosity
contribution (hence low mass-to-light ratio in the NIR) from TP-AGB stars is favored. According to our findings,
the mean TP-AGB star luminosity contribution in nearby disk galaxies may vary with Hubble type. This may be a
consequence of the variation of the TP-AGB mass-loss rate with metallicity, if metal-poor stars begin losing mass
earlier than metal-rich stars, because of a pre-dust wind that precedes the dust-driven wind.

\end{abstract}

\keywords{ 
galaxies: evolution --
galaxies: photometry --
galaxies: stellar content --
stars: AGB and post-AGB. }

\section{Introduction}

The panchromatic analysis of galaxies has become an important tool to understand galaxy formation
and evolution~\citep[e.g.,][]{con13,lara13,bua14,vac16,pou16,nag16,lim17,ma18,rob19,neg20}.
For this purpose, the studies of spatially resolved galaxies, i.e., on a
pixel-by-pixel~\citep[e.g.][]{zcr09,men12,dia15,sor15,abd17,mar17,mar18},
or spaxel-by-spaxel~\citep[with integral field unit spectrographs, IFUs, e.g.][]{ros10,gond14,
gond15,gond16,can16,can19,iba16,deAmo17,sanch19,err19} basis,
provide very important information to complement single spectra analysis of large
samples of objects~\citep[see also,][]{mao12,cav15,san18,san19}.

However, some issues arise in the resolved studies of galaxies that must not be overlooked.
Among common problems are the following:
\begin{enumerate}
\item[(i)] The spatial structure of the recovered stellar mass
is often different from the expected one.
\item[(ii)] The choice of stellar population synthesis models
affects the recovered physical characteristics.
\end{enumerate}
The first issue was investigated by~\citet{mar17}. According to their findings,
the recovered stellar mass map of a nearby galaxy (M51), obtained by fitting stellar
population synthesis models to observations (pixel-by-pixel in this case), results
in an artificial spatial structure, not a real mass structure, with supposed filaments
that spatially coincide with the dust lanes near the spiral arms. The expected mass
structure, on the other hand, resembles the near infrared (NIR) image of the object,
which the recovered mass structure does not reproduce.
This effect also affects the global star formation history (SFH) of the galaxy,
when it is obtained as the sum of the SFHs of all the individual pixels in the object~\citep{mar18}.

\begin{figure*}
\centering
\epsscale{2.0}
\plotone{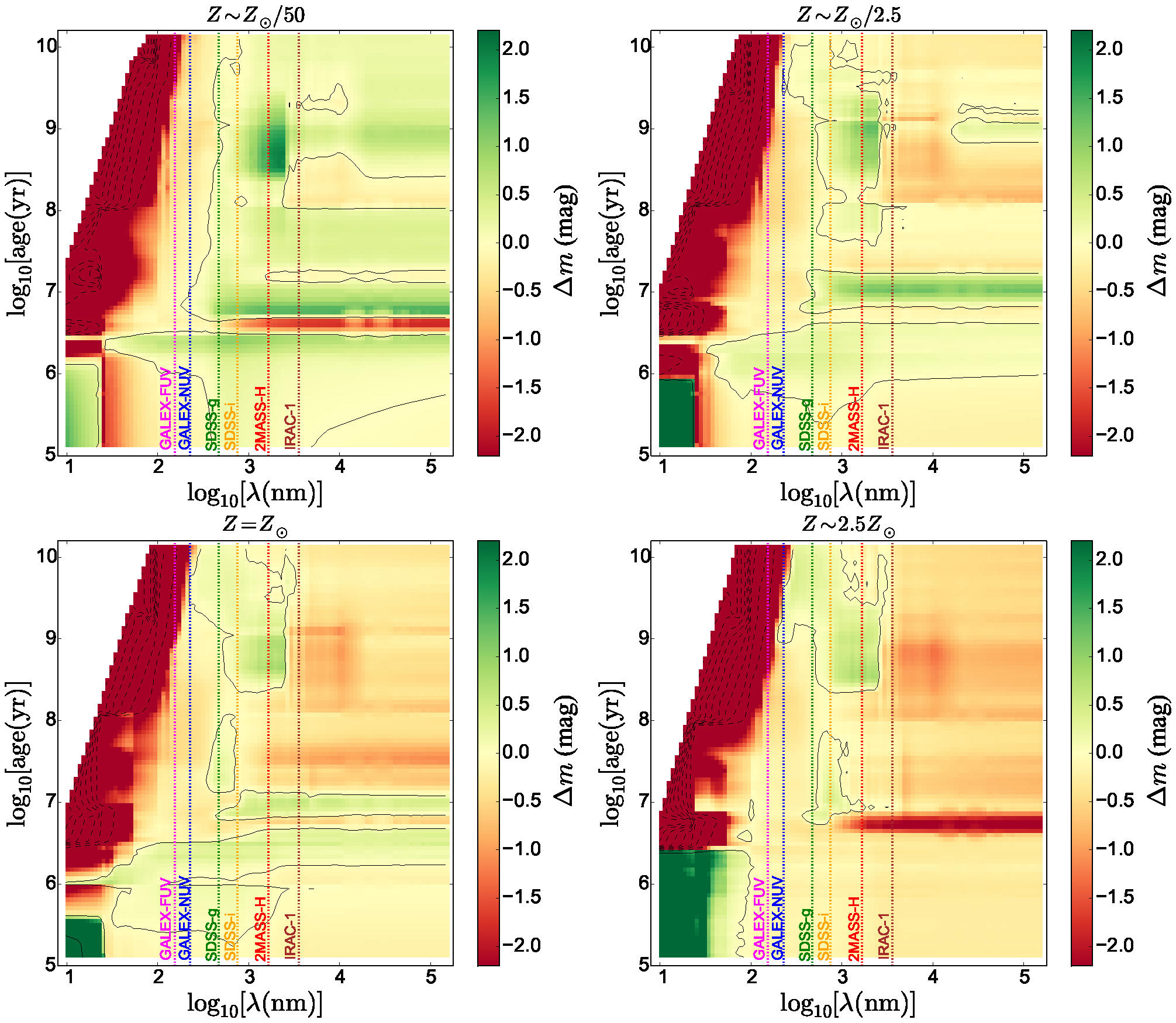}
\caption[f1]{Luminosity ratio expressed as a magnitude difference, $\Delta m$ (see equation~\ref{eq_delmag1}),
between~\citet{mara05} and~\citet{bru03} models, for simple stellar populations (SSP) of different metallicities,
$Z$, where $Z_{\sun}$ is the solar value.
Horizontal axis is $\log_{10}$ of wavelength, $\lambda$, in nm. Vertical
axis is $\log_{10}$ of stellar age in yr. Vertical dotted lines indicate effective wavelengths of 
Galaxy Evolution Explorer~\citep[GALEX,][]{bia99,martin05} $FUV$ and $NUV$ bands,
SDSS $g$ and $i$ bands, 2MASS $H$-band,
and Infrared Array Camera~\citep{faz04} $3.6\micron$ band (IRAC-1)
of the Spitzer Space Telescope~\citep{wer04},
as reference. Negative values (redder colors and dashed contours) indicate Bruzual \& Charlot (2003) models
are more luminous than Maraston (2005) ones.
~\label{fig1}}
\end{figure*}

The second issue is mainly due to the fact that stellar population synthesis models from different authors,
even with the same input parameters, differ in luminosity ($L$) at various wavelengths~\citep[e.g., ][their Figure 2]{boq19}.
For instance, we show in Figure~\ref{fig1} the luminosity ratio between~\citet[][hereafter bc03]{bru03}
and~\citet[][hereafter m2005]{mara05} models, for simple stellar populations (SSP), i.e.,
an instantaneous burst or a Dirac delta function for the SFH.
The luminosity ratio is given as a magnitude difference,
\begin{equation}~\label{eq_delmag1}
  \Delta m = -2.5\log_{10}\frac{L_{\rm{bc03}}}{L_{\rm{m2005}}},
\end{equation}
in the wavelength ($\lambda$) vs.\ stellar age plane. All calculations were done with the {\tt{GALAXEV}} software~\citep{bru03}
for four metallicities:~$Z_{\sun}$ (solar value), $\sim~Z_{\sun}$/50,~$\sim~Z_{\sun}$/2.5, and $\sim~2.5Z_{\sun}$, with no dust attenuation.
In the region of the extreme ultraviolet emission ($EUV$, 10-100 nm), $\Delta m$ has negative values, indicating
a higher luminosity for bc03, compared with m2005.\footnote{This effect is important when fitting the
models to high-redshift objects, where the rest $EUV$ is displaced towards redder $\lambda$.}
In the far and near ultraviolet regions of the spectrum ($FUV$ and $NUV$, respectively), the luminosity is higher for bc03 at older ages.
At optical wavelengths (see the Sloan Digital Sky Survey, SDSS,\footnote{\citet{doi10}.} $g$ and $i$-bands in the plot),
the luminosity is similar for m2005 and bc03.
In the NIR range (see the Two Micron All Sky Survey, 2MASS,\footnote{\citet{skr06}.} $H$-band in the plot),
after $\approx2\times10^{8}$ yr the luminosity is higher in m2005 than in bc03. This is the stellar age where
the thermally pulsing asymptotic giant branch (TP-AGB) stars dominate the stellar population luminosity.
Also, according to the plot for $\lambda\ga2.5\micron$, the luminosity is higher
in bc03 than in m2005 at ages $\approx1\times10^{8}$ yr.

The investigations of AGB stars by several authors have led to a better understanding of this
important phase of stellar evolution~\citep[e.g.,][]{hab96,lan99,pio03,her05,woi06,kar07,poe08,ram08,wei09,kar10,ven10,ven12,
dic13,gar13,gir13,pla13,kal14,cris15,bat16,gol17,brun19,gol19,mcd19,gir20,pas20,wie20}.
However, the overall impact of TP-AGB\footnote{The AGB evolution can be divided into the early AGB (E-AGB) and the TP-AGB phases.}
stars on stellar populations of galaxies is not well understood,
and has been a matter of debate during the last fifteen years~\citep[e.g.,][]{mari10,mari15}.
\citet{mara06} analyzed seven high-redshift galaxies, and concluded
that the contribution of TP-AGB stars played an important role in the interpretation of their data.
The TP-AGB stars are cool giants with low-to-intermediate stellar mass~\citep[$\sim$0.5-6.4~$M_{\sun}$,][]{mari17}.
They have lifetimes of a few Myr~\citep[e.g.,][]{mari07}, and their contribution to the luminosity
of SSP is maximized in the age range 0.2-2 Gyr~\citep{mou02,mara06}.
The TP-AGB phase involves processes that are not straightforward to calibrate independently,
such as envelope convection, mixing~\citep[dredge-up, e.g.,][]{fro96,her07,wag20}, and mass loss~\citep[e.g.,][]{gon10,ros14,ros16,gon18}.
Needless to say, processes like rotation and magnetic fields do not make things easier~\citep{ren15}.
For convenience purposes (that involve the stellar age),
TP-AGB studies are typically conducted with samples of post-starburst (the contribution of these stars is maximum $\sim2$ Gyr after the burst ends),
or high-$z$ ($1\lesssim z \lesssim3$) galaxies. However, TP-AGB stars are present in normal nearby galaxies as well.
The luminosity contribution of TP-AGB stars to the spectral energy distribution (SED) of galaxies
is more prominent in the NIR, ranging from $\sim$40\%~\citep{bru07a,bru13} to $\sim$80\%~\citep{mara06,ton09}, and is significantly
lower at optical wavelengths. 

In m2005 models the TP-AGB contribution to the SSP is computed by means of the
the {\it fuel consumption theorem}~\citep{ren81,ren86,mar98}.
This theorem states that the contribution of any post-main sequence star to the
total luminosity of a SSP is proportional to the amount of nuclear fuel
(hydrogen and helium) consumed through nuclear burning.
In these models, the TP-AGB is calibrated with Large Magellanic Cloud (MC) globular clusters.

The bc03 models include a TP-AGB treatment using the
{\it isochrone synthesis}~\citep{cha91,bru93} technique.
In this method the luminosity of a SSP, at a given time and metallicity,
is obtained by assuming an initial mass function (IMF)
and integrating the contributions of all stars in each mass bin on a certain isochrone.
\citet{bru03} adopt the stellar parameters (effective temperatures, bolometric luminosities, and lifetimes)
for TP-AGB stars from the multimetallicity models of~\citet{vas93}.
The transition from oxygen-rich to carbon-rich stars is computed with the models of~\citet{gro93} and~\citet{gro95},
which give similar ratios of carbon- to oxygen-rich stars as those observed in
the Large MC and in the Galaxy. The bc03 models include only one evolutionary stage
in each of the oxygen-rich, carbon-rich, and super wind phases.

Due to their relatively low contribution of TP-AGB stars to the NIR luminosity,
bc03 models are considered as TP-AGB ``light'' models.
On the other hand, m2005 are considered as TP-AGB ``heavy'' models.\footnote{Also PEGASE~\citep{fio97} and Starburst99~\citep{lei99}
are considered TP-AGB ``light'' models. On the other hand, the flexible stellar population synthesis
({\tt{FSPS}}) package~\citep{con09,con10} allows for adjustable TP-AGB parameters.}

The 2007 version of Charlot \& Bruzual models~\citep[][hereafter cb07]{bru07a} are identical to the
bc03 models in all aspects, except for the treatment of the TP-AGB phase.
cb07 use the TP-AGB prescription of~\citet{mari07} to account for the changes in the chemical
composition of the envelopes, resulting in significantly redder NIR colors, and 
hence lower masses and younger ages,\footnote{The mass turned into stars until time $t$ can
be obtained as $M(t) = {\int_{0}^{t} \Psi(t') {\rm d}t'}$, where $\Psi(t)$ is the star formation rate.
Therefore, a lower age would result in a lower mass for a certain $\Psi(t)$.}
for young and intermediate-age stellar populations~\citep{bru07a}.
These models comprise six evolutionary stages in each of the oxygen-rich
and carbon-rich phases, and three in the super wind phase (15 stages in total).
The models are calibrated by using carbon-rich star
luminosity functions in the MCs, and TP-AGB lifetimes (through star counts) in MC clusters.
The cb07 models can be considered as TP-AGB ``heavy''~\citep{bru07b}, given their comparable results to those of m2005.

\begin{figure*}
\centering
\epsscale{2.0}
\plotone{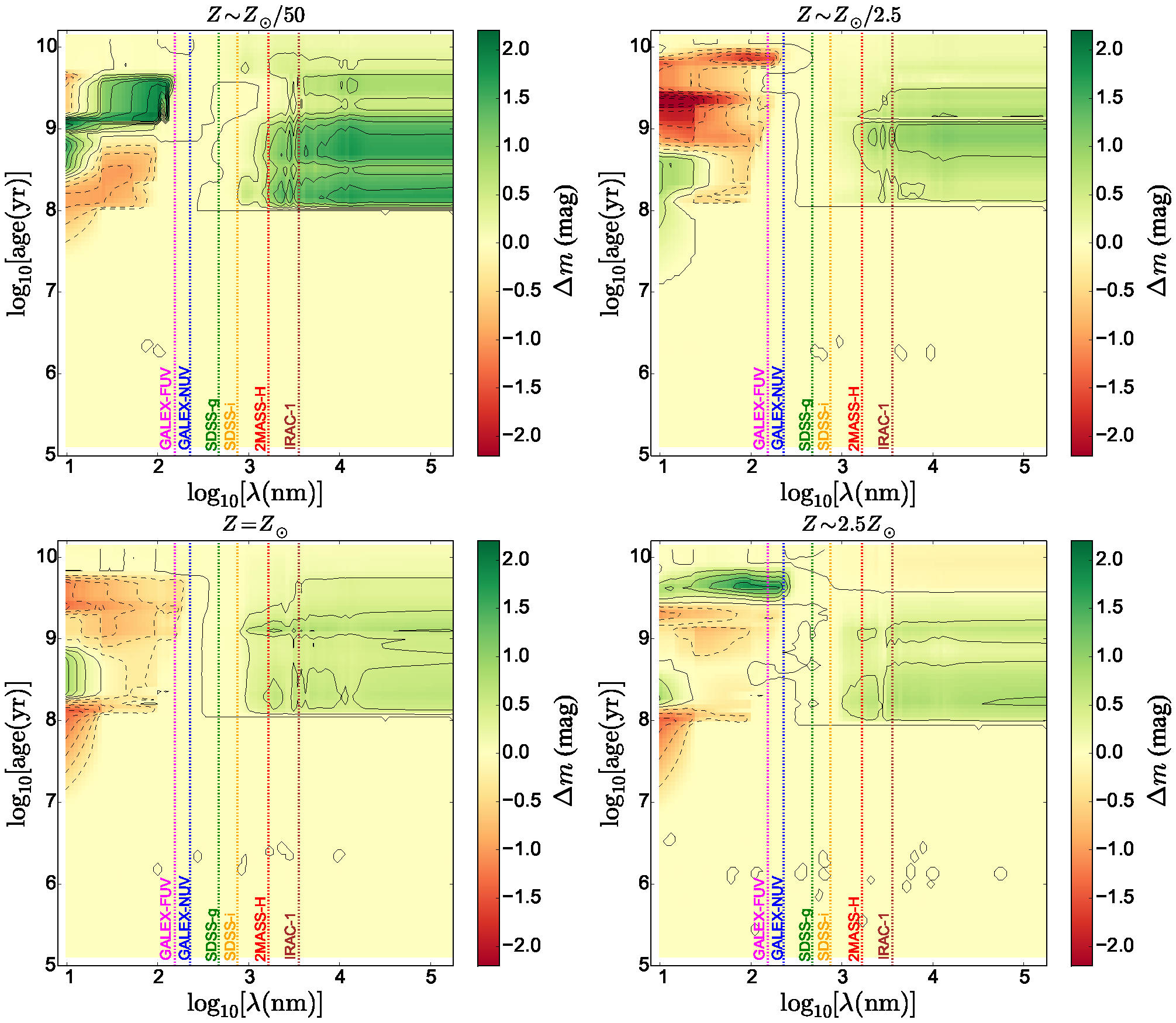}
\caption[f2]{Same as Figure~\ref{fig1}, for cb07-2016 vs.\ bc03-2016 SSP; $\Delta m$ in mag, as in equation~\ref{eq_delmag2}.
Negative values (redder colors and dashed contours) indicate bc03-2016 models are more luminous than cb07-2016 ones.
~\label{fig2}}
\end{figure*}

The recalculated bc03 models (2016 version\footnote{\url{http://bruzual.org/}.}, hereafter bc03-2016)
are again similar to the bc03 models, except that the TP-AGB phase is modeled with
the~\citet{mari08} prescription, which is a revision of the~\citet{mari07} recipe.
This results in a very similar behavior to bc03, i.e., bluer NIR colors compared with cb07,
but the luminosity and color evolution in the NIR
is smoother than in bc03, since the TP-AGB phase is sampled
with a higher time resolution, similar to that of cb07.
Similarily to the former models (except bc03), bc03-2016 comprises fifteen
evolutionary stages among the oxygen-rich, carbon-rich, and super wind phases.
The 2016 version of the cb07 models (hereafter cb07-2016)
employs a TP-AGB treatment similar to cb07. bc03-2016 and cb07-2016 are categorized as
TP-AGB ``light'' and ``heavy'', respectively.
In Figure~\ref{fig2}, we show the luminosity ratio between bc03-2016 and cb07-2016, for SSP.
In this plot, the luminosity ratio is quantified as:
\begin{equation}~\label{eq_delmag2}
  \Delta m = -2.5\log_{10}\frac{L_{\rm{bc03}}}{L_{\rm{cb07}}}.
\end{equation}
The luminosity ratio is $\Delta m\sim 0$ for ages lower than~$\sim1\times10^{8}$ yr at all wavelengths and metallicities.
For ages greater than $1\times10^{8}$ yr, the TP-AGB stars contribute with more luminosity for cb07-2016 models
(positive $\Delta m$, green hues), in the NIR. We also discern
luminosity differences in the $EUV$ region of the spectra for ages $\ga1\times10^{8}$ yr,
but practically no luminosity differences in the optical at all ages.

From comparisons to observations, some authors favor the use of a ``light'' contribution of TP-AGB stars~\citep[e.g.,][]{kri10,mel12,zib13},
while others support a ``heavy'' contribution~\citep[e.g.,][]{mara06,mac10,cap16}.
The differences in stellar masses and ages can be as high as $\sim$50\%~\citep[e.g.,][]{mara06,kan07,mara11};
however, the impact on the SFH may be less significant~\citep{bal18}.


\subsection{Aims of this work}

In this paper we aim to investigate how different stellar population synthesis models fit the photometric data of resolved
nearby disk galaxies, and hence affect their derived properties. For this purpose we build libraries that contain TP-AGB ``light''
stellar population synthesis models, and separate libraries with TP-AGB ``heavy''
models. We use these libraries to fit a sample of nearby disk galaxies on a
pixel-by-pixel basis and compare the results.

This work is developed in the following way.
In Section~\ref{sec_build_libs} we give a detailed explanation of how we build the libraries of models;
in Section~\ref{sec_examining_libraries} we examine the libraries;
in Section~\ref{sec_sample_objects} we describe our sample of objects and photometric data;
in Section~\ref{sec_fits_photometry} we show the results of the fits to the observed photometry;
a discussion of the main results is given in Section~\ref{sec_discussion};
finally, in Section~\ref{sec_conclusions} we present our conclusions.

\section{Building the libraries}~\label{sec_build_libs}

In order to build the composite stellar population (CSP, i.e., with a SFH different from an
instantaneous burst) libraries we use the Code Investigating GALaxy Emission~\citep[{\tt{CIGALE}},][]{bur05,nol09,boq19},
in the model creator mode.
{\tt{CIGALE}} works under the premise that all the absorbed radiation in the UV and optical parts of the
spectrum is re-emitted in the infrared. The two main input ingredients are the SSP
and dust emission models. We build five CSP libraries (bc03, m2005, cb07, bc03-2016, and cb07-2016)\footnote{
The default {\tt{CIGALE}} code only includes bc03 and m2005 as options.
The code was slightly modified to use cb07, bc03-2016, and cb07-2016.
To avoid introducing any other differences in the model behavior, in all cases we use the BaSeL~3.1 stellar
library~\citep{wes02} to compute the model SED's.}
of $5\times10^{4}$ models each, with similar parameters for all libraries.
For the bc03, cb07, bc03-2016, and cb07-2016 libraries we use the~\citet{cha03} IMF,
and for m2005 the~\citet{kro01} IMF,\footnote{The available IMFs in {\tt{CIGALE}}
are the~\citet{sal55} and~\citet{cha03} for bc03, and~\citet{sal55} and~\citet{kro01}
for m2005.~\citet{cha03} and~\citet{kro01} IMFs are very similar to each other.
Variations in the IMF are discussed in Section~\ref{var_IMF}.}
with mass cutoffs 0.1-100 $M_{\sun}$.
Each spectrum is computed by randomly drawing the model parameters: SFH, metallicity,
dust attenuation, and dust emission.
The comparison between libraries is carried out by fitting separately each library and then
comparing the fits on a pixel-by-pixel basis (see Section~\ref{libs_conf}).
In the following, we describe in more detail the parameters adopted for the libraries.

\subsection{Star formation history}~\label{sec_SFH}

\begin{figure}
\centering
\epsscale{1.0}
\plotone{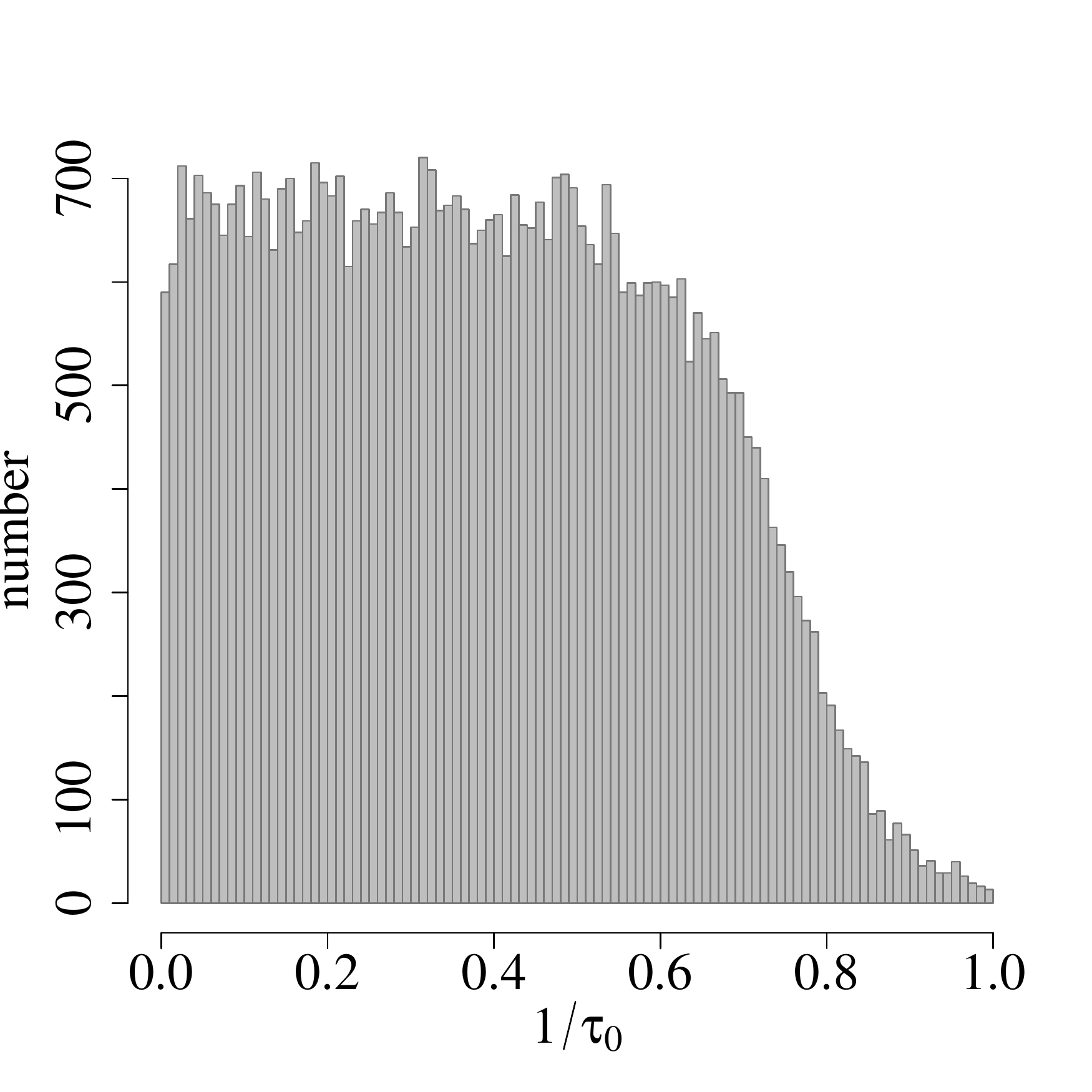}
\caption[f3]{Probability density function $P(\gamma)=1-\tanh(8\gamma-6)$~\citep{dac08},
where $\tau_{0}=1/\gamma$ (Gyr) is the $e$-folding timescale of the main star formation burst.
~\label{fig3}}
\end{figure}

\begin{figure}
\centering
\epsscale{1.0}
\plotone{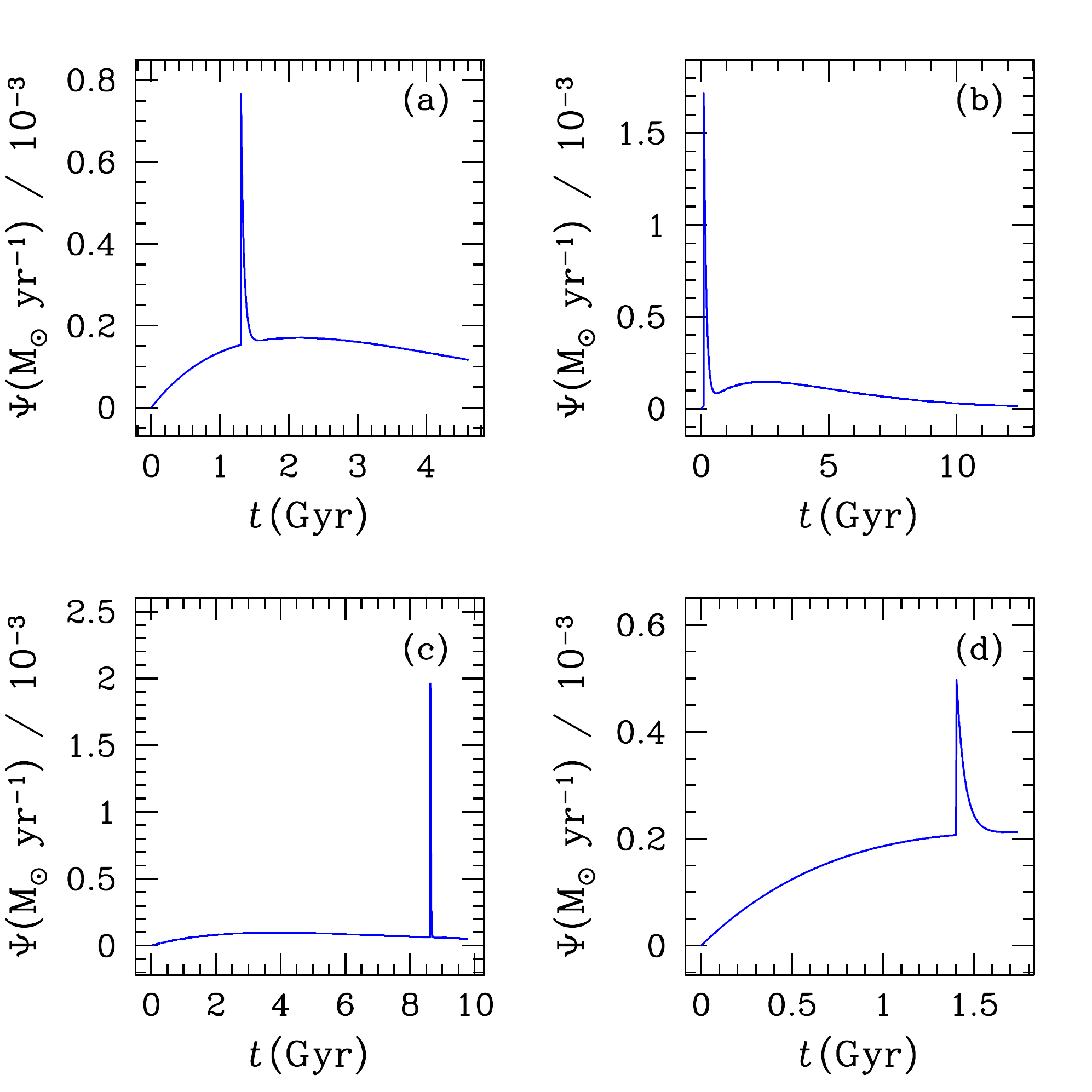}
\caption[f4]{Examples of star formation histories included in our CSP libraries (see equation~\ref{SFH_delayed}).
Parameters are given in Table~\ref{tbl-1}.
~\label{fig4}}
\end{figure}

We model the SFH of our CSP libraries with a delayed, or \`a la~\citet{san86}, prescription,
including an additional burst~\citep[e.g.,][]{mal18,boq19}:
\begin{equation}~\label{SFH_delayed}
  \Psi(t)= t/\tau_{0}^2 \exp(-t/\tau_{0}) + k\exp(-t/\tau_{1}),
\end{equation}
where $\Psi(t)$ is the star formation rate (SFR),
and $t$ is the age of the stellar population, with a maximum value of T$_{\rm form}$, the
age of the oldest stars at the present time. T$_{\rm form}$ is uniformly distributed between
0.1 and 13.7 Gyr, with a precision of 1 Myr.
$\tau_{0}$ is the $e$-folding timescale of the main stellar population, which is distributed
according to the probability density function $P(\gamma)=1-\tanh(8\gamma-6)$~\citep{dac08},
where $\gamma=1/\tau_{0}$, and $0<\gamma<1$ Gyr$^{-1}$ (see Figure~\ref{fig3}).
The constant $k$ is defined as $k\propto\frac{f}{1-f}$, where $f$ if
the fraction of stellar mass formed in the second burst, relative to the total stellar mass ever
formed~\citep[cf.][]{boq19}. $f$ is distributed according to
the probability density function $P(f)=\exp(-f/c)$, where $c=-1/\ln(0.001)$ and $0<f<1$.
$\tau_{1}$ is the $e$-folding timescale of the late burst of star formation, uniformly distributed between 10 and 100 Myr.
The age of this second burst, B$_{\rm form}$, is uniformly distributed between 1 Myr and T$_{\rm form}$,
i.e., with the constraint B$_{\rm form}<$ T$_{\rm form}$ and a precision of 1 Myr.
When $t <$ (T$_{\rm form}$ - B$_{\rm form})$, we have $k=0$.
With these distributions, $\sim40\%$ of the galaxies experience the second burst in the last 2 Gyr.
In Figure~\ref{fig4} and Table~\ref{tbl-1}, we show four examples of the SFHs in our libraries.

\begin{deluxetable}{cccccc}
\tabletypesize{\scriptsize}
\tablecaption{Examples of SFH parameters~\label{tbl-1}}
\tablewidth{0pt}
\tablehead{
\colhead{Panel label} &
\colhead{$\tau_{0}$} &
\colhead{$\tau_{1}$} &
\colhead{$f$} &
\colhead{T$_{\rm form}$} &
\colhead{B$_{\rm form}$}\\
(see Figure~\ref{fig4}) & (Myr) & (Myr) & ~~ & (Myr) & (Myr)
}
\startdata

(a) & 2155.17 & 42.36  & 0.04  & ~4614  & ~3310  \\
(b) & 2506.27 & 94.12  & 0.144 & 12417  & 12310  \\
(c) & 3846.15 & 12.48  & 0.033 & ~9807  & ~1172  \\
(d) & 1739.13 & 45.60  & 0.048 & ~1745  & ~~341  \\

\enddata
\tablecomments{
Col. 2: $e$-folding timescale of the main (delayed) burst of star formation.
Col. 3: $e$-folding timescale of the second (decaying exponential) burst.
Col. 4: mass fraction of the second burst.
Col. 5: age of the main stellar population.
Col. 6: age of the second burst.
}
\end{deluxetable}

\subsection{Metallicity}~\label{sec_metallicity}

For the stellar metallicity, {\tt{CIGALE}} allows to choose between four possible values,
$Z=$ 0.001, 0.01, 0.02, and 0.04, when using the m2005 SSP. On the other hand, bc03
SSP include six values ($Z=$ 0.0001, 0.0004, 0.004, 0.008, 0.02,
and 0.05). To build our libraries, we chose the 4 metallicities of the bc03
SSP closest to the m2005 $Z$ values, i.e., $Z=$ 0.004, 0.008, 0.02, and 0.05.
The same $Z$ values were taken for cb07, bc03-2016, and cb07-2016.
The metallicity is constant for a given SFH~\citep[cf.][their equation 1]{bru03}.
We distribute the four $Z$ possibilities uniformly among the $5\times10^{4}$ models of each library.

\subsection{Dust attenuation}~\label{sec_dust_att}

Dust attenuation was treated as in the two-component model of~\citet{cha00}.\footnote{
Modifications to this model and comparisons with other attenuation laws used to fit the SED of galaxies
can be found in~\citet{lof17,bua18,bua19}.}
In this model, the starlight of the stellar populations with ages below an age threshold, $t_{0}$,
is attenuated by the dust in the stellar birth clouds (BC), and also by the dust in the diffuse interstellar medium (ISM).
The radiation of stellar populations with ages above $t_{0}$ is only attenuated by dust in the diffuse ISM.
Usually $t_{0}=1\times10^{7}$~yr, the typical lifetime of a molecular cloud.
\citet{bru03} adopt the~\citet{cha00} model with attenuation curves of the form $\hat{\tau}_{\lambda}\propto\lambda^{\delta}$,
where $\hat{\tau}_{\lambda}$ is the effective absorption optical depth, 
$\delta_{\rm{ISM}}=-0.7$ for the ISM, and $\delta_{\rm{BC}}=-0.7$ for the birth clouds.
In this work we use $\delta_{\rm{ISM}}=-0.7$ and $\delta_{\rm{BC}}=-1.3$,
the latter being the mean value of the Milky Way, the Small MC,
and the Large MC extinction curves~\citep{cha00}.\footnote{These are the default values adopted in {\tt{CIGALE}},
and in the {\tt{MAGPHYS}} model package~\citep{dac08}.}

In {\tt{CIGALE}}, the main input parameters for the~\citet{cha00} model are the
$V$-band attenuation in the ISM, $A_{V}^{\rm{ISM}}$;
the $\mu$ parameter, defined as $\mu=A_{V}^{\rm{ISM}}/(A_{V}^{\rm{BC}} + A_{V}^{\rm{ISM}})$,
where $A_{V}^{\rm{BC}}$ is the $V$-band attenuation in the birth clouds;
and $\delta_{\rm{ISM}}$, $\delta_{\rm{BC}}$.
In order to compute the $A_{V}^{\rm{ISM}}$ distribution, we first calculate the distribution
of the total effective $V$-band optical depth seen by young stars, $\hat{\tau}_{V}$. This is
distributed according to the probability density function $P(\hat{\tau}_{V})=1-\tanh(1.5\hat{\tau}_{V}-6.7)$~\citep{dac08}.
Subsequently, we calculate the $\mu$ distribution by using $P(\mu)=1-\tanh(8\mu-6)$~\citep{dac08}.
The $A_{V}^{\rm{ISM}}$ probability distribution is then computed as $A_{V}^{\rm{ISM}}=1.086\mu\hat{\tau}_{V}$,\footnote{
This equation can easily be obtained from $L_{\lambda}=L_{\lambda}^{0}\exp^{-\hat{\tau}_{\lambda}}$,
$A_{\lambda} = -2.5\log_{10}\frac{L_{\lambda}}{L_{\lambda}^{0}}$
(where $L_{\lambda}^{0}$ is the unattenuated luminosity at wavelength $\lambda$),
and $\hat{\tau}_{V}^{\rm{ISM}}=\mu\hat{\tau}_{V}$.}
for each $\mu$ and $\hat{\tau}_{V}$ pair of values.

\subsection{Dust emission}

To take into account the dust emission, we select the~\citet{dal14} model implemented in {\tt{CIGALE}}.
This model has two parameters. The first parameter, $\alpha_{\rm SF}$, controls the infrared spectral shapes observed
for star forming galaxies. The $\alpha_{\rm SF}$ parameter is defined from $dM_{\rm{d}}\propto~U^{-\alpha_{\rm SF}}dU$,
where $M_{\rm{d}}$ is the dust mass heated by a radiation field of intensity $U$. The larger the $\alpha_{\rm SF}$ value,
the lower the dust grain temperatures~\citep{nol09}. We distribute $\alpha_{\rm SF}$ uniformly in the range
$0.0625~\leq~\alpha_{\rm SF}~\leq~4.0$, in steps of $\Delta\alpha_{\rm SF}=0.0625$.

The second parameter in the~\citet{dal14} model quantifies the fractional contribution
of an active galactic nucleus (AGN) to the mid-infrared emission. 
Any AGN luminosity contribution to these bands, however, 
would be limited to one or two pixels in the center of the disks.
These central pixels would have a bigger photometric uncertainty compared with the average disk pixels,
and can be excluded by quantifying and applying a cut in the errors~\citep[see, e.g.,][]{mar17}.
For these reasons, we set the AGN fraction to zero.

\section{Examining the libraries}~\label{sec_examining_libraries}

\begin{figure*}
\centering
\epsscale{2.0}
\plotone{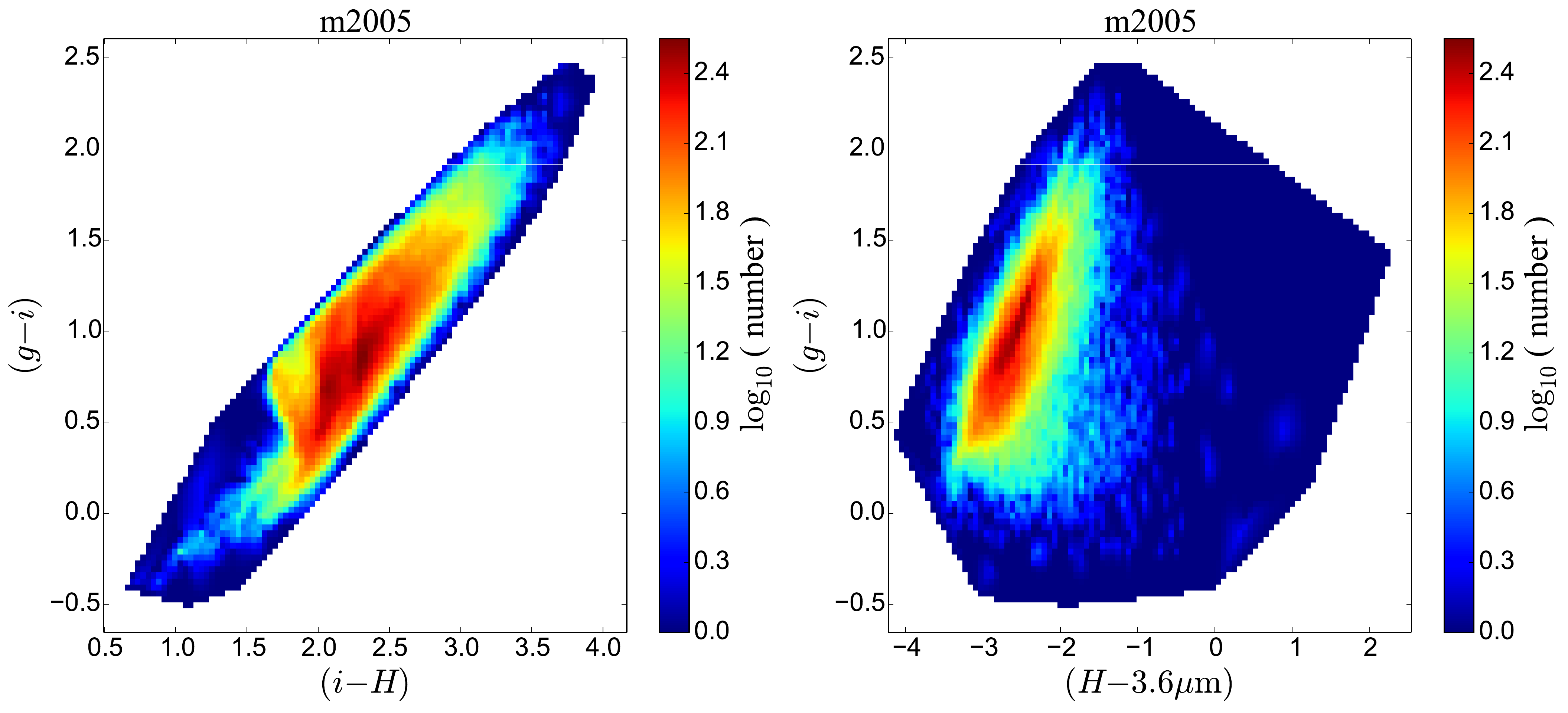}
\caption[f5]{Color-color diagrams for the m2005 CSP library.
{\it Left panel}: $(g-i)$ vs.\ $(i-H)$.
{\it Right panel}: $(g-i)$ vs.\ $(H-3.6\micron)$. 
The colors indicate the number of models in each color bin (0.05 mag$^{2}$).
$g$, $i$, and $3.6\micron$ bands in AB mag, $H$-band referenced to Vega.
~\label{fig5}}
\end{figure*}

\begin{figure*}
\centering
\epsscale{2.0}
\plotone{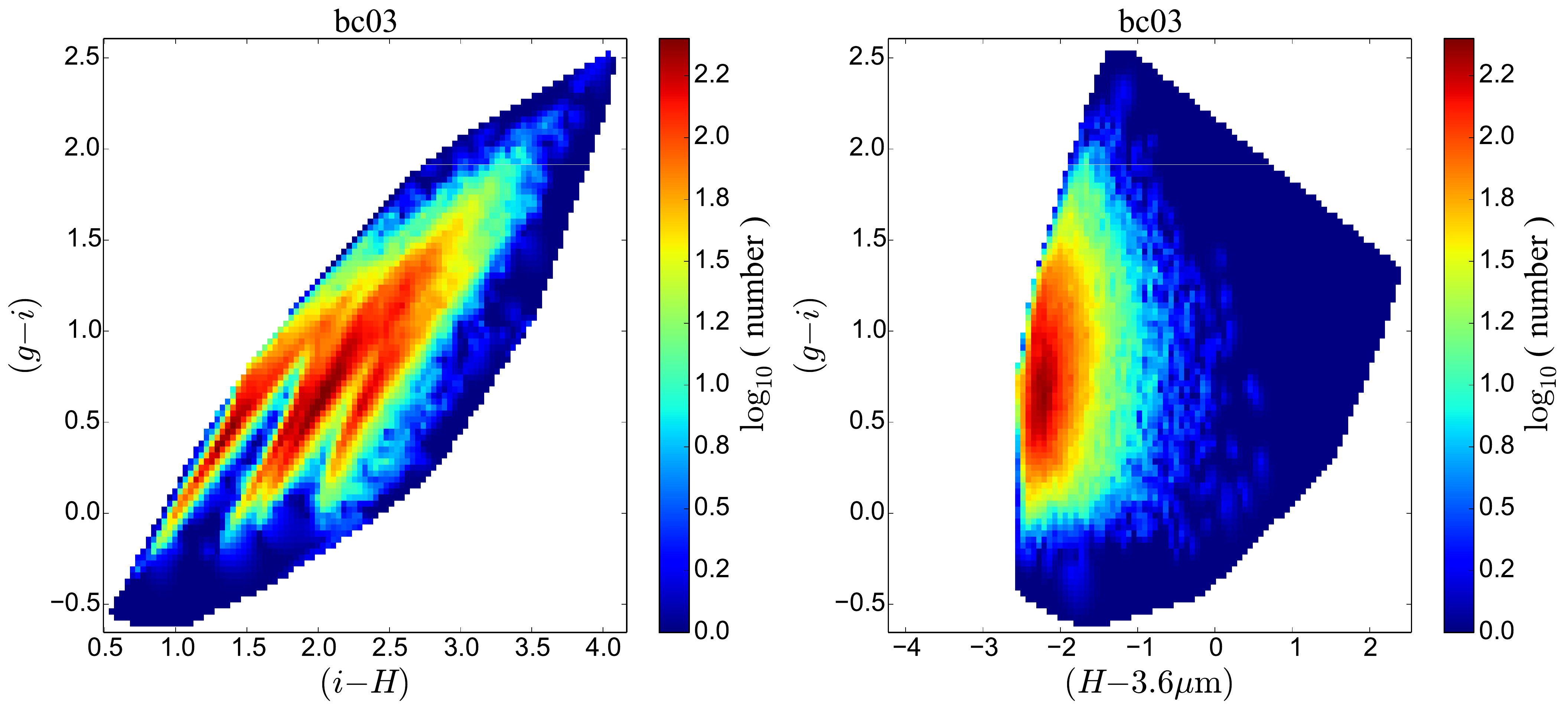}
\caption[f6]{Color-color diagrams for the bc03 CSP library.
Same labels as in Figure~\ref{fig5}.
~\label{fig6}}
\end{figure*}

\begin{figure*}
\centering
\epsscale{2.0}
\plotone{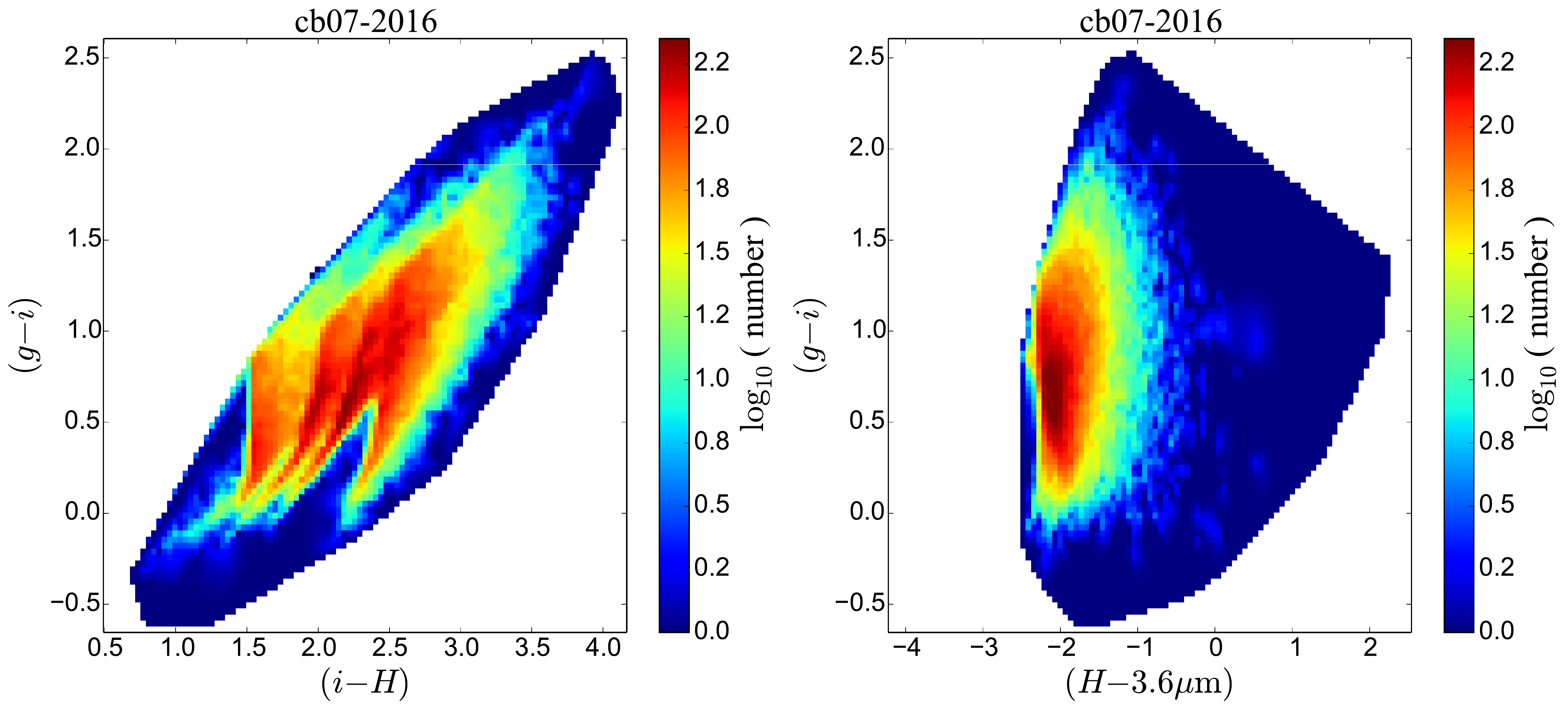}
\caption[f7]{Color-color diagrams for the cb07-2016 CSP library.
Same labels as in Figure~\ref{fig5}.
~\label{fig7}}
\end{figure*}

\begin{figure*}
\centering
\epsscale{2.0}
\plotone{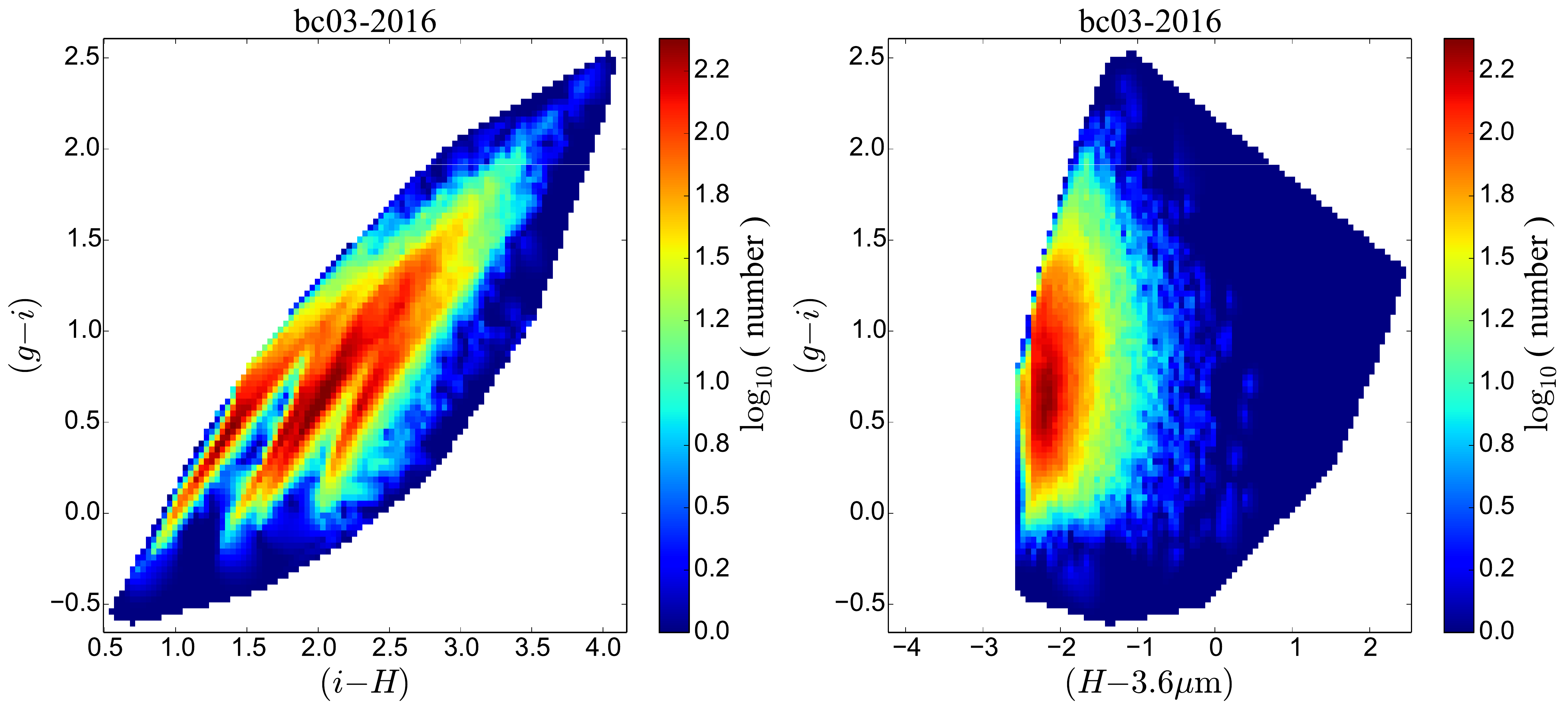}
\caption[f8]{Color-color diagrams for the bc03-2016 CSP library.
Same labels as in Figure~\ref{fig5}.
~\label{fig8}}
\end{figure*}

In this section we examine the output libraries from Section~\ref{sec_build_libs}.
In Figures~\ref{fig5},~\ref{fig6},~\ref{fig7}, and~\ref{fig8}, we show color-color diagrams
for the m2005, bc03, cb07-2016, and bc03-2016 CSP libraries, respectively.
We plot $(g-i)$ vs.\ $(i-H)$, and $(g-i)$ vs.\ $(H-3.6\micron)$.
In these plots, the major difference between m2005 and bc03 is in the $(H-3.6\micron)$ color. 
The cb07-2016 and bc03-2016 color-color diagrams show patterns similar to bc03.

\begin{figure}
\centering
\epsscale{1.0}
\plotone{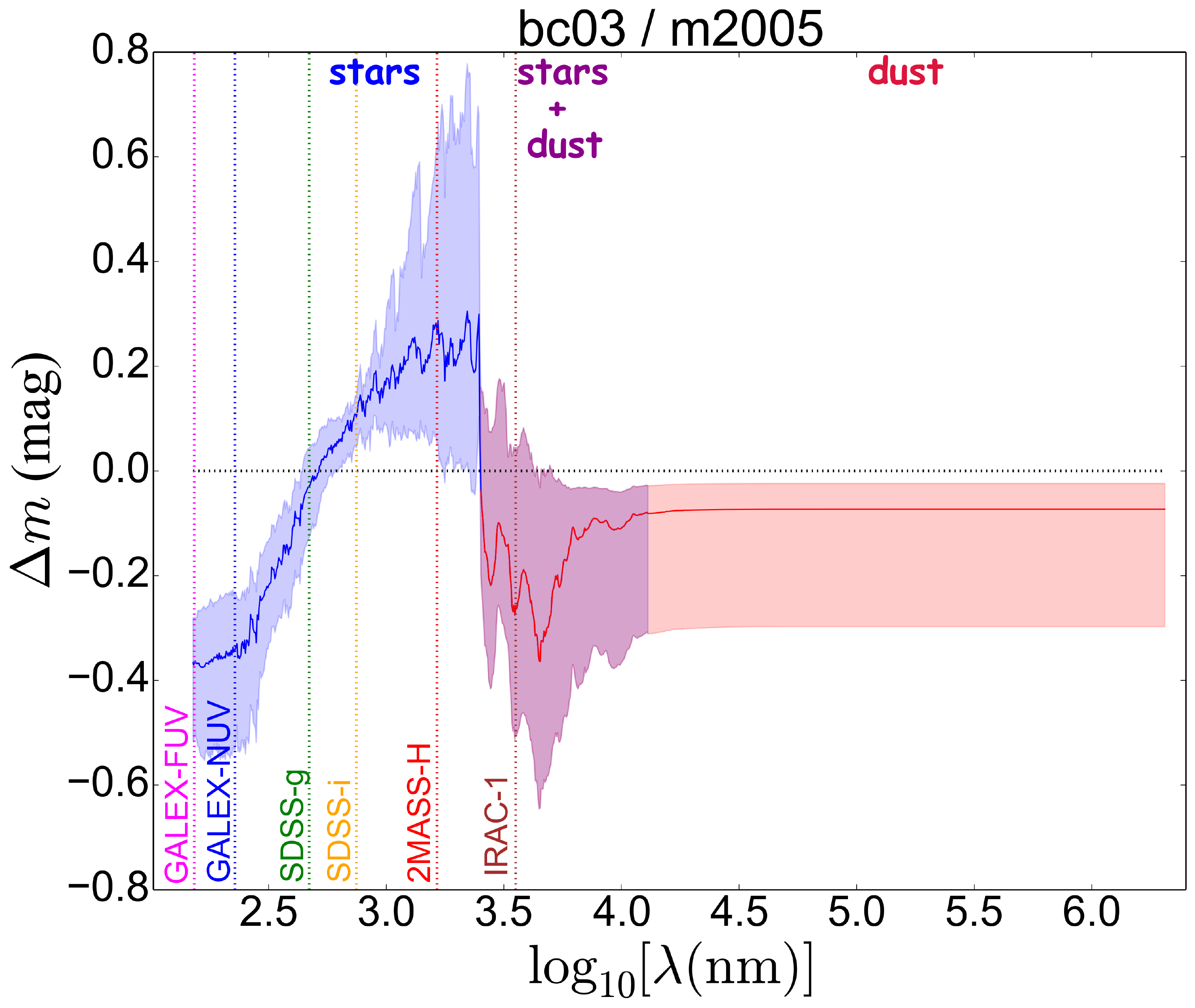}
\caption[f9]{Median luminosity ratio, $\Delta m$, in mag, as a function of $\log_{10}$ wavelength,
$\lambda$, in nm, between bc03 and m2005 spectra of the CSP libraries, as obtained with {\tt{CIGALE}} (see Section~\ref{sec_build_libs}).
The 16th and 84th percentiles are indicated as shaded regions, below and above the 50th percentile (continuous line), respectively.
Only stellar emission in blue, star plus dust region in magenta, pure dust emission in red.
Vertical dotted lines mark effective wavelengths of various filters as reference.
~\label{fig9}}
\end{figure}

\begin{figure}
\centering
\epsscale{1.0}
\plotone{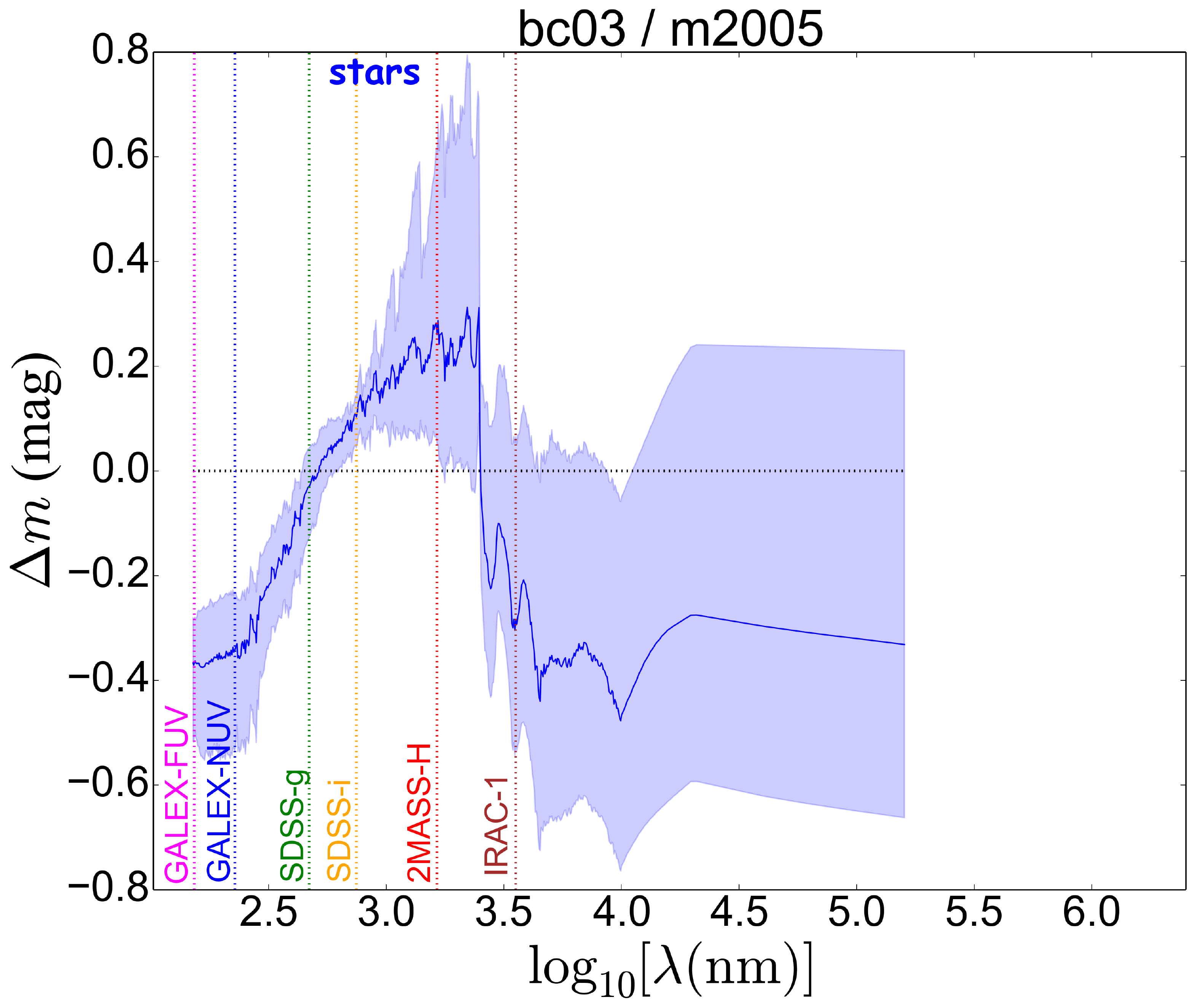}
\caption[f10]{Median luminosity ratio without dust emission between bc03 and m2005 spectra.
Compare with Figure~\ref{fig9}.
~\label{fig10}}
\end{figure}

In Figure~\ref{fig9} we show the median luminosity ratio between the bc03 and m2005 spectra.
To make this plot, we take the luminosity of each bc03 spectrum and divide it
by the luminosity of the m2005 spectrum with similar parameters.\footnote{
The parameters are quantitatively identical with the exception of the metallicity.
As mentioned in Section~\ref{sec_metallicity},
the metallicities of the m2005 and bc03 models are slightly different.}
We then use equation~\ref{eq_delmag1} to compute $\Delta m$ at each $\lambda$.
We end up with $5\times10^{4}$ $\Delta m$ vs.\ $\lambda$ curves,
and compute the median (P$_{50}$), the 16th (P$_{16}$), and the 84th (P$_{84}$) percentiles\footnote{
P$_{50}$, P$_{16}$, and P$_{84}$ are equivalent to the mean ($\bar{x}$), $-1\sigma$ and $1\sigma$,
respectively, of a normal distribution.} of this set of curves.
In the plot, the solid line indicates the median value, P$_{16}$ is shown as the shaded region below P$_{50}$,
and P$_{84}$ is the shaded region above P$_{50}$.
We also indicate the spectral regions where we have only stars (blue region, $\lambda\lesssim~2.5\micron$),
star plus dust (magenta region,  $2.5\micron\lesssim\lambda\lesssim~13\micron$),
and pure dust emission (red region, $\lambda\gtrsim~13\micron$).
As expected, due to TP-AGB stars, in the NIR (see 2MASS $H$-band in the plot) the median luminosity ratio indicates a higher
luminosity for m2005 models than for bc03. We can also appreciate that for the $FUV$ and $NUV$-bands, the m2005 models
produce less luminosity than bc03, and the same happens for the ``stars plus dust'' region ($2.5\micron\lesssim\lambda\lesssim~13\micron$).
In the pure dust region the median luminosity ratio is $\Delta m\sim-0.073$ mag.
For comparison purposes, we show in Figure~\ref{fig10} the median luminosity ratio,
without the dust emission, between the bc03 and m2005 spectra.
An identical plot is obtained with the {\tt{GALAXEV}} software~\citep{bru03},
adopting $\delta_{\rm{ISM}}=-0.7$ and $\delta_{\rm{BC}}=-1.3$ (see Section~\ref{sec_dust_att}). 

\begin{figure}
\centering
\epsscale{1.0}
\plotone{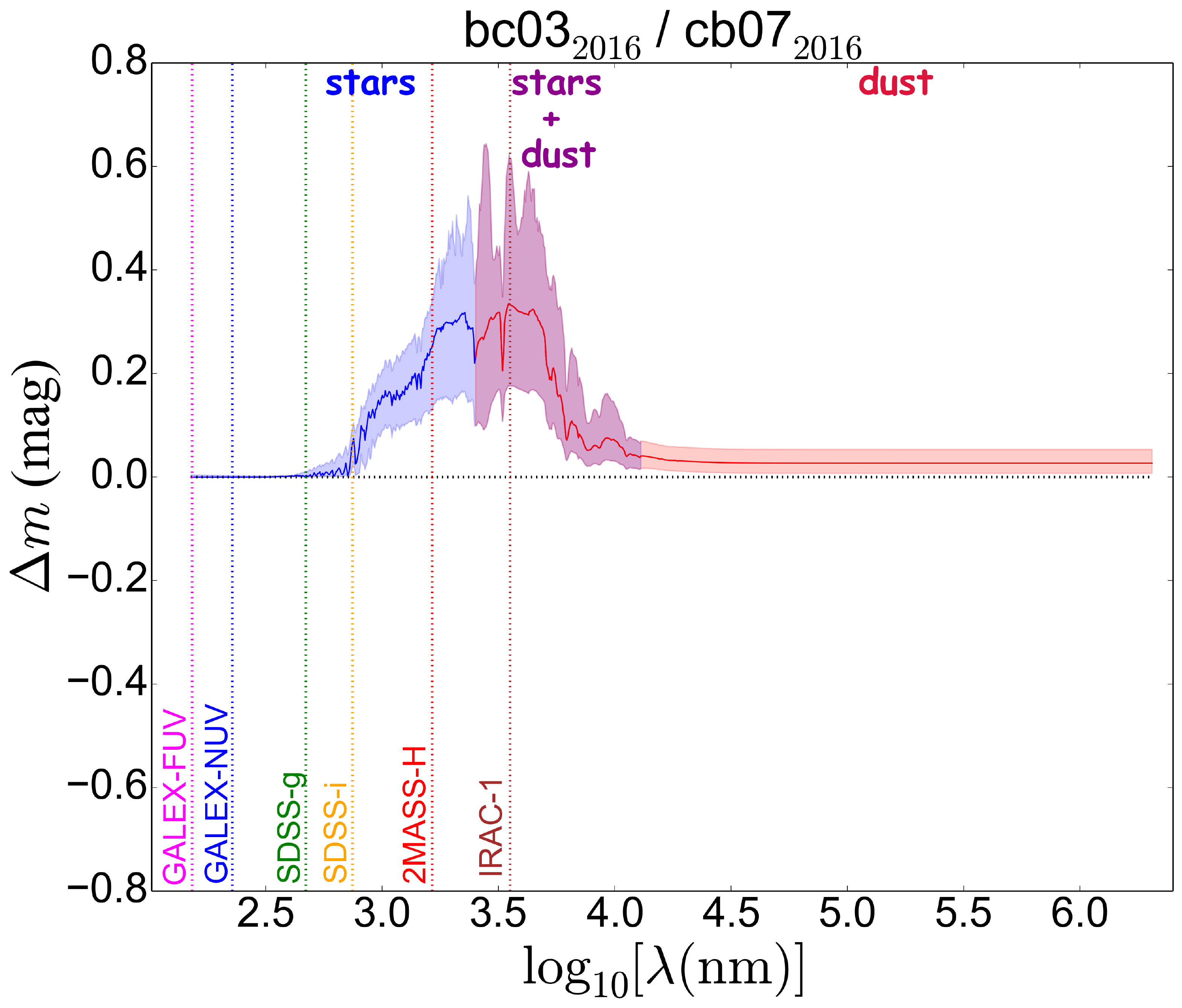}
\caption[f11]{Median luminosity ratio between bc03-2016 and cb07-2016 spectra, as obtained with {\tt{CIGALE}}.
~\label{fig11}}
\end{figure}

\begin{figure}
\centering
\epsscale{1.0}
\plotone{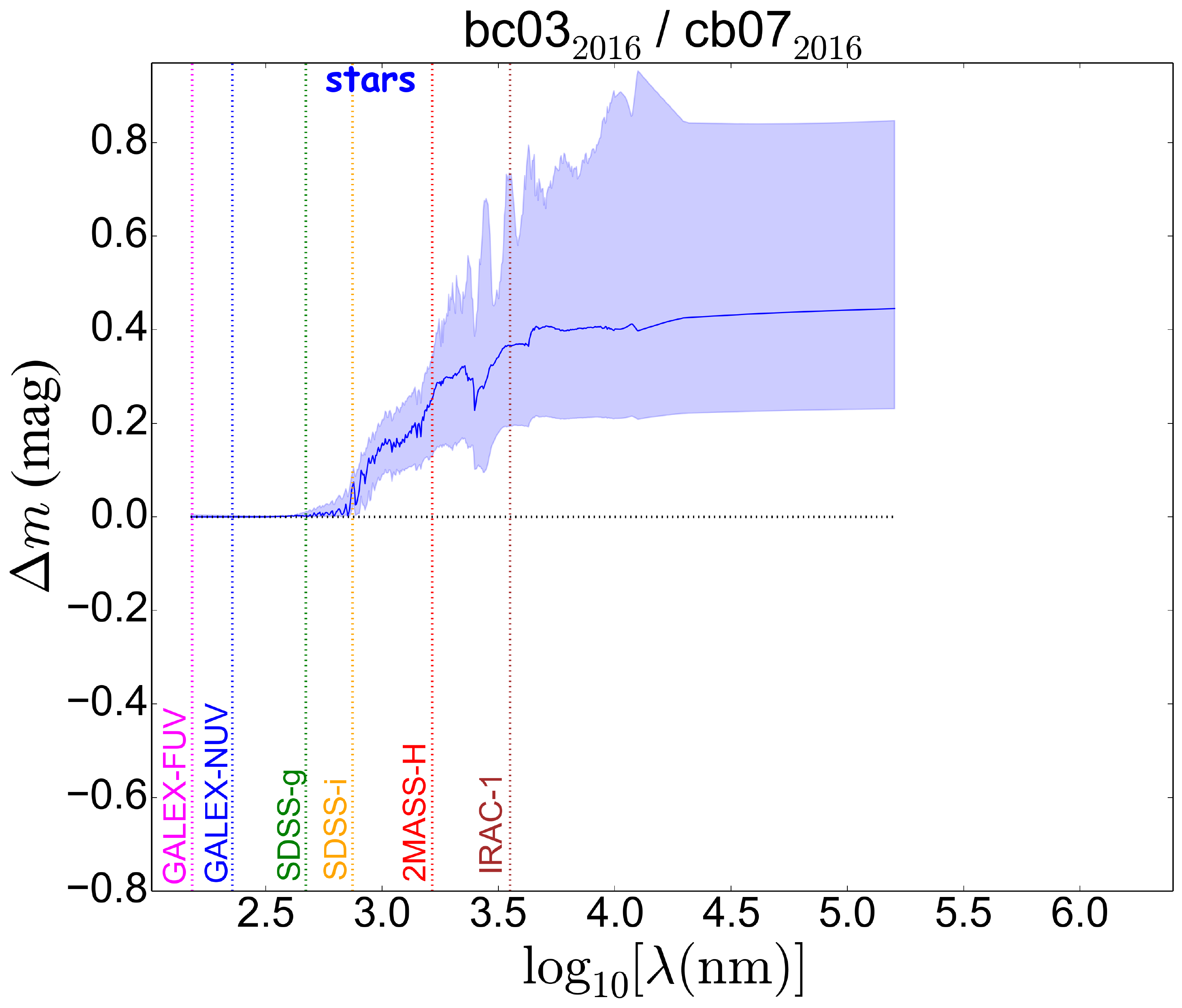}
\caption[f12]{Median luminosity ratio without the dust emission between bc03-2016 and cb07-2016 spectra.
Compare with Figure~\ref{fig11}.
~\label{fig12}}
\end{figure}

Figure~\ref{fig11} shows the median luminosity ratio between bc03-2016 and cb07-2016 spectra.
We can appreciate the TP-AGB stars luminosity contribution in the NIR, where cb07-2016 models radiate more
than bc03-2016. Conversely from bc03 versus m2005, there is no luminosity difference between the models
in the $FUV$ and $NUV$-bands. Also dissimilarly to Figure~\ref{fig9}, in the ``stars plus dust'' region,
the median luminosity ratio indicates that cb07-2016 is brighter than bc03-2016.
In the pure dust region ($\lambda\gtrsim~13\micron$),
the ratio has a positive value of $\Delta m\sim0.027$ mag.
Figure~\ref{fig12} shows the median luminosity ratio, without the dust emission,
between bc03-2016 and cb07-2016 spectra.
Now, cb07-2016 is brighter than bc03-2016 for all NIR and mid-IR wavelengths.
Once again, an identical plot is obtained with {\tt{GALAXEV}}.

\begin{figure}
\centering
\epsscale{1.0}
\plotone{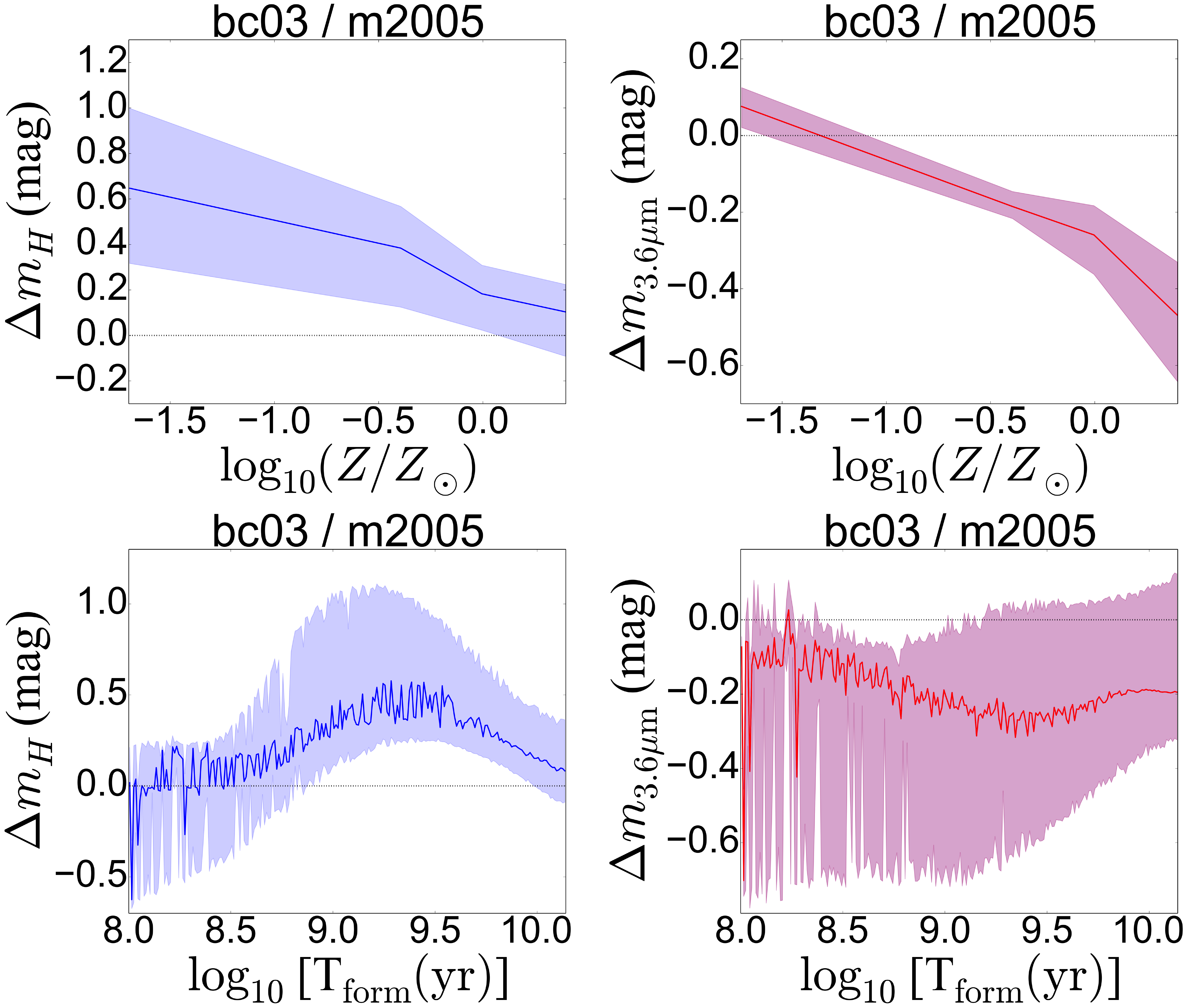}
\caption[f13]{Median luminosity ratio between bc03 and m2005 spectra, obtained with {\tt{CIGALE}}.
{\it Left}: $H$-band, $\Delta m_{H}$ (blue shaded region);
{\it right}: $3.6\micron$ band, $\Delta m_{3.6\micron}$ (magenta shaded region).
{\it Top}: $\Delta m$ as a function of $\log_{10}$($Z/Z_{\sun}$);
{\it bottom}: as a function of $\log_{10}$[T$_{\rm form}$(yr)].
~\label{fig13}}
\end{figure}

\begin{figure}
\centering
\epsscale{1.0}
\plotone{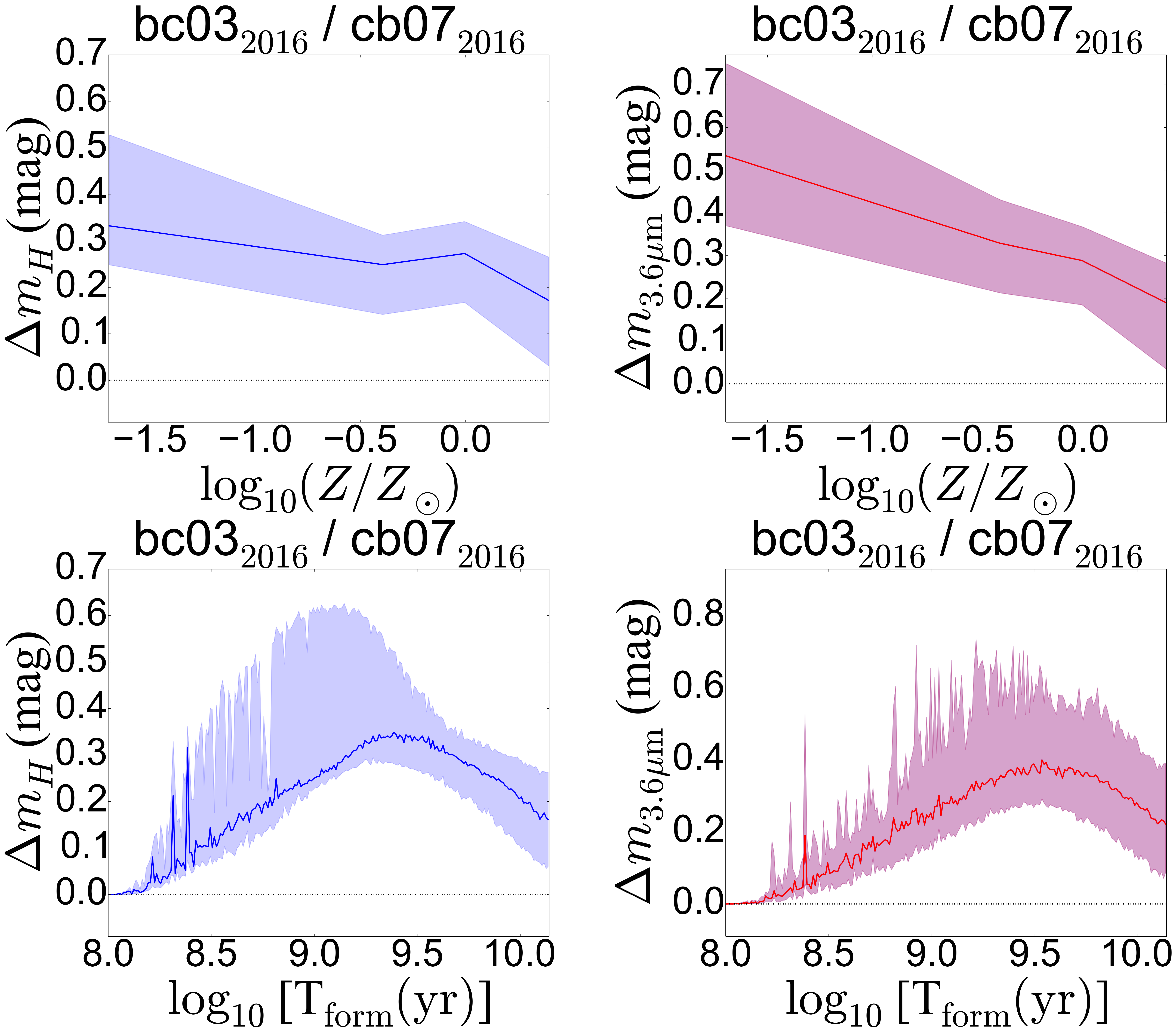}
\caption[f14]{Same as Figure~\ref{fig13} for the median luminosity ratio between
bc03-2016 and cb07-2016 spectra, as obtained with {\tt{CIGALE}}.
~\label{fig14}}
\end{figure}

In Figure~\ref{fig13} we show the median luminosity ratio between the bc03 and m2005 spectra,
for the $H$-band, $\Delta m_{H}$ (left panels), and the $3.6\micron$ band, $\Delta m_{3.6\micron}$ (right panels),
as a function of metallicity ($\log_{10}[Z/Z_{\sun}]$, top panels) and age ($\log_{10}$[T$_{\rm form}$(yr)], bottom panels).
$\Delta m_{H}$ has the highest value for stellar populations with ages $\sim1$ Gyr.
As mentioned earlier, TP-AGB stars are expected to be brighter in ``heavy'' models for
stellar populations in the age range 0.2-2 Gyr, being $\sim1$ Gyr the age of maximum contribution~\citep{mou02,mara06}.
This behavior is more prominent at low metallicities, as shown in the top left panel of Figure~\ref{fig13}.\footnote
{For SSP, a higher contribution of carbon stars is found in metal-poor stellar populations~\citep{mou03}.}
Figure~\ref{fig14} displays $\Delta m_{H}$ and $\Delta m_{3.6\micron}$ (left and right panels, respectively)
for the bc03-2016 and cb07-2016 spectra. The metallicity plots (top panels) show a behavior similar
to the bc03/m2005 case (Figure~\ref{fig13}, top panels), in the sense that lower metallicities
have higher $\Delta m$. The $\Delta m_{H}$ vs.\ $\log_{10}$[T$_{\rm form}$(yr)] plot (bottom left panel)
is also similar to the bc03/m2005 case, in the sense that $\Delta m_{H}$ has a maximum at $\sim1$ Gyr,
and has lower values for older ages. The only difference is seen in the $\Delta m_{3.6\micron}$ vs.
$\log_{10}$[T$_{\rm form}$(yr)] plot (bottom right panel), where for the bc03/m2005
case (see Figure~\ref{fig13}) there is no clear maximum value.

\clearpage

A comparison of the current stellar mass between TP-AGB ``light'' and ``heavy'' models
with the same parameters yields a ratio of M$^{\rm{light}}_{*}$/M$^{\rm{heavy}}_{*}\approx1$.
Also, the total stellar mass ever formed, i.e., the integral of the SFH,
is the same for both models. This is not the case for the current stellar mass-to-light ratio
in the NIR, $\Upsilon_{*}^{\rm{NIR}}$,
where ``light'' models have on average a 20\%-40\% larger $\Upsilon_{*}^{\rm{NIR}}$~\citep[see also][]{int13}.

\section{Sample of objects}~\label{sec_sample_objects}

\begin{figure*}
\centering
\epsscale{2.0}
\plotone{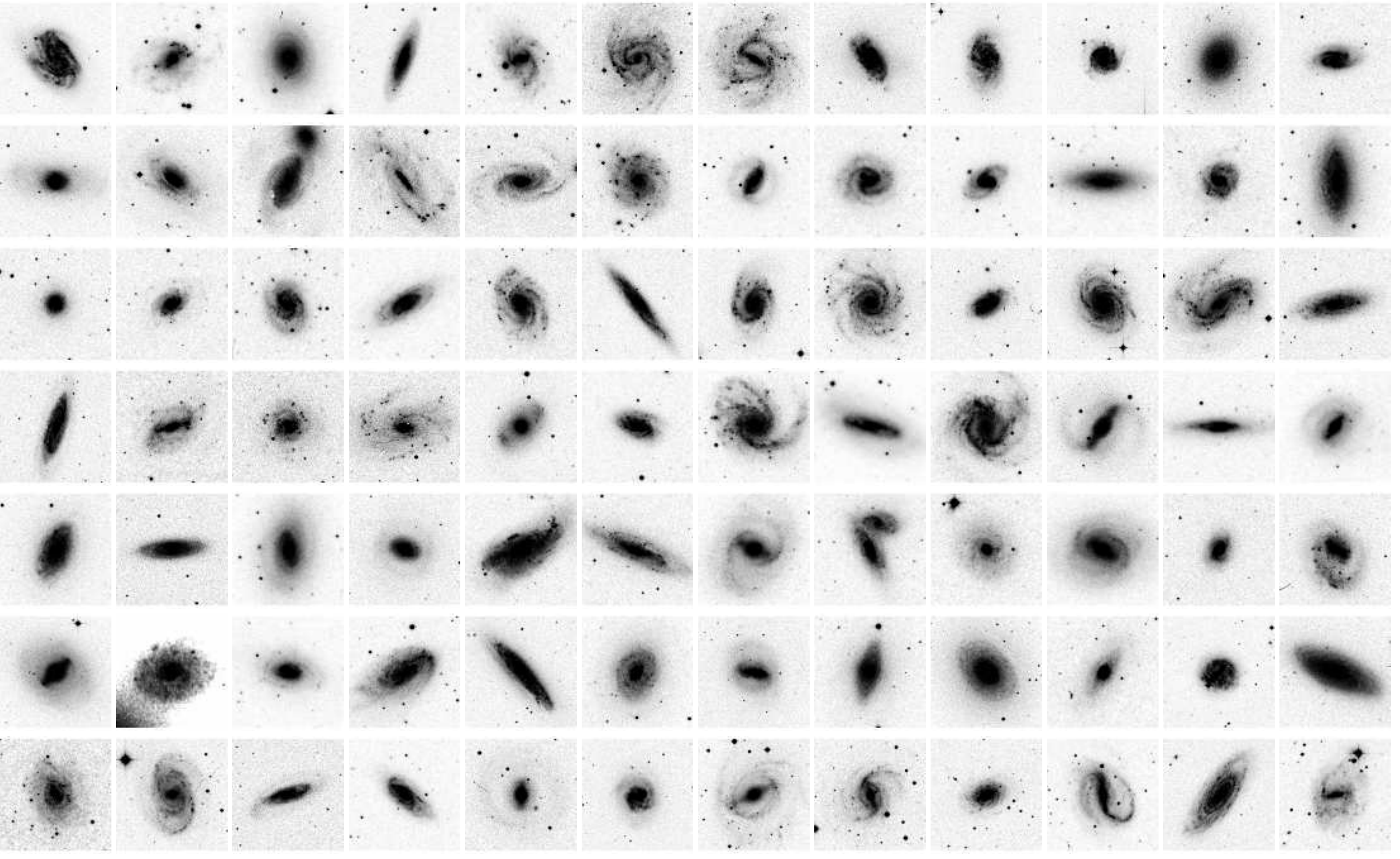}
\caption[f15]{Sample, 84 spiral galaxies. Images from the Digitized Sky Survey, DSS (blue).
Objects follow the same order as in Table~\ref{tbl-2}, in such a way that the top left image is NGC~157,
NGC~428 is to the right of NGC~157, and NGC~7741 is in the bottom right.
Foreground and background objects were masked for the analysis.
~\label{fig15}}
\end{figure*}

\begin{figure}
\centering
\epsscale{1.0}
\plotone{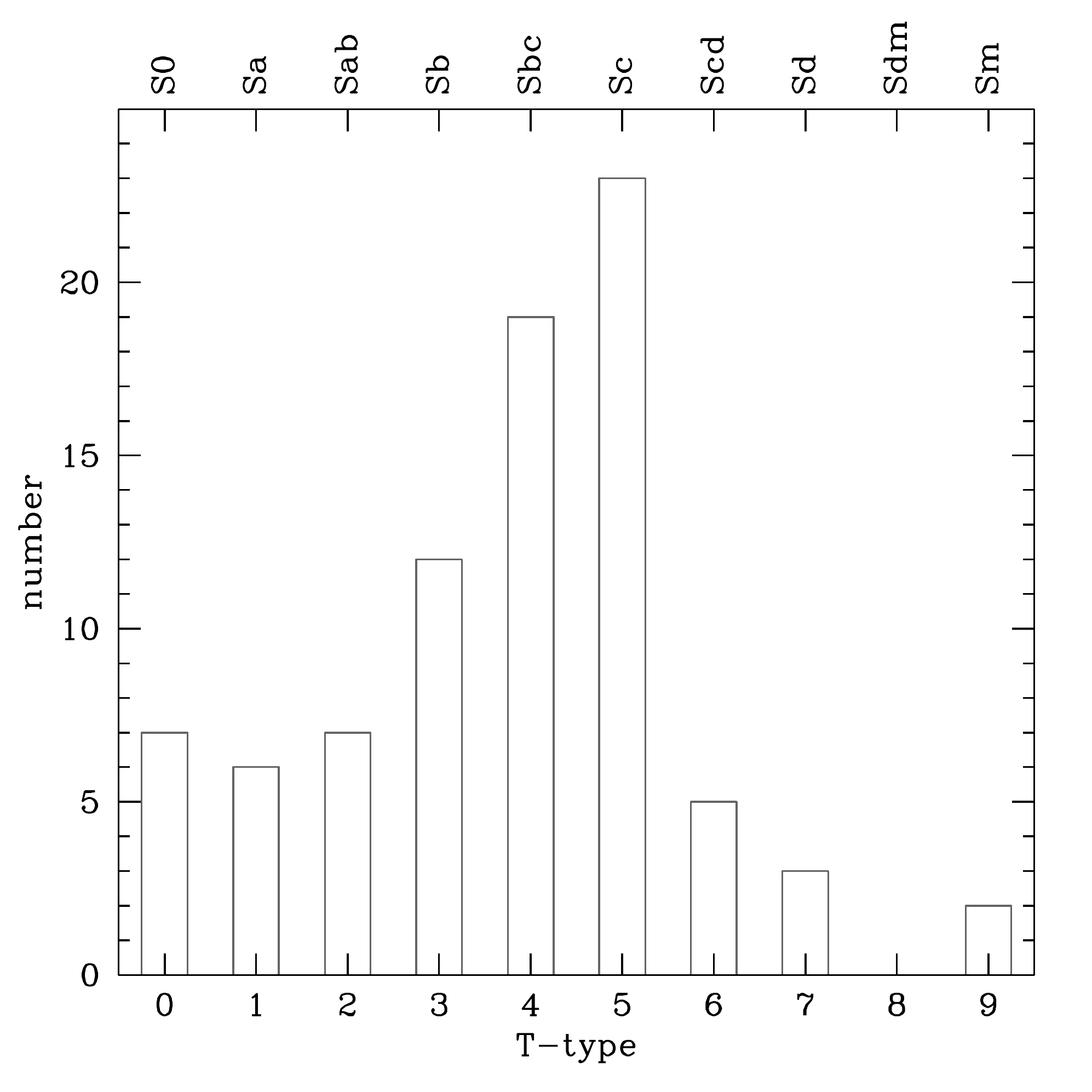}
\caption[f16]{Bar chart of the Hubble types in our sample of galaxies (see Table~\ref{tbl-2}).
~\label{fig16}}
\end{figure}

For this investigation, we use a sample of 84 nearby disk galaxies listed in 
Table~\ref{tbl-2} and shown in Figure~\ref{fig15}.
In Figure~\ref{fig16} we present a bar chart of the Hubble types in our sample.
For our analysis we use photometric images in the $g$ and $i$-bands from the Sloan Digital Sky Survey~\citep[SDSS DR8,][]{aih11},
$H$-band images from the Ohio State University Bright Spiral Galaxy Survey~\citep[OSUBSGS,][]{esk02},
and $3.6\micron$ images from the Spitzer Survey of Stellar Structure in Galaxies~\citep[S$^4$G,][]{she10}.

\begin{deluxetable}{llcrrr}
\tabletypesize{\scriptsize}
\tablecaption{Galaxy sample~\label{tbl-2}}
\tablewidth{0pt}
\tablehead{
\colhead{Name} &
\colhead{RC3~type} &
\colhead{T-type} &
\colhead{Dist~(Mpc)} &
\colhead{$q=b/a$} &
\colhead{P.A.} 
}
\startdata

NGC~157		&	SAB(rs)bc	&	4.0	&	22.6	$\pm$	1.6 &        0.724 	$\pm$	0.008	&         36.5  	$\pm$  1.2	  \\
NGC~428		&	SAB(s)m		&	9.0	&	15.9	$\pm$	1.1 &	     0.750 	$\pm$	0.013	&        100.5  	$\pm$  1.8	  \\
NGC~488		&	SA(r)b		&	3.0	&	30.4	$\pm$	2.1 &        0.770 	$\pm$	0.010	&          5.6  	$\pm$  1.9	  \\
NGC~779		&	SAB(r)b		&	3.0	&	18.5	$\pm$	1.3 & $\star$0.333 	$\pm$	0.020	& $\star$164.0          $\pm$  2.0 	  \\
NGC~864		&	SAB(rs)c	&	5.0	&	20.9	$\pm$	1.5 &        0.842 	$\pm$	0.008	&         28.7  	$\pm$  4.7	  \\
NGC~1042	&	SAB(rs)cd	&	6.0	&	18.1	$\pm$	1.3 &        0.781 	$\pm$	0.013	&          4.5 	        $\pm$  1.9	  \\
NGC~1073	&	SB(rs)c		&	5.0	&	16.1	$\pm$	1.1 &        0.875 	$\pm$	0.013	&          1.0  	$\pm$  4.2	  \\
NGC~1084	&	SA(s)c		&	5.0	&	18.6	$\pm$	1.3 &        0.753 	$\pm$	0.021	&         57.2  	$\pm$  1.1	  \\
NGC~1087	&	SAB(rs)c	&	5.0	&	20.1	$\pm$	1.4 &        0.609 	$\pm$	0.012	&          2.8  	$\pm$  0.1	  \\
NGC~1309	&	SA(s)bc:	&	4.0	&	28.3	$\pm$	2.0 &        0.926 	$\pm$	0.023	&         65.3  	$\pm$ 10.2	  \\
NGC~2775	&	SA(r)ab		&	2.0	&	21.4	$\pm$	1.5 &        0.801 	$\pm$	0.009	&        163.5  	$\pm$  1.8	  \\
NGC~2964	&	SAB(r)bc:	&	4.0	&	23.2	$\pm$	1.6 &        0.566 	$\pm$	0.004	&         96.8  	$\pm$  0.4	  \\
NGC~3166	&	SAB(rs)0/a	&	0.0	&	22.0	$\pm$	1.5 &        0.586 	$\pm$	0.013	&         82.5  	$\pm$  1.2	  \\
NGC~3169	&	SA(s)a~pec	&	1.0	&	19.9	$\pm$	1.4 &        0.776 	$\pm$	0.020	&         56.4  	$\pm$  2.7	  \\
NGC~3227	&	SAB(s)a~pec	&	1.0	&	20.3	$\pm$	1.4 &        0.661 	$\pm$	0.013	&        153.1  	$\pm$  1.0	  \\
NGC~3319	&	SB(rs)cd	&	6.0	&	~3.29   $\pm$	0.93&        0.554 	$\pm$	0.012	&         32.8  	$\pm$  1.4	  \\
NGC~3338	&	SA(s)c		&	5.0	&	23.2	$\pm$	1.6 &        0.495 	$\pm$	0.008	&         93.6  	$\pm$  1.3	  \\
NGC~3423	&	SA(s)cd		&	6.0	&	14.1	$\pm$	1.0 &        0.769 	$\pm$	0.012	&         31.2  	$\pm$  1.1	  \\
NGC~3504	&	(R)SAB(s)ab	&	2.0	&	27.8	$\pm$	1.9 &        0.980 	$\pm$	0.012	&          0.0 	        $\pm$  0.0	  \\
NGC~3507	&	SB(s)b		&	3.0	&	15.0	$\pm$	1.1 &        0.944 	$\pm$	0.056	&         91.9  	$\pm$  0.9	  \\
NGC~3583	&	SB(s)b		&	3.0	&	35.7	$\pm$	2.5 &        0.744 	$\pm$	0.019	&        119.2  	$\pm$  2.3	  \\
NGC~3593	&	SA(s)0/a	&	0.0	&	~5.55   $\pm$	0.39&        0.486 	$\pm$	0.021	&         86.2  	$\pm$  0.8	  \\
NGC~3596	&	SAB(rs)c	&	5.0	&	22.5	$\pm$	1.6 &        0.829 	$\pm$	0.006	&         92.5  	$\pm$  5.3	  \\
NGC~3675	&	SA(s)b		&	3.0	&	14.3	$\pm$	1.0 &        0.494 	$\pm$	0.003	&        178.1  	$\pm$  0.3	  \\
NGC~3681	&	SAB(r)bc	&	4.0	&	24.9	$\pm$	1.7 &        0.901 	$\pm$	0.017	&         34.9  	$\pm$  5.7	  \\
NGC~3684	&	SA(rs)bc	&	4.0	&	22.8	$\pm$	1.6 &        0.704 	$\pm$	0.008	&        119.1  	$\pm$  1.0	  \\
NGC~3686	&	SB(s)bc		&	4.0	&	22.6	$\pm$	1.6 &        0.753 	$\pm$	0.008	&         18.2  	$\pm$  1.0	  \\
NGC~3705	&	SAB(r)ab	&	2.0	&	13.2	$\pm$	0.9 & $\star$0.479 	$\pm$	0.020	& $\star$120.0          $\pm$  2.0 	  \\
NGC~3810	&	SA(rs)c		&	5.0	&	10.7	$\pm$	0.8 &        0.680 	$\pm$	0.007	&         21.4  	$\pm$  1.1	  \\
NGC~3877	&	SA(s)c:		&	5.0	&	17.8	$\pm$	1.3 & $\star$0.296 	$\pm$	0.020	&  $\star$40.0          $\pm$  2.0 	  \\
NGC~3893	&	SAB(rs)c:	&	5.0	&	19.4	$\pm$	1.4 &        0.595 	$\pm$	0.010	&        170.1  	$\pm$  0.3	  \\
NGC~3938	&	SA(s)c		&	5.0	&	15.5	$\pm$	1.1 &        0.914 	$\pm$	0.020	&         37.2  	$\pm$  0.8	  \\
NGC~3949	&	SA(s)bc:	&	4.0	&	15.8	$\pm$	1.1 &        0.904 	$\pm$	0.020	&        103.4  	$\pm$  9.6	  \\
NGC~4030	&	SA(s)bc		&	4.0	&	26.4	$\pm$	1.8 &        0.729 	$\pm$	0.009	&         26.7  	$\pm$  3.1	  \\
NGC~4051	&	SAB(rs)bc	&	4.0	&	~2.91	$\pm$	0.9 &        0.846 	$\pm$	0.154	&        128.0  	$\pm$  1.3	  \\
NGC~4062	&	SA(s)c		&	5.0	&	10.4	$\pm$	0.7 & $\star$0.515 	$\pm$	0.020	& $\star$100.0          $\pm$  2.0 	  \\
NGC~4100	&	(R')SA(rs)bc	&	4.0	&	21.5	$\pm$	1.5 & $\star$0.286 	$\pm$	0.020	& $\star$165.0          $\pm$  2.0 	  \\
NGC~4123	&	SB(r)c		&	5.0	&	27.3	$\pm$	1.9 &        0.677 	$\pm$	0.017	&        125.7  	$\pm$  1.4	  \\
NGC~4136	&	SAB(r)c		&	5.0	&	~6.72   $\pm$	0.48&        0.958 	$\pm$	0.015	&          0.0  	$\pm$  0.0	  \\
NGC~4145	&	SAB(rs)d	&	7.0	&	20.3	$\pm$	1.4 &        0.572 	$\pm$	0.007	&        101.5  	$\pm$  0.5	  \\
NGC~4151	&	(R')SAB(rs)ab:	&	2.0	&	20.0	$\pm$	1.4 &        0.920 	$\pm$	0.000 	&          0.0  	$\pm$  0.0	  \\
NGC~4212	&	SAc:		&	4.5	&     16.3\tn{a}$\pm$	3.8 &        0.663 	$\pm$	0.017	&         75.7  	$\pm$  0.8	  \\
NGC~4254	&	SA(s)c		&	5.0	&     16.5\tn{b}$\pm$	1.1 &        0.868 	$\pm$	0.012	&         57.4  	$\pm$  5.6	  \\
NGC~4293	&	(R)SB(s)0/a	&	0.0	&	14.1	$\pm$	1.0 &        0.463 	$\pm$	0.007	&         65.1  	$\pm$  0.4	  \\
NGC~4303	&	SAB(rs)bc	&	4.0	&	13.6	$\pm$	1.0 &        0.861 	$\pm$	0.011	&        146.9  	$\pm$  1.8	  \\
NGC~4314	&	SB(rs)a		&	1.0	&	17.8	$\pm$	1.3 &        0.959 	$\pm$	0.019	&         61.8  	$\pm$ 15.1	  \\
NGC~4388	&	SA(s)b:~sp	&	3.0	&	41.4	$\pm$	2.9 & $\star$0.339 	$\pm$	0.020	&  $\star$93.0          $\pm$  2.0 	  \\
NGC~4394	&	(R)SB(r)b	&	3.0	&	14.1	$\pm$	1.0 &        0.902 	$\pm$	0.009	&        103.0  	$\pm$  3.7	  \\
NGC~4414	&	SA(rs)c?	&	5.0	&	~9.03   $\pm$	0.64&        0.644 	$\pm$	0.011	&        160.0  	$\pm$  0.9	  \\
NGC~4448	&	SB(r)ab		&	2.0	&	~6.98   $\pm$	0.5 & $\star$0.390 	$\pm$	0.020	&  $\star$95.0          $\pm$  2.0 	  \\
NGC~4450	&	SA(s)ab		&	2.0	&	14.1	$\pm$	1.0 &        0.720 	$\pm$	0.009	&          2.2  	$\pm$  3.0	  \\
NGC~4457	&	(R)SAB(s)0/a	&	0.0	&	13.6	$\pm$	1.0 &        0.883 	$\pm$	0.017	&         80.8  	$\pm$  2.1	  \\
NGC~4490	&	SB(s)d~pec	&	7.0	&	~9.22   $\pm$	0.65&        0.441 	$\pm$	0.004	&        123.5  	$\pm$  0.5	  \\
NGC~4527	&	SAB(s)bc	&	4.0	&	13.5	$\pm$	0.9 &        0.456 	$\pm$	0.007	&         67.1  	$\pm$  0.6	  \\
NGC~4548	&	SB(rs)b		&	3.0	&	~3.68   $\pm$	0.26&        0.744 	$\pm$	0.009	&        153.2  	$\pm$  1.7	  \\
NGC~4568	&	SA(rs)bc	&	4.0	&	13.9	$\pm$	1.0 & $\star$0.508 	$\pm$	0.020	&  $\star$32.0          $\pm$  2.0 	  \\
NGC~4571	&	SA(r)d		&	6.5	&	~2.58   $\pm$	0.19&        0.821 	$\pm$	0.012	&         34.9  	$\pm$  4.6	  \\
NGC~4579	&	SAB(rs)b	&	3.0	&	13.9	$\pm$	1.0 &        0.783 	$\pm$	0.008	&         94.8  	$\pm$  1.3	  \\
NGC~4580	&	SAB(rs)a~pec	&	1.0	&	13.6	$\pm$	1.0 &        0.712 	$\pm$	0.012	&        163.1  	$\pm$  1.5	  \\
NGC~4618	&	SB(rs)m		&	9.0	&	~8.78   $\pm$	0.61&        0.807 	$\pm$	0.008	&         36.6  	$\pm$  0.0	  \\
NGC~4643	&	SB(rs)0/a	&	0.0	&	27.3	$\pm$	1.9 &        0.818 	$\pm$	0.012	&         56.0  	$\pm$  3.1	  \\
NGC~4647	&	SAB(rs)c	&	5.0	&	13.9	$\pm$	1.0 &        0.663 	$\pm$	0.019	&        119.0  	$\pm$  1.2	  \\
NGC~4651	&	SA(rs)c		&	5.0	&	14.0	$\pm$	1.0 &        0.612 	$\pm$	0.127	&         73.1  	$\pm$  1.0	  \\
NGC~4654	&	SAB(rs)cd	&	6.0	&	13.9	$\pm$	1.0 &        0.563 	$\pm$	0.017	&        123.1  	$\pm$  3.2	  \\
NGC~4666	&	SABc:		&	5.0	&	27.5	$\pm$	1.9 & $\star$0.311 	$\pm$	0.020	&  $\star$40.0          $\pm$  2.0 	  \\
NGC~4689	&	SA(rs)bc	&	4.0	&	14.0	$\pm$	1.0 &        0.734 	$\pm$	0.027	&        167.4  	$\pm$  1.4	  \\
NGC~4691	&	(R)SB(s)0/a~pec	&	0.0	&	17.0	$\pm$	1.2 &        0.842 	$\pm$	0.023	&         41.2  	$\pm$  3.8	  \\
NGC~4698	&	SA(s)ab		&	2.0	&	13.7	$\pm$	1.0 &        0.566 	$\pm$	0.016	&        174.6  	$\pm$  1.6	  \\
NGC~4699	&	SAB(rs)b	&	3.0	&	22.9	$\pm$	1.6 &        0.720 	$\pm$	0.016	&         41.2  	$\pm$  3.4	  \\
NGC~4772	&	SA(s)a		&	1.0	&	13.3	$\pm$	0.9 &        0.503 	$\pm$	0.011	&        144.8  	$\pm$  1.5	  \\
NGC~4900	&	SB(rs)c		&	5.0	&	~9.1	$\pm$	0.6 &        0.925 	$\pm$	0.015	&         96.1  	$\pm$  2.0	  \\
NGC~5005	&	SAB(rs)bc	&	4.0	&	19.3	$\pm$	1.4 &        0.444 	$\pm$	0.023	&         63.5  	$\pm$  0.6	  \\
NGC~5334	&	SB(rs)c		&	5.0	&	24.2	$\pm$	1.7 &        0.760 	$\pm$	0.012	&         10.7  	$\pm$  3.2	  \\
NGC~5371	&	SAB(rs)bc	&	4.0	&	42.8	$\pm$	3.0 & $\star$0.468 	$\pm$	0.020	&   $\star$9.0          $\pm$  2.0 	  \\
NGC~5448	&	(R)SAB(r)a	&	1.0	&	35.2	$\pm$	2.5 & $\star$0.474 	$\pm$	0.020	& $\star$112.0          $\pm$  2.0 	  \\
NGC~5676	&	SA(rs)bc	&	4.0	&	36.5	$\pm$	2.6 &        0.442 	$\pm$	0.005	&         45.6  	$\pm$  0.9	  \\
NGC~5701	&	(R)SB(rs)0/a	&	0.0	&	26.7	$\pm$	1.9 &        0.913 	$\pm$	0.018	&         52.0  	$\pm$  4.2	  \\
NGC~5713	&	SAB(rs)bc~pec	&	4.0	&	31.3	$\pm$	2.2 &        0.863 	$\pm$	0.029	&          3.9  	$\pm$  0.0	  \\
NGC~5850	&	SB(r)b		&	3.0	&	41.6	$\pm$	2.9 &        0.866 	$\pm$	0.024	&        181.6  	$\pm$  6.8	  \\
NGC~5921	&	SB(r)bc		&	4.0	&	26.2	$\pm$	1.8 &        0.705 	$\pm$	0.013	&        130.9  	$\pm$  3.4	  \\
NGC~5962	&	SA(r)c		&	5.0	&	34.2	$\pm$	2.4 &        0.660 	$\pm$	0.029	&        111.5  	$\pm$  3.5	  \\
NGC~7479	&	SB(s)c		&	5.0	&	33.7	$\pm$	2.4 &        0.741 	$\pm$	0.018	&         35.7  	$\pm$  2.5	  \\
NGC~7606	&	SA(s)b		&	3.0	&	31.3	$\pm$	2.2 & $\star$0.357 	$\pm$	0.020	& $\star$148.0          $\pm$  2.0 	  \\
NGC~7741	&	SB(s)cd		&	6.0	&	12.5	$\pm$	0.9 &        0.690 	$\pm$	0.020	&        161.9  	$\pm$  4.1	  \\

\enddata

\tablenotetext{a}{~\citet{sor14}}
\tablenotetext{b}{~\citet{mei07}}
\tablecomments{Col. 1: galaxy name. 
Col. 2: RC3 type~\citep{deV91}.
Col. 3: T Hubble type~\citep{deV91}.
Col. 4: distance to object in Mpc, from NASA/IPAC Extragalactic Database (Virgo + GA + Shapley), unless otherwise indicated.
Col. 5: minor to major axis ratio, $q=b/a=\cos(\alpha)$, where $\alpha$ is the inclination angle of the disk.
Col. 6: position angle (P.A.) of the galaxy.
All $q$ and P.A. values were taken from~\citet{lau04}, with the exception of those marked with $\star$,
which were calculated by fitting ellipses to the outer isophotes of the disks in the $H$-band images.
}
\end{deluxetable}


The optical $g$ and $i$ band frames were mosaicked with the SWarp software~\citep{ber10}.
The $H$-band data has a ``sky offset''~\citep{kas06}, which we subtracted as
a constant or a plane, depending on the object; we then calibrated the images with 2MASS~\citep{skr06}.
The $3.6\micron$ data were sky-subtracted with the values given in~\citet{sal15}.
SDSS mosaics and $3.6\micron$ images were registered and re-sampled to match the
$H$-band data, which has the lowest spatial resolution (the pixel size is $\sim$1.5 arcsec$^2$,
which, at the mean distance to the objects, is equivalent to $\sim132\pm61$ pc$^2$).
Since the point-spread function (PSF) is similar in all bands, no PSF match was done between the images.
We apply the masks created in the S$^4$G Pipeline 2~\citep{mun15}, to mask foreground and background objects
(this includes other galaxies in the images different from the target).
We use the {\tt{Adaptsmooth}} code~\citep{zbt09}
to increase the signal-to-noise (S/N) ratio of the outermost regions of the disk,
while maintaining the relatively higher S/N ratio of the inner disk pixels.

The photometric errors of the images were computed on a pixel-by-pixel basis by using
$\sigma_{\rm mag}\approx\sqrt{\sigma^2_{\rm flux} + \sigma^2_{\rm calib}}$,
where $\sigma_{\rm mag}$ is the photometric error per pixel for a certain band,
$\sigma_{\rm flux}$ is the random error in the flux per pixel,
and $\sigma_{\rm calib}$ is the zero point error.
We assume that the error in the flux is dominated by the uncertainty in the background,
and compute $\sigma_{\rm flux}=1.086\times\frac{\sigma_{\rm back}}{{\rm flux}}$ in mag,
where $\sigma_{\rm back}$ is the standard deviation in the background
of the sky-subtracted image. We compute $\sigma_{\rm back}$ by sampling in boxes near the edges of the images
before using {\tt{Adaptsmooth}}; we then divide $\sigma_{\rm back}$ by $\sqrt{n_{\rm pix}}$, where $n_{\rm pix}$ is
the number of pixels used to increase the S/N after applying {\tt{Adaptsmooth}}.
For the zero point error we assume calibration uncertainties of $\sigma_{\rm calib, SDSS}=0.01$ mag
for the SDSS images~\citep{pad08}, and $\sigma_{\rm calib, IRAC}=0.03$ mag for the $3.6\micron$ band~\citep{rea05}.
The calibration uncertainty of the $H$-band images was computed as
$\sigma_{\rm{calib},\it{H}}=\sqrt{0.03^2 + \sigma^2_{\rm calerr}}$,
where 0.03 is the zero point error in 2MASS data~\citep{jar03},
and $\sigma_{\rm calerr}$ is the error in our calibration of the OSUBSGS $H$-band image with the 2MASS $H$-band image.
The value of $\sigma_{\rm calerr}$ depends on the object and has a mean value of $\sim 0.01$ mag.
To compute the error for a color, e.g., $(g-i)$, we use $\sigma_{\rm col}\approx\sqrt{\sigma^2_{g}+\sigma^2_{i}}$,
where $\sigma^2_{g}$ and $\sigma^2_{i}$ are the $\sigma_{\rm mag}$ errors for the $g$ and $i$-bands, respectively.
We assume no error correlations between bands.

\begin{figure*}
\centering
\epsscale{2.0}
\plotone{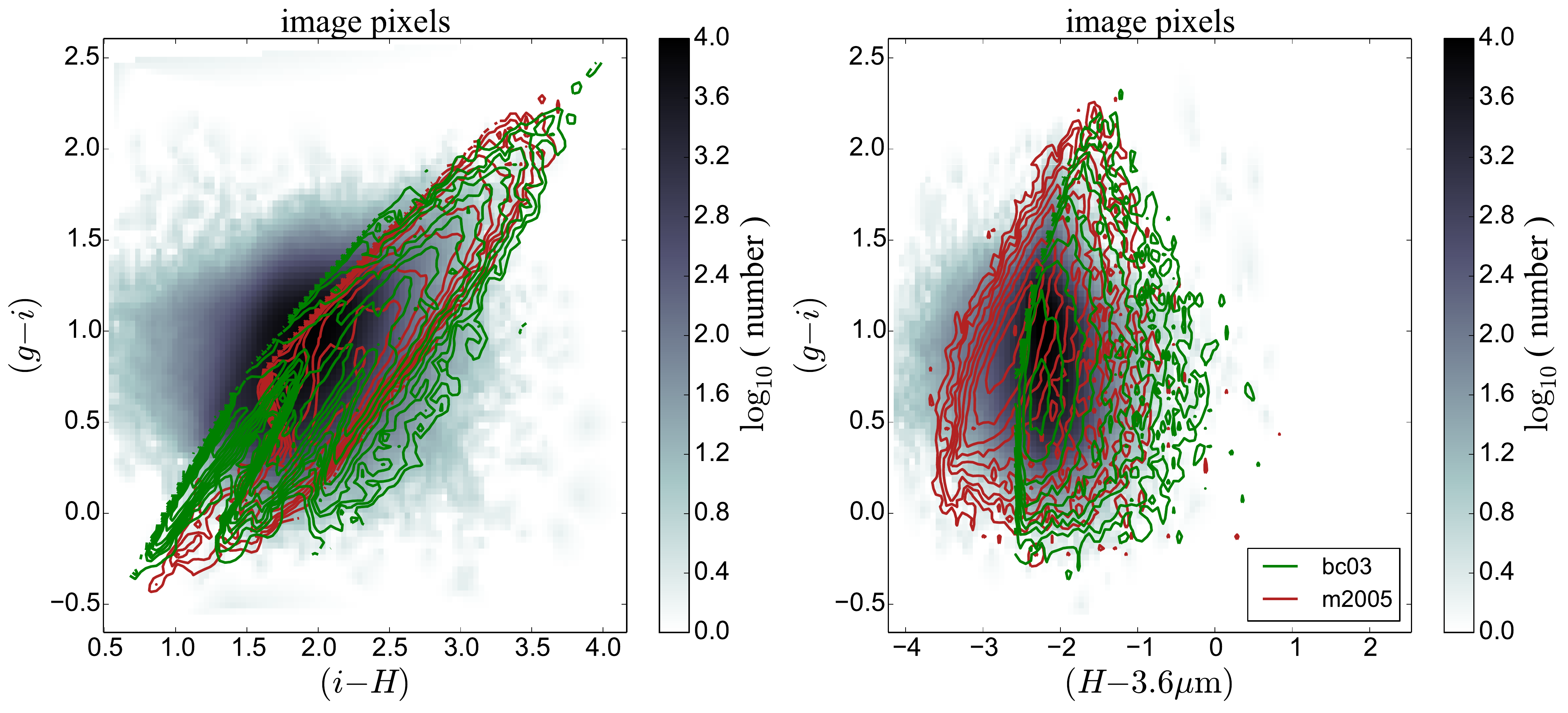}
\caption[f17]{Color-color diagrams of the observed pixels in our sample of objects,
corrected for Galactic extinction~\citep{sch11,cha09}.
{\it Left panel}: $(g-i)$ vs.\ $(i-H)$.
{\it Right panel}: $(g-i)$ vs.\ $(H-3.6\micron)$. 
The color bar indicates the number of pixels in each color bin ($0.05\times0.05$ mag).
Red and green contours correspond to the m2005 (see Figure~\ref{fig5}) and bc03 (see Figure~\ref{fig6})
CSP libraries, respectively.
$g$, $i$, and $3.6\micron$ bands in AB mag, $H$-band referenced to Vega.
~\label{fig17}}
\end{figure*}

\begin{figure*}
\centering
\epsscale{2.0}
\plotone{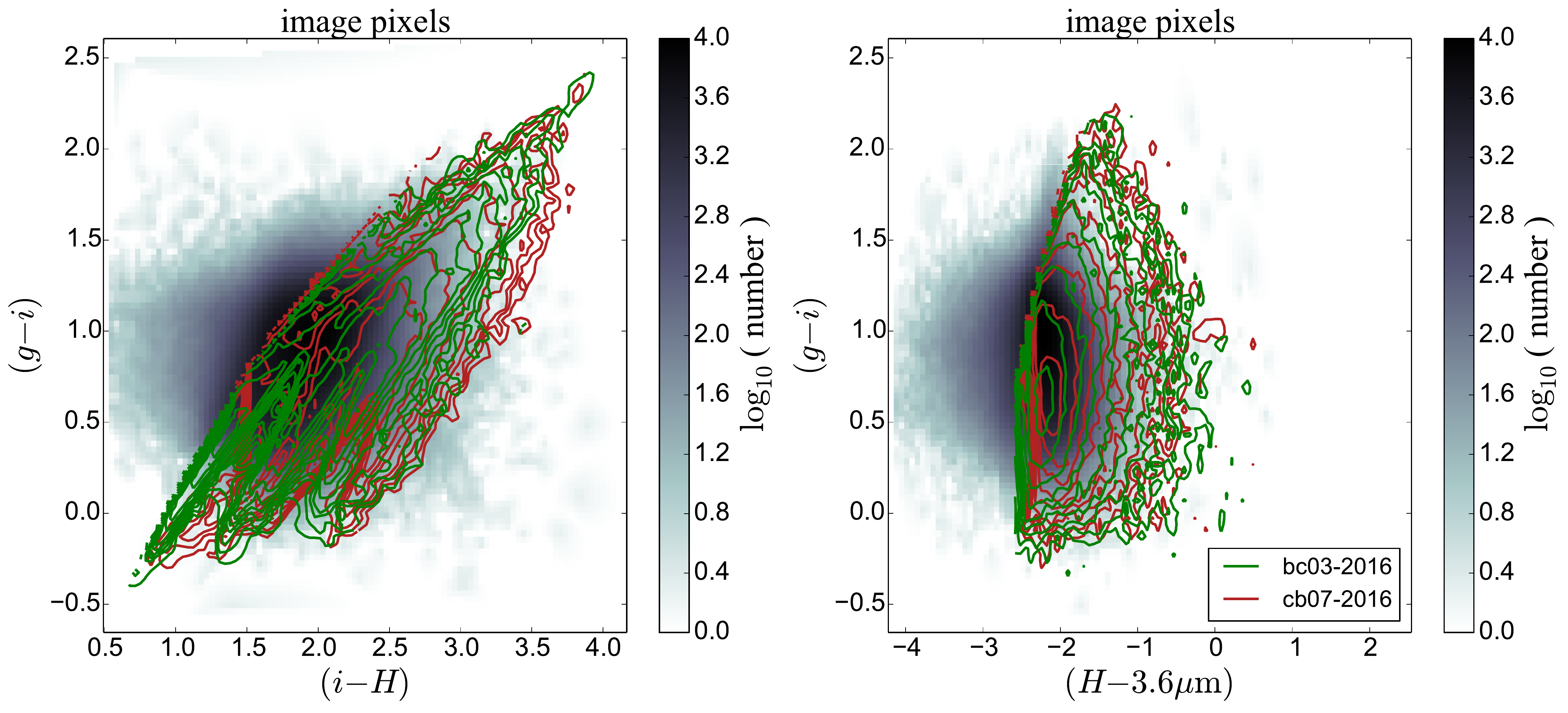}
\caption[f18]{Color-color diagrams of the observed pixels in our sample of objects.
Same labels as in Figure~\ref{fig17}.
Red and green contours correspond to the cb07-2016 (see Figure~\ref{fig7}) and bc03-2016 (see Figure~\ref{fig8})
CSP libraries, respectively.
~\label{fig18}}
\end{figure*}

In Figures~\ref{fig17} and~\ref{fig18} we show color-color diagrams of the 
pixels in our sample of objects. The observed colors were corrected for Galactic extinction~\citep{sch11,cha09}.
We have superimposed the contour plots of the CSP libraries (see Figures~\ref{fig5}-\ref{fig8}) bc03 and m2005
in Figure~\ref{fig17}, and bc03-2016 along with cb07-2016 in Figure~\ref{fig18}. Most of the
observed colors fall within the color space covered by the contours.

\section{Fits to the observed photometry}~\label{sec_fits_photometry}

We use the individual libraries to fit the $(g-i)$, $(i-H)$, and $(H-3.6\micron)$ colors
of the individual pixels in each object of our sample. The total number of pixels for all objects
is $\sim1.6\times10^{6}$ per photometric band.
We apply a maximum likelihood approach, i.e., we
compute the probability of each model to fit the observed colors:

\begin{equation}~\label{maxlike}
  P \propto \frac{1}{\sqrt{2\pi}} \exp \left(-\frac{\chi^{2}}{2}\right),
\end{equation}

\begin{equation}~\label{chi2}
  \chi^{2}=\sum_{n=1}^{3} \left(\frac{C_{n}^{\rm obs}-C_{n}^{\rm model}}{\sigma_{\rm col}} \right)^{2},
\end{equation}
where $C_{n}^{\rm obs}$ is the observed $n_{\rm th}$ color with $\sigma_{\rm col}$ photometric error,
and $C_{n}^{\rm model}$ is the color of the model.
Each model is assigned a probability $P$, the best fitting model being the one with the maximum probability.
A likelihood distribution for each parameter, e.g., $Z/Z_{\sun}$, T$_{\rm form}$,
etc., can be obtained with the $P$ values. From the likelihood distributions we can compute
the percentiles, P$_{16}$ and P$_{84}$, to approximate the $1\sigma$ error of the corresponding parameter.

\subsection{Recovered stellar masses}

\begin{figure}
\centering
\epsscale{1.0}
\plotone{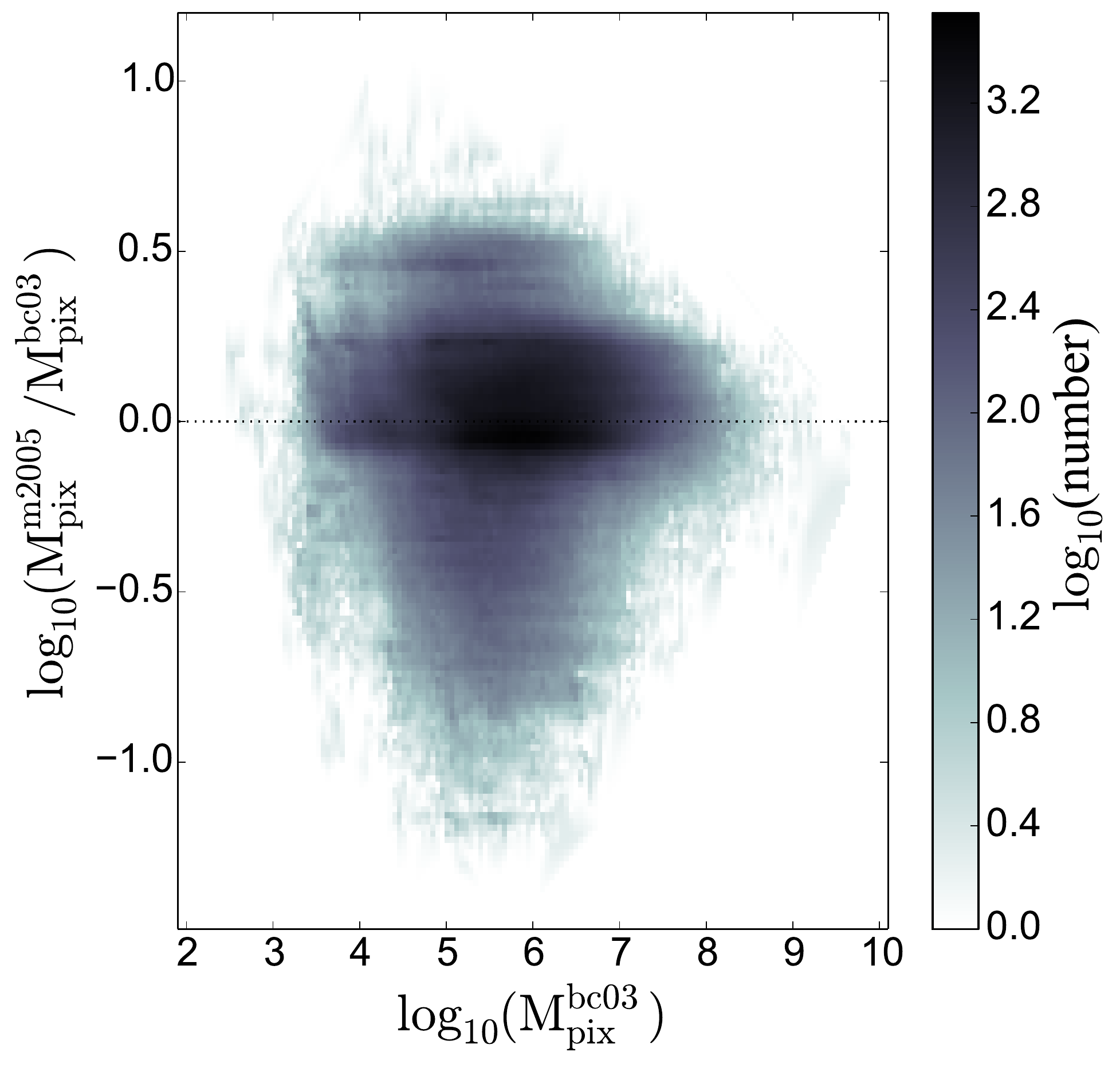}
\caption[f19]{Ratios of recovered stellar masses per pixel.
Fits with individual libraries m2005 and bc03.
M$^{\rm m2005}_{\rm pix}$ is the stellar mass obtained with the m2005 library;
M$^{\rm bc03}_{\rm pix}$ is the stellar mass retrieved with the bc03 library.
Color scale indicates number of pixels in each stellar mass bin.
Stellar masses in $M_{\sun}$.
~\label{fig19}}
\end{figure}

\begin{figure}
\centering
\epsscale{1.0}
\plotone{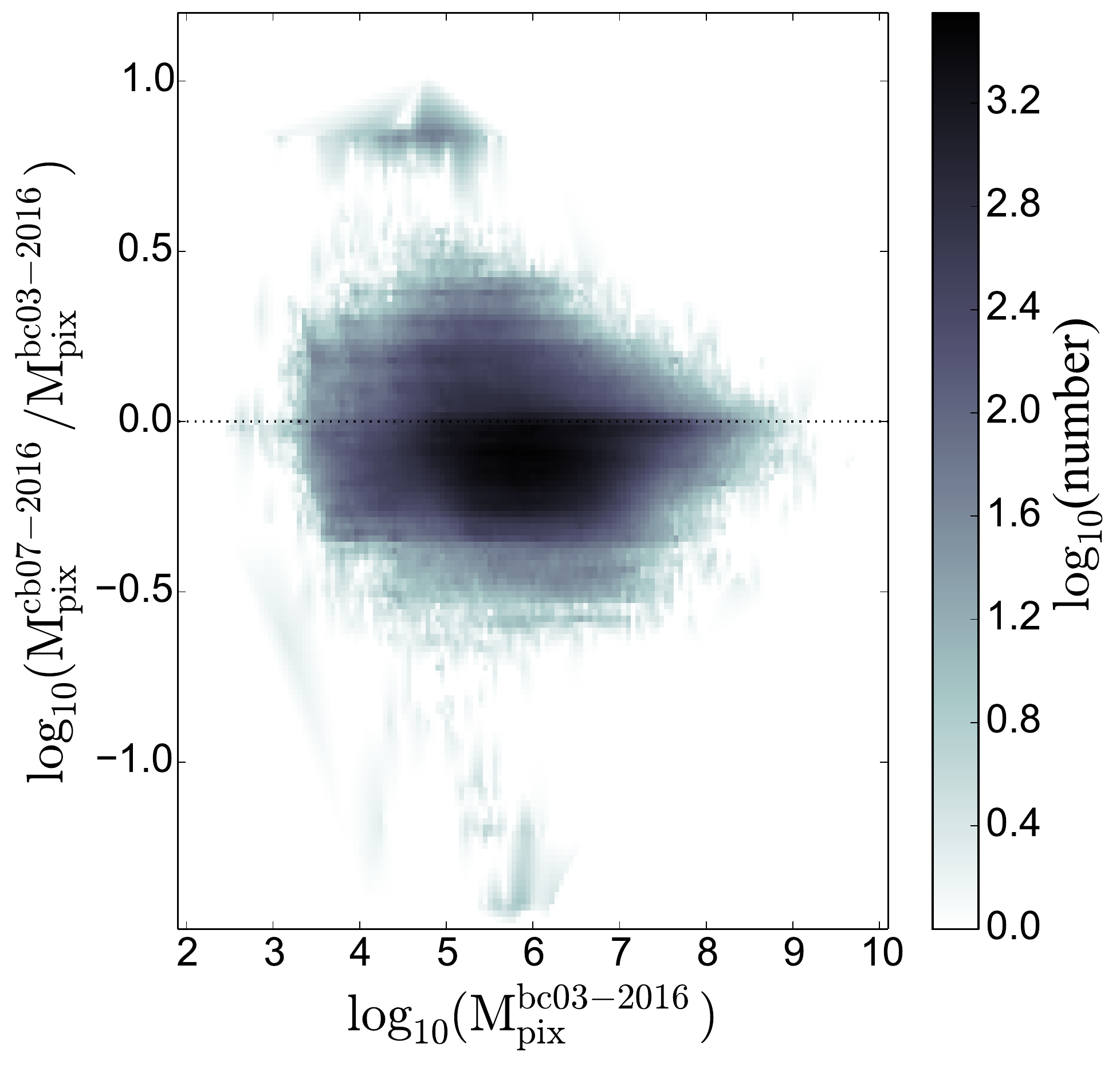}
\caption[f20]{Ratios of stellar masses per pixel.
Fits with individual libraries cb07-2016 and bc03-2016. 
M$^{\rm cb07-2016}_{\rm pix}$ is the stellar mass obtained with the cb07-2016 library;
M$^{\rm bc03-2016}_{\rm pix}$ is the stellar mass retrieved with the bc03-2016 library.
Color scale indicates number of pixels in each stellar mass bin.
Stellar masses are given in $M_{\sun}$.
~\label{fig20}}
\end{figure}

In Figure~\ref{fig19} we show a 2D histogram of the stellar masses
obtained for every pixel from the fits to the observed colors with m2005 on the one hand, and bc03 on the other.
At the maximum peak of the histogram [$\log_{10}($M$^{\rm bc03}_{\rm pix})\sim6$],
there are more pixels where the stellar masses obtained with the bc03 library are larger,
a result consistent with the expectations.
A similar thing happens with the stellar masses obtained with
the cb07-2016 and the bc03-2016 libraries, in Figure~\ref{fig20}: 
at the maximum peak of the histogram, there are more pixels where the bc03-2016 stellar masses are larger.

\subsection{Library Comparison}~\label{libs_conf}

\begin{figure}
\centering
\epsscale{1.0}
\plotone{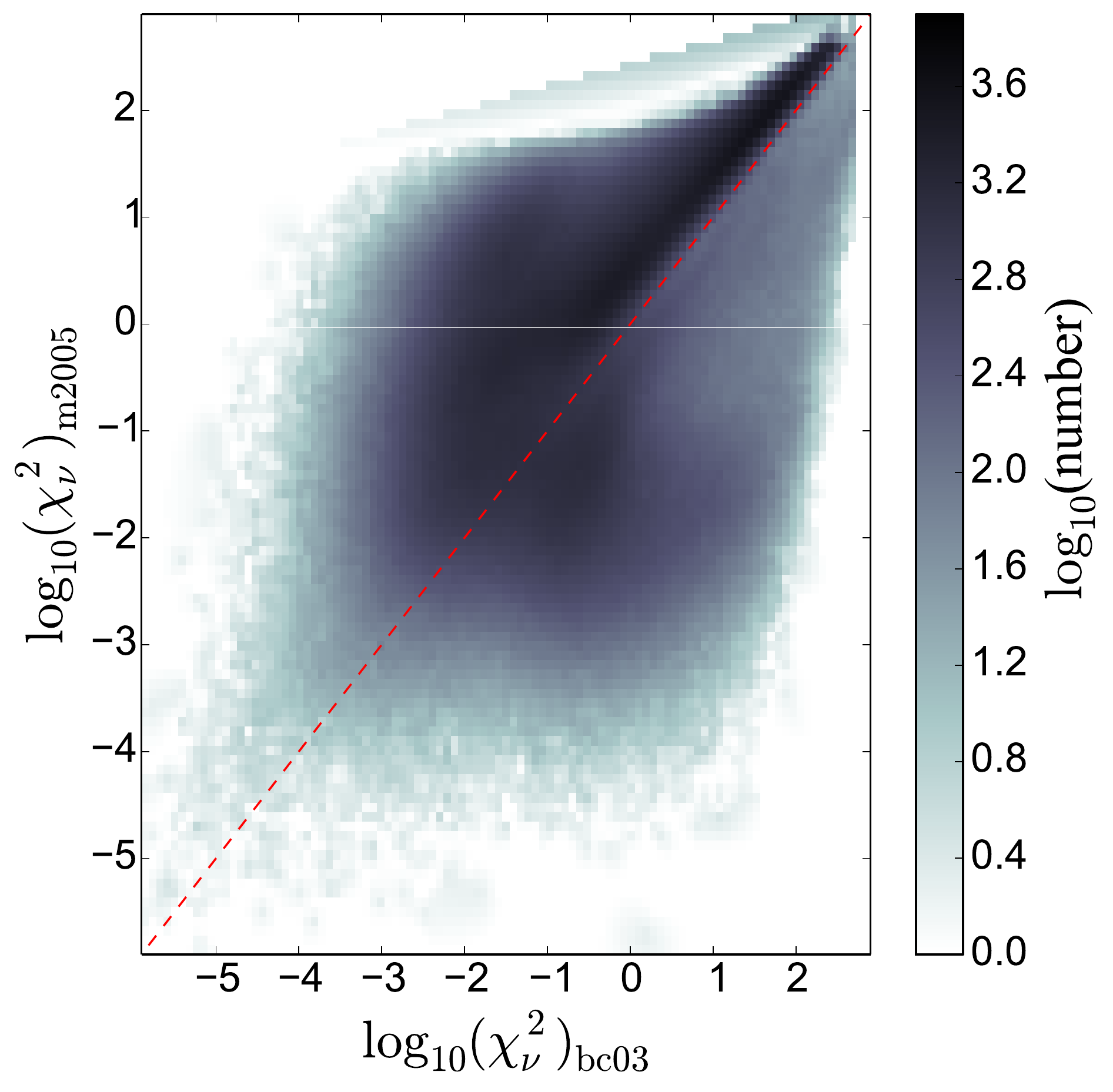}
\caption[f21]{
Reduced $\chi^{2}$ ($\chi^{2}_{\nu}$) for the
fits with the individual libraries bc03 ($x$-axis) and m2005 ($y$-axis).
The dashed red line indicates the 1:1 relation.
Color scale indicates the number of pixels in each bin.
~\label{fig21}}
\end{figure}

\begin{figure}
\centering
\epsscale{1.0}
\plotone{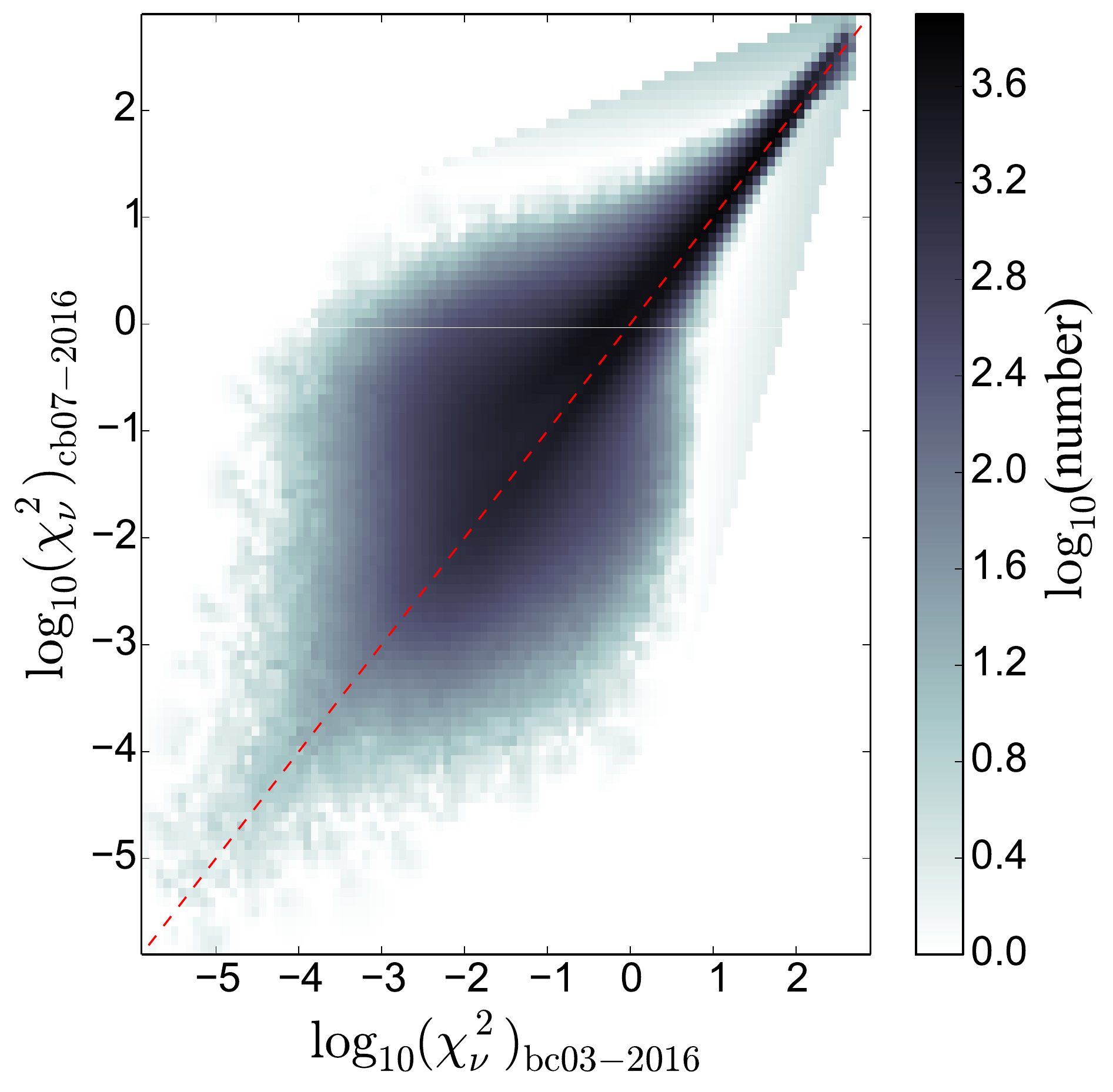}
\caption[f22]{
Same as Figure~\ref{fig21} for the 
fits with the individual libraries bc03-2016 ($x$-axis) and cb07-2016 ($y$-axis).
~\label{fig22}}
\end{figure}

In this section we perform the library comparison.
For this purpose, we compare the probabilities $P$ (equation~\ref{maxlike}) of the libraries
to fit each pixel separately. For a certain pixel, a higher $P$, i.e., lower $\chi^{2}$,
indicates a better fit with the corresponding library.
In Figure~\ref{fig21} we show the reduced $\chi^{2}$ values,
$\chi^{2}_{\nu}=\chi^{2}/3$ (see equation~\ref{chi2}),
after fitting the m2005 and bc03 libraries to the observed colors in the pixels.
In Figure~\ref{fig22} we plot the $\chi^{2}_{\nu}$ values for the
cb07-2016 and bc03-2016 libraries.

We define the pixel percentage with best model fits for the ``light'' TP-AGB models
as $\Lambda$. The pixel percentage with best model fits for the ``heavy'' TP-AGB models
is then $\Gamma=100-\Lambda$.
In Table~\ref{tbl-3} we summarize the values of $\Gamma$ and $\Lambda$ for the library comparisons
(m2005 vs.\ bc03), (m2005 vs.\ bc03-2016), (cb07 vs.\ bc03), and (cb07-2016 vs.\ bc03-2016).
These results are for the global sample of pixels, i.e., they include the pixels for
every galaxy in the same statistic.
For the (m2005 vs.\ bc03) library comparison we find that $\Gamma=28$, i.e., 28\% of the pixels
are better fitted with the m2005 models, and $\Lambda=72$, i.e., the remaining pixels, are better fitted with the bc03 models.
For the (cb07-2016 vs.\ bc03-2016) library comparison we find that $\Gamma=32\%$,
and $\Lambda=68\%$.
The pixel percentage is higher for the ``light'' TP-AGB models in all cases, i.e., $\Lambda>\Gamma$.

\subsubsection{Correlations with projection parameters and Hubble type}~\label{sec_correlations}

We investigate the possible correlation of $\Lambda$ with 
the disk projection parameters (see Table~\ref{tbl-2}), i.e., axial ratio $q=b/a$, 
and position angle (P.A.). For this purpose we calculate the
value of $\Lambda$ within twice the half-light radius in the $g$-band, $2R_{\rm hl}^{g}$.\footnote{
The analyzed region is actually an ellipse in the sky with semi-major axis equal to $2R_{\rm hl}^{g}$.}
$R_{\rm hl}^{g}$ is defined as the radius where the cumulative flux is one half of the total flux.
In Table~\ref{tbl-4} we show the pixel percentages with best model fits per object,
i.e., the pixels for every galaxy are separated into an individual statistic.
The objects were ranked from the highest to the lowest $\Gamma$.
For the (m2005 vs.\ bc03) library comparison, 19 out of 84 objects have $\Gamma>50\%$. 
For the (m2005 vs.\ bc03-2016), (cb07 vs.\ bc03), and (cb07-2016 vs.\ bc03-2016) cases,
we have 25, 8, and 3 out of 84 objects with $\Gamma>50\%$, respectively.

\begin{figure}
\centering
\epsscale{1.0}
\plotone{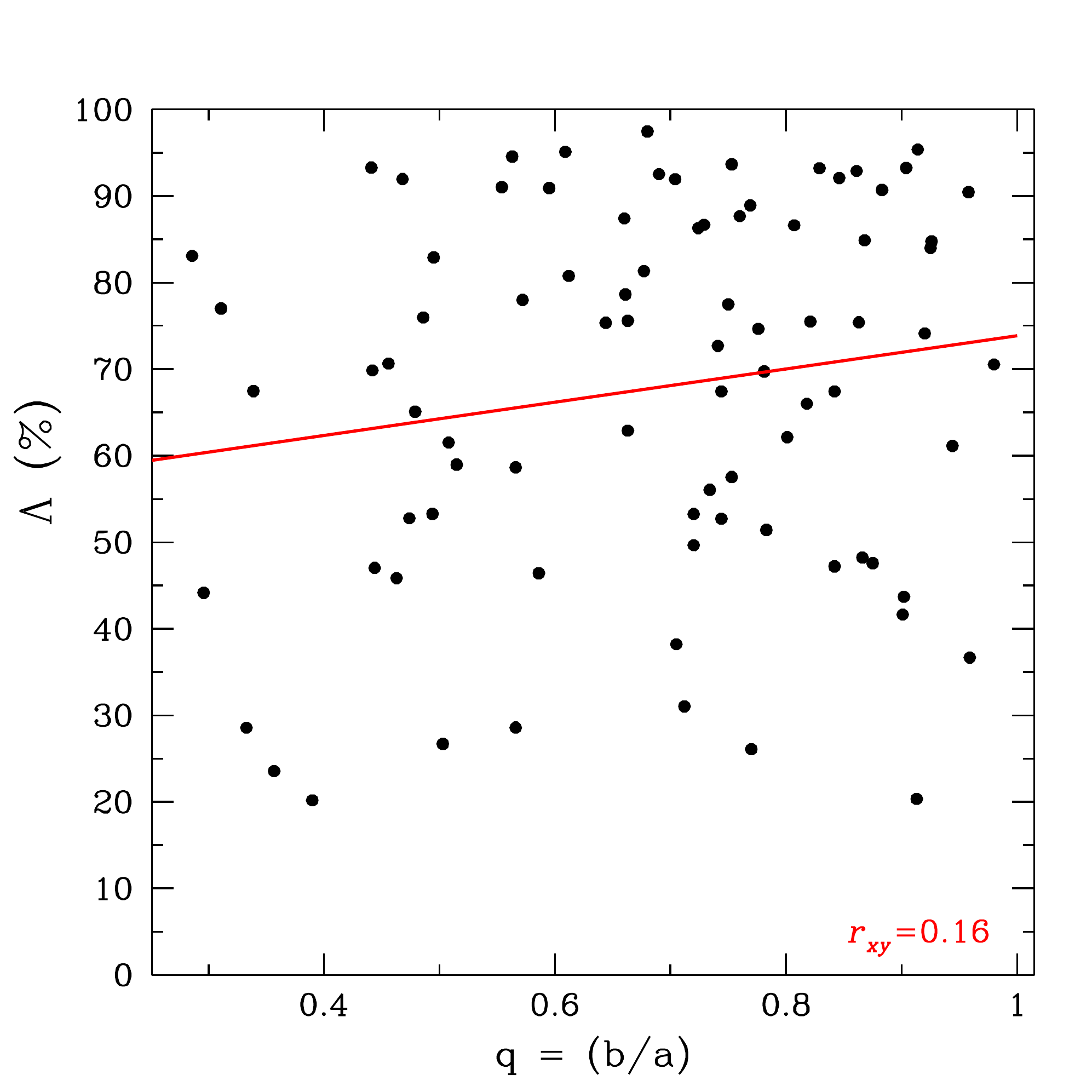}
\caption[f23]{$\Lambda$ vs.\ inclination parameter $q=b/a$ for the fits obtained with the (m2005 vs.\ bc03) library.
{\it Continuous red line}: best linear fit to the data.
The correlation coefficient between the points, $r_{xy}$~\citep{bev69}, is indicated in the lower-right corner.
~\label{fig23}}
\end{figure}

\begin{figure}
\centering
\epsscale{1.0}
\plotone{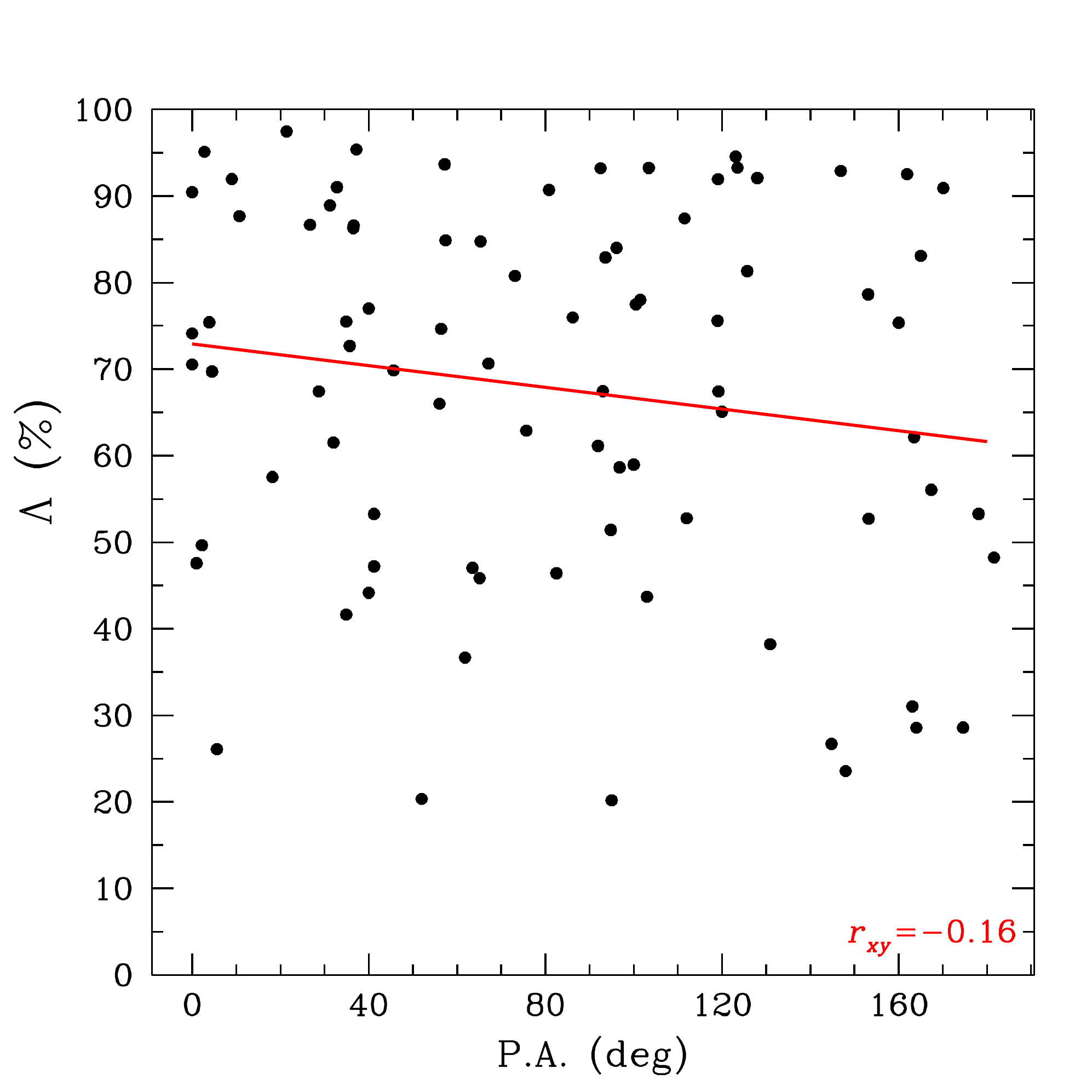}
\caption[f24]{$\Lambda$ vs.\ position angle P.A.
for the fits obtained with the (m2005 vs.\ bc03) library.
~\label{fig24}}
\end{figure}

In Figure~\ref{fig23} we plot $\Lambda$ vs.\ $q$ for the (m2005 vs.\ bc03) case.
The correlation coefficient~\citep{bev69} is $r_{xy}=0.16$, which indicates no correlation.\footnote{
Generally, $1\gtrsim |r_{xy}| \gtrsim 0.7$ is considered a strong correlation,
$|r_{xy}| \approx 0.5$ a moderate correlation, and $0.3\gtrsim |r_{xy}| \gtrsim 0.0$ a weak correlation.}
For the $\Lambda$ vs.\ P.A. plot, there is also no correlation in the (m2005 vs.\ bc03) case,
since $r_{xy}=-0.16$, see Figure~\ref{fig24}.
Similar results are obtained for the (cb07-2016 vs.\ bc03-2016) case, where $r_{xy}=-0.03$
and $r_{xy}=-0.06$, for the $q$ and P.A. plots, respectively.

\begin{figure}
\centering
\epsscale{1.0}
\plotone{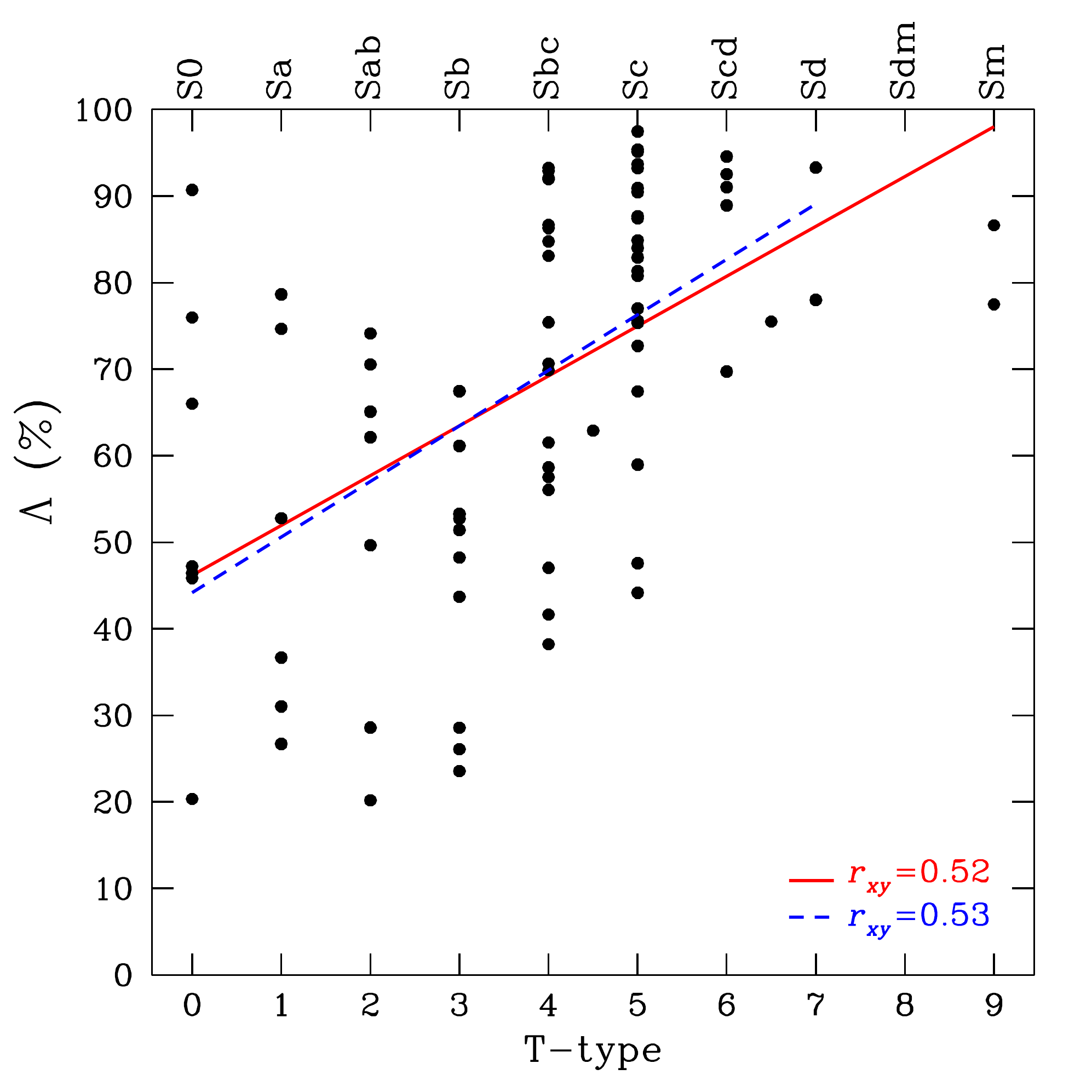}
\caption[f25]{$\Lambda$ vs.\ T Hubble type 
for the fits obtained with the (m2005 vs.\ bc03) library.
{\it Continuous red line}: linear fit to all data.
{\it Dashed blue line}: linear fit, excluding the T=9 points.
The correlation coefficient between the points, $r_{xy}$, is indicated in the lower-right corner.
~\label{fig25}}
\end{figure}

\begin{figure}
\centering
\epsscale{1.0}
\plotone{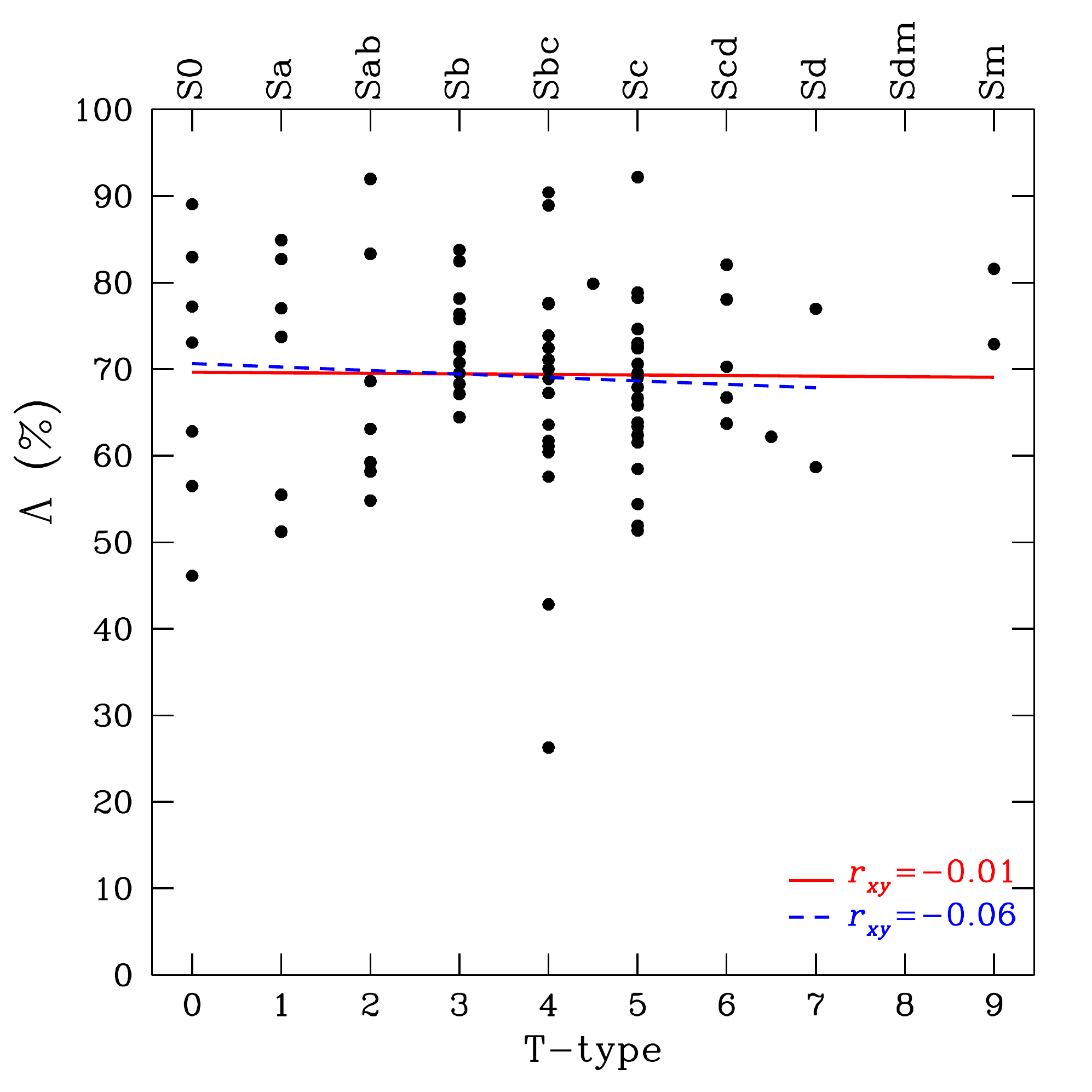}
\caption[f26]{$\Lambda$ vs.\ T Hubble type for the fits obtained with the (cb07-2016 vs.\ bc03-2016) library.
Same labels as in Figure~\ref{fig25}.
~\label{fig26}}
\end{figure}

We also consider the possible correlation with T Hubble type.
In Figures~\ref{fig25} and~\ref{fig26} we plot $\Lambda$ vs.\ T-type, for the (m2005 vs.\ bc03)
and (cb07-2016 vs.\ bc03-2016) cases, respectively. We have computed two different fits to the points.
The continuous (red) lines are the best linear fits, considering all the T-types in our sample.
The dashed (blue) lines are the best fits obtained by dismissing the last two points, where T-type=9.
In the (m2005 vs.\ bc03) case we obtain a moderate correlation with $r_{xy}\sim0.5$,
and for (cb07-2016 vs.\ bc03-2016) we get no correlation, since $r_{xy}\sim0.0$.

\subsubsection{The radial dependence of $\Lambda$}~\label{radial_dependence}

\begin{figure}
\centering
\epsscale{1.0}
\plotone{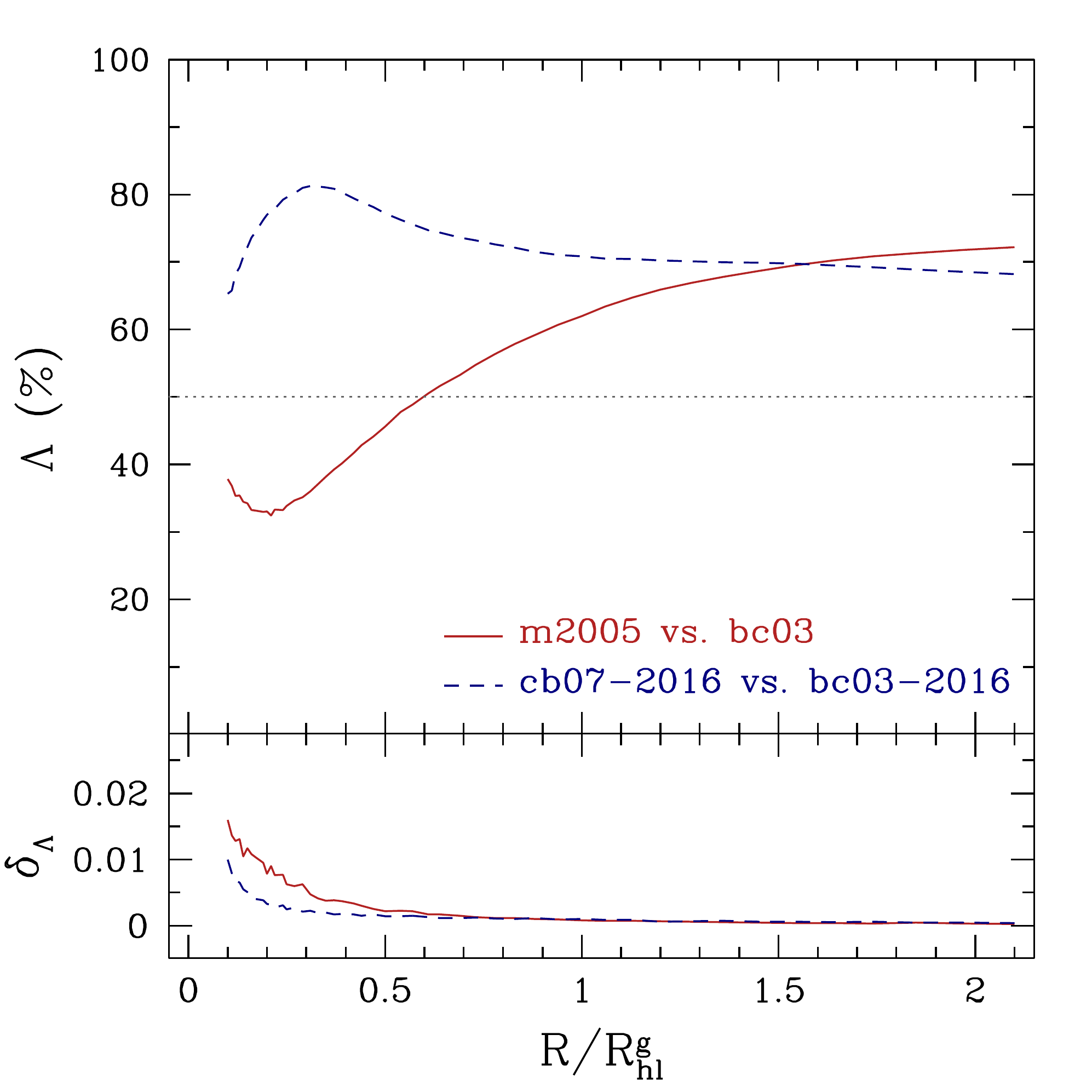}
\caption[f27]{
{\it Top panel}: Global pixel percentage best fitted by ``light'' TP-AGB models,
$\Lambda$, as a function of $R/R_{\rm hl}^{g}$.
{\it Bottom panel}: Relative error, $\delta_{\Lambda}=\frac{\sigma_{\Lambda}}{\Lambda}$.
Solid red line: (m2005 vs.\ bc03);
dashed blue line: (cb07-2016 vs.\ bc03-2016);
dotted black line: $\Lambda = 50\%$.
~\label{fig27}}
\end{figure}

In this section we analyze the radial dependence of $\Lambda$ in our sample of galaxies.
We compute the global pixel percentages with best model fits ($\Gamma$ and $\Lambda$) at different radii,
$R/R_{\rm hl}^{g}$.
The results of this exercise for the (m2005 vs.\ bc03) and (cb07-2016 vs.\ bc03-2016) cases are shown in Figure~\ref{fig27}.
For the (m2005 vs.\ bc03) case (continuous red line),
the regions where $R/R_{\rm hl}^{g}\lesssim0.6$ are mostly dominated by 
TP-AGB ``heavy'' models, i.e., $\Lambda < 50\%$, while in the outer parts of the disks we find $\Lambda > 50\%$.
For the (cb07-2016 vs.\ bc03-2016) case (dashed blue line), at all radii we have  $\Lambda > 50\%$, i.e.,
the regions are dominated by TP-AGB ``light'' models.
By comparing the two curves for $\Lambda$ for the innermost regions,
$0.1\lesssim R/R_{\rm hl}^{g} \lesssim 0.5$,
we notice a different behavior of $\frac{d (\Lambda)}{d (R/R_{\rm hl}^{g})}$.
This indicates that the fits obtained for the bulge regions of the disks depend on the adopted models.

\section{Discussion of results}~\label{sec_discussion}

As shown in Section~\ref{radial_dependence} and Figure~\ref{fig27},
$\Lambda$ (the pixel percentage with best model fits for the ``light'' TP-AGB models)
may vary with location in the disk. In this sense, the innermost regions
of the galaxy disks may show a higher percentage of pixels that are best fitted with the ``heavy'' TP-AGB models.

\begin{figure}
\centering
\epsscale{1.0}
\plotone{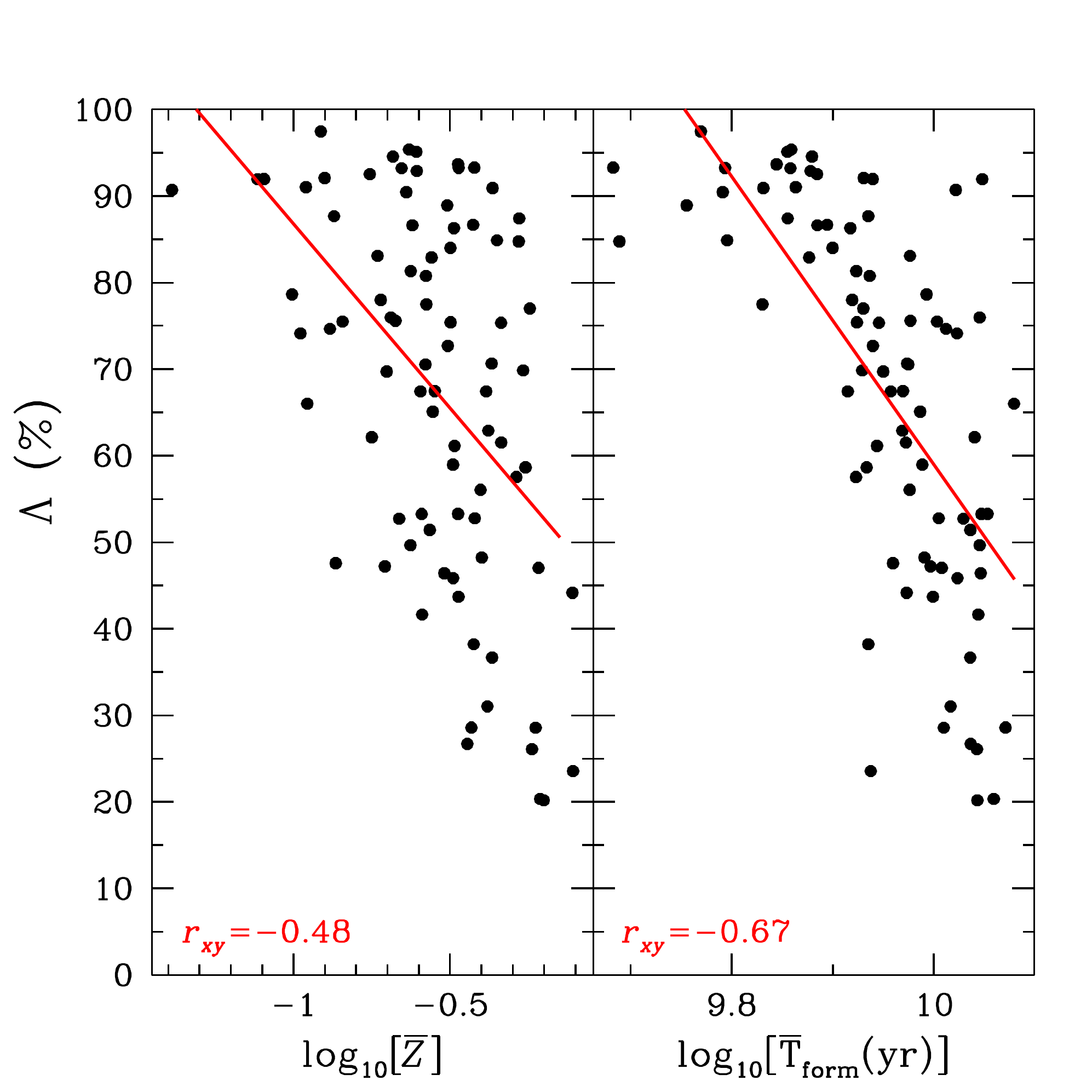}
\caption[f28]{
{\it Left panel}: $\Lambda$ vs.\ mean stellar metallicity ($\overline{Z}$, see equation~\ref{mean_Z}).
{\it Right panel}: $\Lambda$ vs.\ mean stellar age ($\overline{\rm{T}}_{\rm form}$, see equation~\ref{mean_Tform}).
Both panels correspond to the (m2005 vs.\ bc03) case.
The correlation coefficient between the points, $r_{xy}$, is indicated in the lower-left corner.
~\label{fig28}}
\end{figure}

Also, for the (m2005 vs.\ bc03) and (m2005 vs.\ bc03-2016) cases,
we have found indications of a possible correlation of $\Lambda$
with Hubble type, i.e., late-type spirals (Sc) tend to be better fitted with ``light'' models.
A possible physical reason for this is that the contribution of TP-AGB stars depends on
other parameters that also correlate with Hubble type, e.g., age and stellar metallicity.
The mean age and metallicity of disk galaxies are lower for late-type spirals~\citep[e.g.,][]{zar94,gond15},
suggesting that morphology is strongly correlated with the shutdown of star formation.
In Figure~\ref{fig28}, we show $\Lambda$ vs.\ mean metallicity ($\overline{Z}$),
\begin{equation}~\label{mean_Z}
 \overline{Z} = \frac{1}{n}\sum\limits_{p=1}^{n} (Z/Z_{\sun})_{p},
\end{equation}
\noindent where $n$ is the total number of pixels within $2R_{\rm hl}^{g}$ for a certain object,\footnote{
The $\overline{Z}$ values for our sample of galaxies are in the range [0.041,0.78].
\citet{mou10} obtained the characteristic (i.e., globally averaged) nebular oxygen abundances
of a sample of 55 nearby galaxies, which can be related to the stellar metallicity
assuming $\log(Z)\simeq~1.43+\log(O/H)$. By using the~\citet[][KK04]{kk04},
and~\citet[][PT05]{pt05} calibrations,~\citet{mou10} $Z/Z_{\sun}$ values
are in the range [0.135,2.183], and [0.047,0.536], respectively.}
and vs.\ mean stellar age ($\overline{\rm{T}}_{\rm form}$),
\begin{equation}~\label{mean_Tform}
 \overline{\rm{T}}_{\rm form} = \frac{1}{n}\sum\limits_{p=1}^{n} (\rm{T}_{\rm form})_{{\it p}},
\end{equation}
on the left and right panels, respectively.
There is indeed a moderate correlation in both panels, in the sense that $\Lambda$ has a higher
value for lower $\overline{Z}$ and lower $\overline{\rm{T}}_{\rm form}$.

The TP-AGB mass-loss rate ($\dot{M}$) may play an important role
in explaining the trend in Figure~\ref{fig25}.
The TP-AGB $\dot{M}$ would need to be more efficient (or higher) for low metallicity stars.
If this were the case, metal-poor stellar populations would be better fitted by ``light''
models because their TP-AGB stars would have shorter lifetimes, and hence contribute less light.
This is counterintuitive. It is widely accepted that AGB winds are driven by radiative pressure
on dust after the gas is levitated by stellar pulsations~\citep[e.g.,][]{lil18}.
Hence, one would expect mass-loss rate to increase with $Z$.
However,~\citet{van06} has pointed out that the chromospherically-driven $\dot{M}$ formula given
by~\citet{sch05}\footnote{$\dot{M}=\eta\frac{L_*R_*}{M_*}
{\left(\frac{T_{\rm eff}}{4000{\rm K}}\right)}^{3.5}
\left(1+\frac{g_{\odot}}{4300~g_*}\right)~[M_{\odot}{\rm yr}^{-1}]$,
where $\eta$ is a fitting parameter;
$L_*$, $R_*$, and $M_*$ are the stellar luminosity, radius, and mass, given in solar units;
$T_{\rm eff}$ is the stellar effective temperature;
and $g_\odot$ and $g_*$ are the solar and stellar surface gravity, respectively.} 
could result in a higher mass loss for metal-poor than for metal-rich stars.
In Schr\"oder \& Cuntz's model, the winds are produced by the spillover of
the chromosphere due to magnetoacoustic (or Alfv\'en) waves. Their formula is similar to Reimers'
law~\citep{rei75,rei77}, but includes a dependence of $\dot{M}$ with the effective temperature and
the surface gravity of the star.
\citet{gir10} have proposed a scenario that divides AGB mass loss into two main regimes.
The first one involves the use of the~\citet{sch05} mass-loss rate to model a {\it pre-dust} wind in metal poor AGB stars.
The second regime consists in a {\it dust-driven} wind that activates only when
a balance between the radiation pressure on dust and the inward gravitational force is reached.
A third regime could be included that implicates a short {\it super wind}~\citep[e.g.,][]{ros14}.
The inclusion of a {\it pre-dust} wind effectively reduces the lifetimes of low metallicity
AGB stars~\citep{gir10,ros14}. The {\it pre-dust} $\dot{M}$
is not necessarily more efficient than the {\it dust-driven} or the {\it super wind} regimes,
but occurs for a larger portion of the TP-AGB lifetime in metal-poor stars~\citep{ros16}.
This framework may explain the trend between Hubble type
with mean TP-AGB contribution in Figure~\ref{fig25}, as a consequence of a dependence
of TP-AGB mass-loss rate with metallicity.

\subsection{Variations in the SFH, IMF, and stellar metallicity}

In this section we discuss how our main results may differ
if we change the SFH, the IMF, and the stellar metallicity of the
models in our libraries. A summary of the results obtained with these variations
is given in Table~\ref{tbl-5}; in the following, we discuss each modification individually.

\subsubsection{SFH}~\label{var_SFH}

\begin{figure}
\centering
\epsscale{1.0}
\plotone{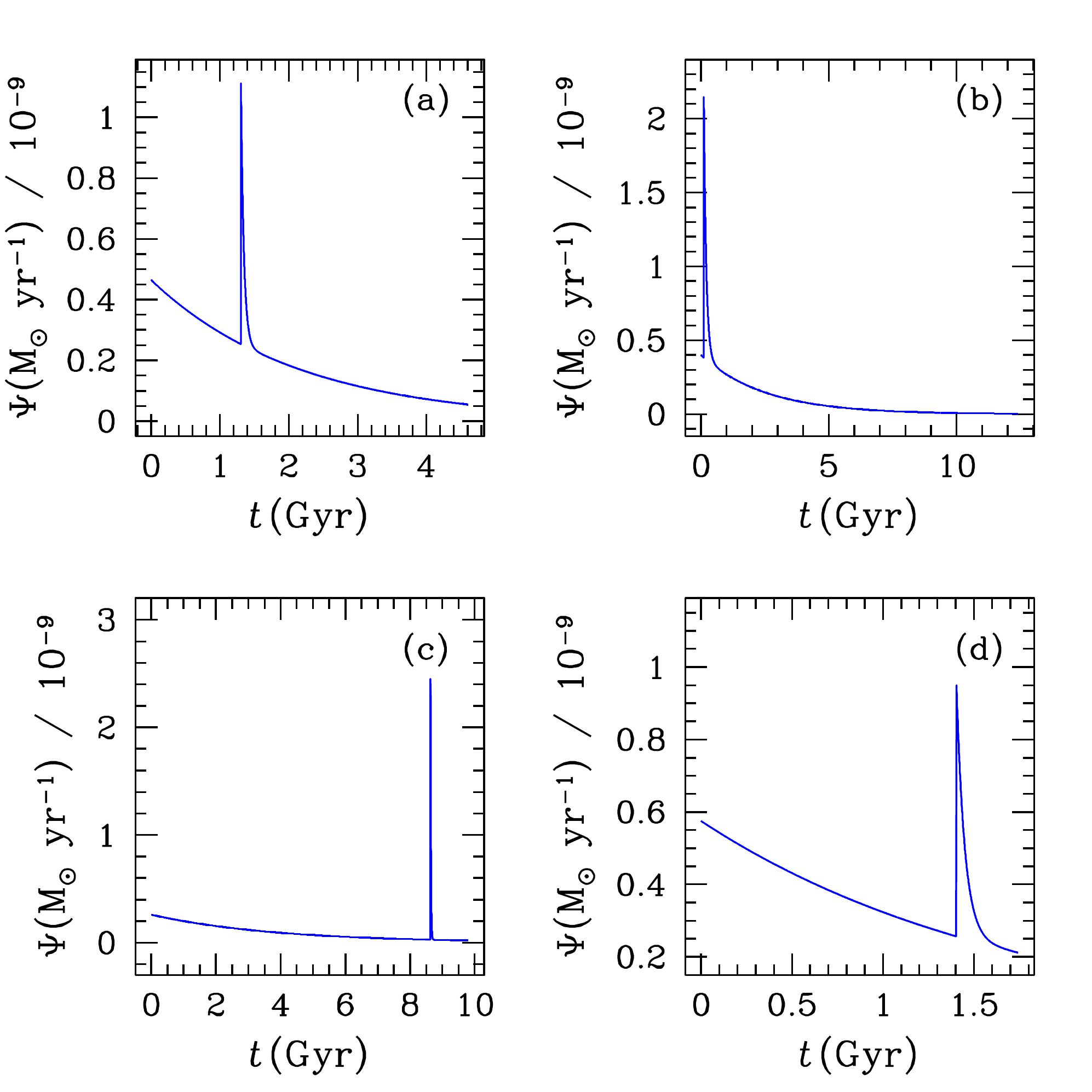}
\caption[f29]{Star formation histories obtained from equation~\ref{SFH_exp}.
Same parameters as in Table~\ref{tbl-1}. Compare with Figure~\ref{fig4}.
~\label{fig29}}
\end{figure}

In order to explore the effect that a different SFH may have in our results,
we model the SFH of our CSP libraries with two decaying exponentials.
The first exponential models the long-term star formation,
and the second one the most recent burst~\citep[e.g.,][]{boq19}:
\begin{equation}~\label{SFH_exp}
  \Psi(t)= \Psi_{(t=0)}\left[\exp(-t/\tau_{0}) + k\exp(-t/\tau_{1})\right],
\end{equation}
where $t$, $\tau_{0}$, $k$, and $\tau_{1}$, are the same as in equation~\ref{SFH_delayed},
and $\Psi_{(t=0)}$ is the SFR at $t=0$, computed as $\Psi_{(t=0)}=1/\tau_{0}$, such that
$\int_{t=0}^{t=\infty} \Psi(t) dt = 1$~\citep{bru03} when $k$=0.
In Figure~\ref{fig29}, we show four examples of the SFHs obtained with equation~\ref{SFH_exp},
adopting the same parameters as in Table~\ref{tbl-1}.
We use this form of the SFH to compute our 5 libraries (bc03, m2005, cb07, bc03-2016, and cb07-2016),
leaving the other parameters unchanged.
As discussed in Section~\ref{sec_fits_photometry},
we then fit the colors of the individual pixels for the objects in our sample.

\begin{figure}
\centering
\epsscale{1.0}
\plotone{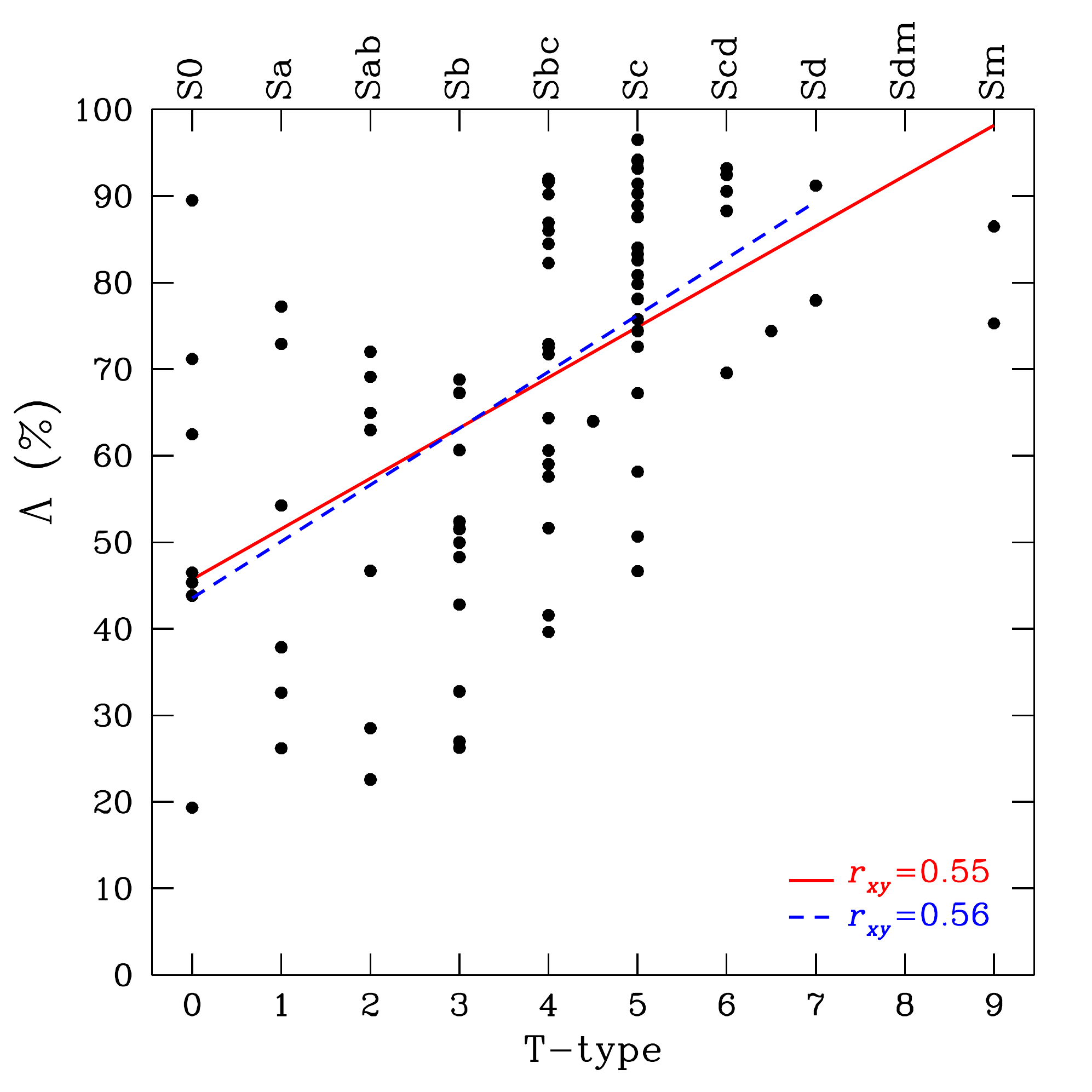}
\caption[f30]{$\Lambda$ vs.\ T Hubble type 
for the fits obtained with the (m2005 vs.\ bc03) library, with
a SFH modeled as two decaying exponentials (see section~\ref{var_SFH}).
Same labels as in Figure~\ref{fig25}.
~\label{fig30}}
\end{figure}

The global results for $\Gamma$ and $\Lambda$ (see Table~\ref{tbl-5}) are very similar to
our previous calculations (see Table~\ref{tbl-3}).
In Figure~\ref{fig30} we plot $\Lambda$ vs.\ T-type, for the (m2005 vs.\ bc03) case.
The correlation is similar to that found in the case of a delayed SFH (see equation~\ref{SFH_delayed},
in Section~\ref{sec_SFH}, and Figure~\ref{fig25}).

\subsubsection{IMF}~\label{var_IMF}

\begin{figure}
\centering
\epsscale{1.0}
\plotone{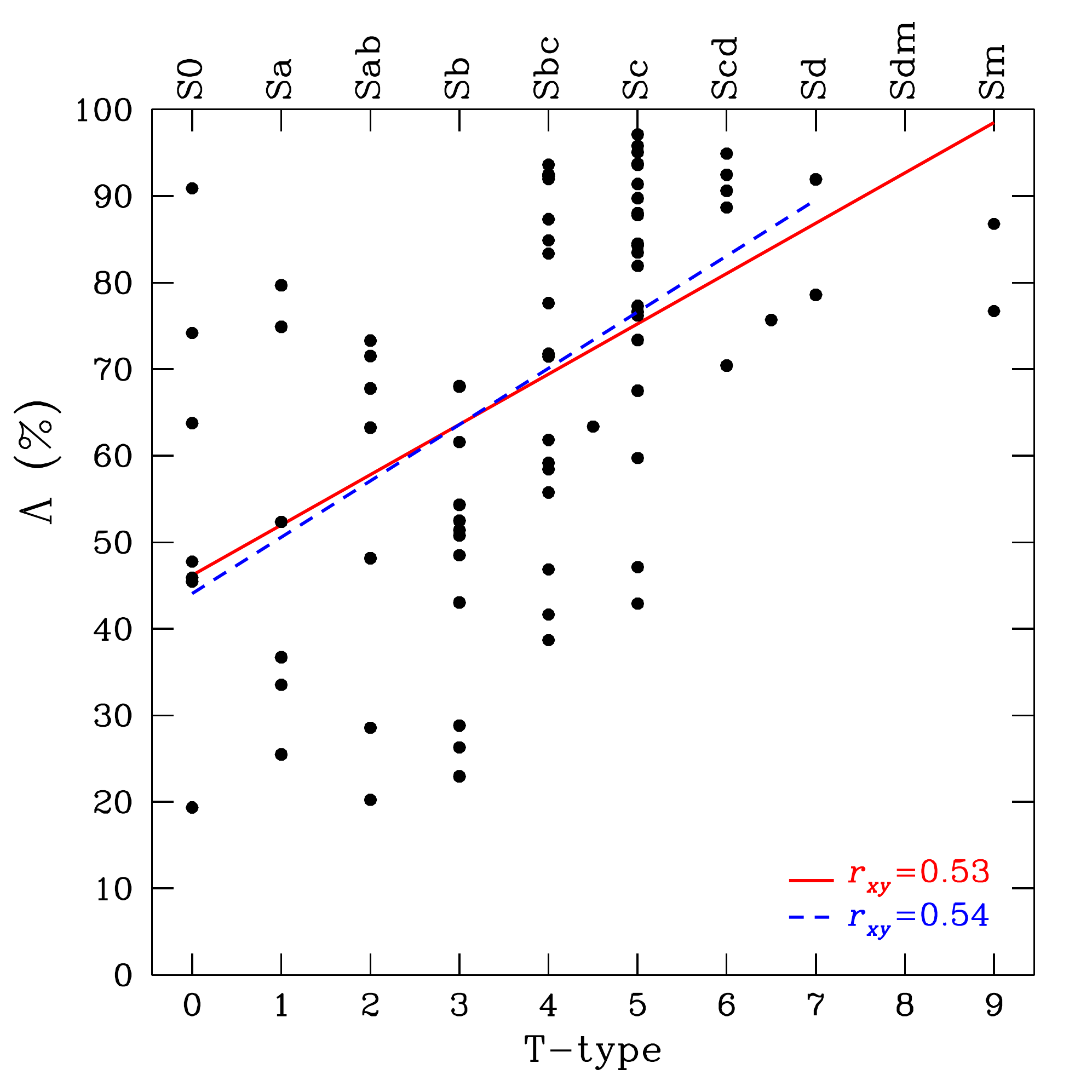}
\caption[f31]{$\Lambda$ vs.\ T Hubble type 
for the fits obtained with the (m2005 vs.\ bc03) library and
a~\citet{sal55} IMF (see section~\ref{var_IMF}).
Same labels as in Figure~\ref{fig25}.
~\label{fig31}}
\end{figure}

In order to explore the effect of a different IMF,
we compute the 5 libraries (bc03, cb07, bc03-2016, cb07-2016, and m2005) with the~\citet{sal55} IMF.
We keep the SFH and metallicity as previously described in Section~\ref{sec_build_libs},
and then fit the observed pixels of the galaxies in our sample.
As shown in Table~\ref{tbl-5}, $\Lambda$ and $\Gamma$ display similar values to the case
with the~\citet{cha03} and~\citet{kro01} IMFs (see Table~\ref{tbl-3}).
Also, the plot for $\Lambda$ vs.\ T-type, for the (m2005 vs.\ bc03) case (see Figure~\ref{fig31}),
has a similar behavior (see Figure~\ref{fig25}).

\subsubsection{Stellar metallicity}~\label{var_metallicity}

\begin{figure}
\centering
\epsscale{1.0}
\plotone{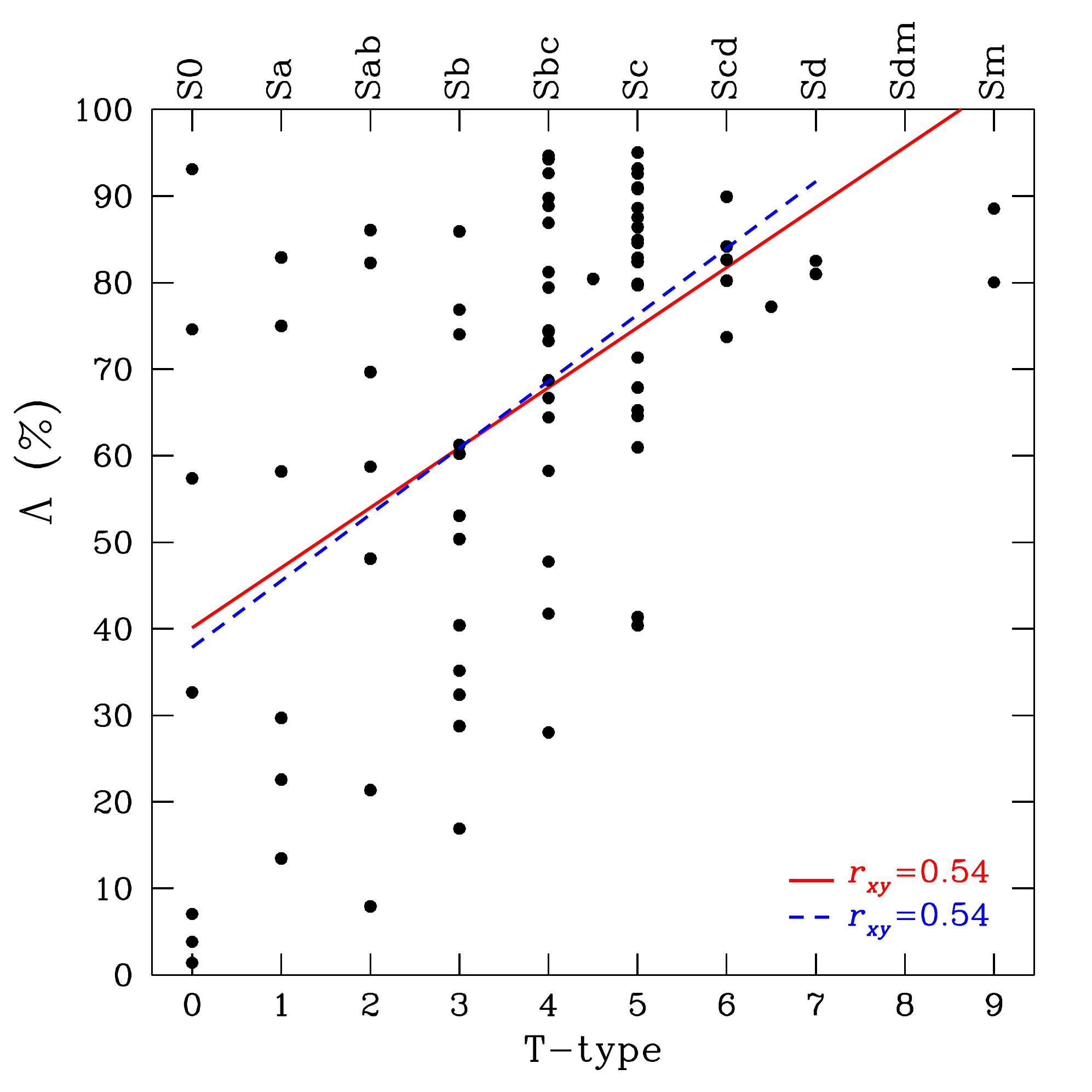}
\caption[f32]{$\Lambda$ vs.\ T Hubble type 
for the fits obtained with the (m2005 vs.\ bc03) library and
a constant solar metallicity ($Z_{\sun}$, see section~\ref{var_metallicity}).
Same labels as in Figure~\ref{fig25}.
~\label{fig32}}
\end{figure}

As mentioned in Section~\ref{sec_metallicity}, the chemical enrichment history (ChEH)
for our libraries is a constant for a given SFH.\footnote{A constant ChEH is the default in the current version of {\tt{CIGALE}}.}
In order to explore the effect of different ChEHs on our results, we compute
5 libraries (bc03, cb07, bc03-2016, cb07-2016, and m2005) with the same SFH,
and IMF described in Section~\ref{sec_build_libs},
and a single metallicity value, $Z=Z_{\sun}$.
We then fit the colors of the pixels in our objects with these libraries.
For the (m2005 vs.\ bc03) and (m2005 vs.\ bc03-2016) cases, the results are similar
(see Table~\ref{tbl-5}) to our previous calculations (see Table~\ref{tbl-3}).
On the other hand, for the (cb07 vs.\ bc03) and (cb07-2016 vs.\ bc03-2016) cases,
we find a $\Lambda$ value of $96\%$ (see Table~\ref{tbl-5}), which differs from our previous result (see Table~\ref{tbl-3}).
This signifies that for the cb07, and cb07-2016 libraries, a wider metallicity range results in better fits to the data.
Regarding the $\Lambda$ vs.\ T-type plot however, for the (m2005 vs.\ bc03) case we find
that the previously mentioned correlation (see Figure~\ref{fig25}) persists (see Figure~\ref{fig32}).

We conclude that a different SFH, IMF, or ChEH has a small impact on our main results.

\subsection{Heavy vs.\ heavy}

We also investigate how two different versions of the TP-AGB ``heavy'' models,
i.e., (m2005 vs.\ cb07) or (m2005 vs.\ cb07-2016), compare with each other. 
For (m2005 vs.\ cb07), we find that $28\%$ of the global sample
of pixels are better fitted with the m2005 library, and $72\%$ with the cb07 library.
For the case of (m2005 vs.\ cb07-2016), $40\%$ and $60\%$ of the pixels are better fitted
with the m2005 and cb07-2016 libraries, respectively.
The cb07 and cb07-2016 libraries give better fits in the outer regions of the disks (larger $R/R_{\rm hl}^{g}$),
and the m2005 library results in better fits in the inner (smaller $R/R_{\rm hl}^{g}$)
regions (this is similar to the result in Figure~\ref{fig27}).

\subsection{Which models to use?}

As a final note, one might wonder about which models to use when fitting
the optical and NIR luminosities of resolved stellar populations in nearby disk galaxies.
Most of the results we have obtained in this analysis indicate that the best fits
are obtained with the TP-AGB-poor or ``light'' models. In a conservative analysis, using
these models will result in better fits of the pixels, or resolved stellar populations, under test.
However, TP-AGB-rich ``heavy'' models may yield better fits for
certain galaxies, most of them early-type spirals, as well as in the inner regions of the disks
(this is the case for the m2005 library).
In these cases, a combined library (``heavy'' +``light'') may better match the observations.


\begin{deluxetable}{ccccc}
\tabletypesize{\scriptsize}
\tablecaption{Global pixel percentages with best model fits~\label{tbl-3}}
\tablewidth{0pt}
\tablehead{
\colhead{TP-AGB model} &
\colhead{(m2005 vs.\ bc03)} &
\colhead{(m2005 vs.\ bc03-2016)} &
\colhead{(cb07 vs.\ bc03)} &
\colhead{(cb07-2016 vs.\ bc03-2016)}\\
}
\startdata
$\Gamma$  (``heavy'')                &  $28\%$   &   $30\%$   &   $39\%$  &  $32\%$  \\
$\Lambda$ (``light'')                &  $72\%$   &   $70\%$   &   $61\%$  &  $68\%$  \\

\enddata
\tablecomments{
$\Gamma$ represents the pixel percentage better fitted by the ``heavy'' TP-AGB models,
and $\Lambda$ the percentage better fitted by the ``light'' TP-AGB models.
The percentages of this table result from the whole sample of
pixels, i.e., without differentiating between individual galaxies.
The adopted SFH, and IMF are described in Section~\ref{sec_build_libs}.
}

\end{deluxetable}


\begin{deluxetable}{cccccccccccc}
\tabletypesize{\scriptsize}
\tablecaption{Pixel percentages with best model fits per object~\label{tbl-4}}
\tablewidth{0pt}
\tablehead{
\colhead{}    & 
\multicolumn{2}{c}{(m2005 vs.\ bc03)} &  
\colhead{}    & 
\multicolumn{2}{c}{(m2005 vs.\ bc03-2016)} &
\colhead{}    & 
\multicolumn{2}{c}{(cb07 vs.\ bc03)} &  
\colhead{}    & 
\multicolumn{2}{c}{(cb07-2016 vs.\ bc03-2016)} \\
\cline{2-3} \cline{5-6} \cline{8-9} \cline{11-12} \\
\colhead{Object}      &
\colhead{$\Gamma$}    &  
\colhead{$\Lambda$}   &  
\colhead{Object}      &
\colhead{$\Gamma$}    &
\colhead{$\Lambda$}   &
\colhead{Object}      &
\colhead{$\Gamma$}    &
\colhead{$\Lambda$}   &
\colhead{Object}      &
\colhead{$\Gamma$}    &
\colhead{$\Lambda$} \\
}
\startdata																						
																							
NGC~4448	& $	80	\%$  &  $	20	\%$ &	NGC~4448	& $	84	\%$  &  $	16	\%$ &	NGC~4651	& $	54	\%$  &  $	46	\%$ &	NGC~3684	& $	74	\%$  &  $	26	\%$ \\
NGC~5701	& $	80	\%$  &  $	20	\%$ &	NGC~5701	& $	81	\%$  &  $	19	\%$ &	NGC~4691	& $	54	\%$  &  $	46	\%$ &	NGC~4051	& $	57	\%$  &  $	43	\%$ \\
NGC~7606	& $	76	\%$  &  $	24	\%$ &	NGC~7606	& $	81	\%$  &  $	19	\%$ &	NGC~4772	& $	54	\%$  &  $	46	\%$ &	NGC~4457	& $	54	\%$  &  $	46	\%$ \\
NGC~0488	& $	74	\%$  &  $	26	\%$ &	NGC~0488	& $	77	\%$  &  $	23	\%$ &	NGC~7606	& $	54	\%$  &  $	46	\%$ &	NGC~3227	& $	49	\%$  &  $	51	\%$ \\
NGC~4772	& $	73	\%$  &  $	27	\%$ &	NGC~4772	& $	77	\%$  &  $	23	\%$ &	NGC~5713	& $	53	\%$  &  $	47	\%$ &	NGC~4647	& $	49	\%$  &  $	51	\%$ \\
NGC~0779	& $	71	\%$  &  $	29	\%$ &	NGC~0779	& $	76	\%$  &  $	24	\%$ &	NGC~1309	& $	51	\%$  &  $	49	\%$ &	NGC~4666	& $	48	\%$  &  $	52	\%$ \\
NGC~4698	& $	71	\%$  &  $	29	\%$ &	NGC~4698	& $	76	\%$  &  $	24	\%$ &	NGC~4448	& $	51	\%$  &  $	49	\%$ &	NGC~3810	& $	46	\%$  &  $	54	\%$ \\
NGC~4580	& $	69	\%$  &  $	31	\%$ &	NGC~4580	& $	74	\%$  &  $	26	\%$ &	NGC~5962	& $	51	\%$  &  $	49	\%$ &	NGC~2775	& $	45	\%$  &  $	55	\%$ \\
NGC~4314	& $	63	\%$  &  $	37	\%$ &	NGC~5921	& $	68	\%$  &  $	32	\%$ &	NGC~3507	& $	50	\%$  &  $	50	\%$ &	NGC~3169	& $	45	\%$  &  $	55	\%$ \\
NGC~5921	& $	62	\%$  &  $	38	\%$ &	NGC~4314	& $	67	\%$  &  $	33	\%$ &	NGC~3877	& $	50	\%$  &  $	50	\%$ &	NGC~3593	& $	43	\%$  &  $	57	\%$ \\
NGC~3681	& $	58	\%$  &  $	42	\%$ &	NGC~3681	& $	63	\%$  &  $	37	\%$ &	NGC~4030	& $	49	\%$  &  $	51	\%$ &	NGC~1084	& $	42	\%$  &  $	58	\%$ \\
NGC~3877	& $	56	\%$  &  $	44	\%$ &	NGC~4394	& $	62	\%$  &  $	38	\%$ &	NGC~3166	& $	48	\%$  &  $	52	\%$ &	NGC~3504	& $	42	\%$  &  $	58	\%$ \\
NGC~4394	& $	56	\%$  &  $	44	\%$ &	NGC~3166	& $	60	\%$  &  $	40	\%$ &	NGC~3686	& $	48	\%$  &  $	52	\%$ &	NGC~4527	& $	42	\%$  &  $	58	\%$ \\
NGC~3166	& $	54	\%$  &  $	46	\%$ &	NGC~3877	& $	60	\%$  &  $	40	\%$ &	NGC~4293	& $	48	\%$  &  $	52	\%$ &	NGC~4151	& $	41	\%$  &  $	59	\%$ \\
NGC~4293	& $	54	\%$  &  $	46	\%$ &	NGC~4293	& $	60	\%$  &  $	40	\%$ &	NGC~4579	& $	48	\%$  &  $	52	\%$ &	NGC~4490	& $	41	\%$  &  $	59	\%$ \\
NGC~4691	& $	53	\%$  &  $	47	\%$ &	NGC~5005	& $	59	\%$  &  $	41	\%$ &	NGC~4666	& $	48	\%$  &  $	52	\%$ &	NGC~4100	& $	40	\%$  &  $	60	\%$ \\
NGC~5005	& $	53	\%$  &  $	47	\%$ &	NGC~4691	& $	57	\%$  &  $	43	\%$ &	NGC~4698	& $	48	\%$  &  $	52	\%$ &	NGC~4030	& $	39	\%$  &  $	61	\%$ \\
NGC~1073	& $	52	\%$  &  $	48	\%$ &	NGC~4450	& $	56	\%$  &  $	44	\%$ &	NGC~5676	& $	48	\%$  &  $	52	\%$ &	NGC~1309	& $	38	\%$  &  $	62	\%$ \\
NGC~5850	& $	52	\%$  &  $	48	\%$ &	NGC~5850	& $	55	\%$  &  $	45	\%$ &	NGC~5850	& $	48	\%$  &  $	52	\%$ &	NGC~3877	& $	38	\%$  &  $	62	\%$ \\
NGC~4450	& $	50	\%$  &  $	50	\%$ &	NGC~1073	& $	54	\%$  &  $	46	\%$ &	NGC~3583	& $	47	\%$  &  $	53	\%$ &	NGC~4254	& $	38	\%$  &  $	62	\%$ \\
NGC~4579	& $	49	\%$  &  $	51	\%$ &	NGC~4579	& $	53	\%$  &  $	47	\%$ &	NGC~3893	& $	47	\%$  &  $	53	\%$ &	NGC~4571	& $	38	\%$  &  $	62	\%$ \\
NGC~3675	& $	47	\%$  &  $	53	\%$ &	NGC~3675	& $	52	\%$  &  $	48	\%$ &	NGC~4394	& $	47	\%$  &  $	53	\%$ &	NGC~1087	& $	37	\%$  &  $	63	\%$ \\
NGC~4548	& $	47	\%$  &  $	53	\%$ &	NGC~5448	& $	52	\%$  &  $	48	\%$ &	NGC~4580	& $	47	\%$  &  $	53	\%$ &	NGC~3705	& $	37	\%$  &  $	63	\%$ \\
NGC~4699	& $	47	\%$  &  $	53	\%$ &	NGC~4548	& $	51	\%$  &  $	49	\%$ &	NGC~4689	& $	47	\%$  &  $	53	\%$ &	NGC~4643	& $	37	\%$  &  $	63	\%$ \\
NGC~5448	& $	47	\%$  &  $	53	\%$ &	NGC~4699	& $	51	\%$  &  $	49	\%$ &	NGC~2964	& $	46	\%$  &  $	54	\%$ &	NGC~4303	& $	36	\%$  &  $	64	\%$ \\
NGC~4689	& $	44	\%$  &  $	56	\%$ &	NGC~4689	& $	49	\%$  &  $	51	\%$ &	NGC~3705	& $	46	\%$  &  $	54	\%$ &	NGC~4414	& $	36	\%$  &  $	64	\%$ \\
NGC~3686	& $	42	\%$  &  $	58	\%$ &	NGC~3686	& $	47	\%$  &  $	53	\%$ &	NGC~4062	& $	46	\%$  &  $	54	\%$ &	NGC~4548	& $	36	\%$  &  $	64	\%$ \\
NGC~2964	& $	41	\%$  &  $	59	\%$ &	NGC~2964	& $	44	\%$  &  $	56	\%$ &	NGC~4212	& $	46	\%$  &  $	54	\%$ &	NGC~7741	& $	36	\%$  &  $	64	\%$ \\
NGC~4062	& $	41	\%$  &  $	59	\%$ &	NGC~4062	& $	44	\%$  &  $	56	\%$ &	NGC~4254	& $	46	\%$  &  $	54	\%$ &	NGC~3938	& $	34	\%$  &  $	66	\%$ \\
NGC~3507	& $	39	\%$  &  $	61	\%$ &	NGC~3507	& $	42	\%$  &  $	58	\%$ &	NGC~4490	& $	46	\%$  &  $	54	\%$ &	NGC~3949	& $	33	\%$  &  $	67	\%$ \\
NGC~2775	& $	38	\%$  &  $	62	\%$ &	NGC~4212	& $	42	\%$  &  $	58	\%$ &	NGC~5005	& $	46	\%$  &  $	54	\%$ &	NGC~4136	& $	33	\%$  &  $	67	\%$ \\
NGC~4568	& $	38	\%$  &  $	62	\%$ &	NGC~4568	& $	42	\%$  &  $	58	\%$ &	NGC~0779	& $	45	\%$  &  $	55	\%$ &	NGC~4654	& $	33	\%$  &  $	67	\%$ \\
NGC~4212	& $	37	\%$  &  $	63	\%$ &	NGC~2775	& $	39	\%$  &  $	61	\%$ &	NGC~1087	& $	45	\%$  &  $	55	\%$ &	NGC~4689	& $	33	\%$  &  $	67	\%$ \\
NGC~3705	& $	35	\%$  &  $	65	\%$ &	NGC~4643	& $	38	\%$  &  $	62	\%$ &	NGC~3681	& $	45	\%$  &  $	55	\%$ &	NGC~7606	& $	33	\%$  &  $	67	\%$ \\
NGC~4643	& $	34	\%$  &  $	66	\%$ &	NGC~3583	& $	36	\%$  &  $	64	\%$ &	NGC~4314	& $	45	\%$  &  $	55	\%$ &	NGC~3893	& $	32	\%$  &  $	68	\%$ \\
NGC~0864	& $	33	\%$  &  $	67	\%$ &	NGC~3705	& $	36	\%$  &  $	64	\%$ &	NGC~4450	& $	45	\%$  &  $	55	\%$ &	NGC~4699	& $	32	\%$  &  $	68	\%$ \\
NGC~3583	& $	33	\%$  &  $	67	\%$ &	NGC~4388	& $	36	\%$  &  $	64	\%$ &	NGC~4548	& $	45	\%$  &  $	55	\%$ &	NGC~2964	& $	31	\%$  &  $	69	\%$ \\
NGC~4388	& $	33	\%$  &  $	67	\%$ &	NGC~0864	& $	35	\%$  &  $	65	\%$ &	NGC~5448	& $	45	\%$  &  $	55	\%$ &	NGC~4450	& $	31	\%$  &  $	69	\%$ \\
NGC~1042	& $	30	\%$  &  $	70	\%$ &	NGC~1042	& $	33	\%$  &  $	67	\%$ &	NGC~5921	& $	45	\%$  &  $	55	\%$ &	NGC~4651	& $	31	\%$  &  $	69	\%$ \\
NGC~5676	& $	30	\%$  &  $	70	\%$ &	NGC~5676	& $	33	\%$  &  $	67	\%$ &	NGC~3504	& $	44	\%$  &  $	56	\%$ &	NGC~3423	& $	30	\%$  &  $	70	\%$ \\
NGC~3504	& $	29	\%$  &  $	71	\%$ &	NGC~3504	& $	31	\%$  &  $	69	\%$ &	NGC~3675	& $	44	\%$  &  $	56	\%$ &	NGC~3675	& $	30	\%$  &  $	70	\%$ \\
NGC~4527	& $	29	\%$  &  $	71	\%$ &	NGC~4527	& $	31	\%$  &  $	69	\%$ &	NGC~3949	& $	44	\%$  &  $	56	\%$ &	NGC~5334	& $	30	\%$  &  $	70	\%$ \\
NGC~7479	& $	27	\%$  &  $	73	\%$ &	NGC~7479	& $	30	\%$  &  $	70	\%$ &	NGC~4414	& $	44	\%$  &  $	56	\%$ &	NGC~5371	& $	30	\%$  &  $	70	\%$ \\
NGC~4151	& $	26	\%$  &  $	74	\%$ &	NGC~4151	& $	28	\%$  &  $	72	\%$ &	NGC~4568	& $	44	\%$  &  $	56	\%$ &	NGC~4579	& $	29	\%$  &  $	71	\%$ \\
NGC~3169	& $	25	\%$  &  $	75	\%$ &	NGC~3169	& $	27	\%$  &  $	73	\%$ &	NGC~4647	& $	44	\%$  &  $	56	\%$ &	NGC~5676	& $	29	\%$  &  $	71	\%$ \\
NGC~4414	& $	25	\%$  &  $	75	\%$ &	NGC~4571	& $	27	\%$  &  $	73	\%$ &	NGC~1084	& $	43	\%$  &  $	57	\%$ &	NGC~5962	& $	29	\%$  &  $	71	\%$ \\
NGC~5713	& $	25	\%$  &  $	75	\%$ &	NGC~0428	& $	26	\%$  &  $	74	\%$ &	NGC~7479	& $	43	\%$  &  $	57	\%$ &	NGC~0157	& $	28	\%$  &  $	72	\%$ \\
NGC~3593	& $	24	\%$  &  $	76	\%$ &	NGC~3593	& $	26	\%$  &  $	74	\%$ &	NGC~2775	& $	42	\%$  &  $	58	\%$ &	NGC~4568	& $	28	\%$  &  $	72	\%$ \\
NGC~4571	& $	24	\%$  &  $	76	\%$ &	NGC~5713	& $	26	\%$  &  $	74	\%$ &	NGC~4388	& $	42	\%$  &  $	58	\%$ &	NGC~4900	& $	28	\%$  &  $	72	\%$ \\
NGC~4647	& $	24	\%$  &  $	76	\%$ &	NGC~4414	& $	25	\%$  &  $	75	\%$ &	NGC~4527	& $	42	\%$  &  $	58	\%$ &	NGC~5850	& $	28	\%$  &  $	72	\%$ \\
NGC~0428	& $	23	\%$  &  $	77	\%$ &	NGC~4647	& $	25	\%$  &  $	75	\%$ &	NGC~4699	& $	42	\%$  &  $	58	\%$ &	NGC~3338	& $	27	\%$  &  $	73	\%$ \\
NGC~4666	& $	23	\%$  &  $	77	\%$ &	NGC~4145	& $	24	\%$  &  $	76	\%$ &	NGC~0157	& $	40	\%$  &  $	60	\%$ &	NGC~3583	& $	27	\%$  &  $	73	\%$ \\
NGC~4145	& $	22	\%$  &  $	78	\%$ &	NGC~4666	& $	24	\%$  &  $	76	\%$ &	NGC~0488	& $	40	\%$  &  $	60	\%$ &	NGC~4062	& $	27	\%$  &  $	73	\%$ \\
NGC~3227	& $	21	\%$  &  $	79	\%$ &	NGC~3227	& $	22	\%$  &  $	78	\%$ &	NGC~3227	& $	40	\%$  &  $	60	\%$ &	NGC~4293	& $	27	\%$  &  $	73	\%$ \\
NGC~4123	& $	19	\%$  &  $	81	\%$ &	NGC~4123	& $	21	\%$  &  $	79	\%$ &	NGC~3423	& $	40	\%$  &  $	60	\%$ &	NGC~4618	& $	27	\%$  &  $	73	\%$ \\
NGC~4651	& $	19	\%$  &  $	81	\%$ &	NGC~4651	& $	20	\%$  &  $	80	\%$ &	NGC~4151	& $	39	\%$  &  $	61	\%$ &	NGC~7479	& $	27	\%$  &  $	73	\%$ \\
NGC~3338	& $	17	\%$  &  $	83	\%$ &	NGC~3338	& $	18	\%$  &  $	82	\%$ &	NGC~4303	& $	39	\%$  &  $	61	\%$ &	NGC~5448	& $	26	\%$  &  $	74	\%$ \\
NGC~4100	& $	17	\%$  &  $	83	\%$ &	NGC~4100	& $	17	\%$  &  $	83	\%$ &	NGC~4900	& $	39	\%$  &  $	61	\%$ &	NGC~5713	& $	26	\%$  &  $	74	\%$ \\
NGC~4900	& $	16	\%$  &  $	84	\%$ &	NGC~4900	& $	17	\%$  &  $	83	\%$ &	NGC~1042	& $	38	\%$  &  $	62	\%$ &	NGC~3596	& $	25	\%$  &  $	75	\%$ \\
NGC~1309	& $	15	\%$  &  $	85	\%$ &	NGC~1309	& $	16	\%$  &  $	84	\%$ &	NGC~3338	& $	38	\%$  &  $	62	\%$ &	NGC~0488	& $	24	\%$  &  $	76	\%$ \\
NGC~4254	& $	15	\%$  &  $	85	\%$ &	NGC~4254	& $	16	\%$  &  $	84	\%$ &	NGC~3938	& $	38	\%$  &  $	62	\%$ &	NGC~4394	& $	24	\%$  &  $	76	\%$ \\
NGC~0157	& $	14	\%$  &  $	86	\%$ &	NGC~4618	& $	16	\%$  &  $	84	\%$ &	NGC~4136	& $	38	\%$  &  $	62	\%$ &	NGC~3166	& $	23	\%$  &  $	77	\%$ \\
NGC~4030	& $	13	\%$  &  $	87	\%$ &	NGC~0157	& $	15	\%$  &  $	85	\%$ &	NGC~4643	& $	38	\%$  &  $	62	\%$ &	NGC~4145	& $	23	\%$  &  $	77	\%$ \\
NGC~4618	& $	13	\%$  &  $	87	\%$ &	NGC~4030	& $	14	\%$  &  $	86	\%$ &	NGC~3169	& $	37	\%$  &  $	63	\%$ &	NGC~4314	& $	23	\%$  &  $	77	\%$ \\
NGC~5962	& $	13	\%$  &  $	87	\%$ &	NGC~5334	& $	14	\%$  &  $	86	\%$ &	NGC~4654	& $	37	\%$  &  $	63	\%$ &	NGC~1042	& $	22	\%$  &  $	78	\%$ \\
NGC~5334	& $	12	\%$  &  $	88	\%$ &	NGC~5962	& $	14	\%$  &  $	86	\%$ &	NGC~5701	& $	37	\%$  &  $	63	\%$ &	NGC~3681	& $	22	\%$  &  $	78	\%$ \\
NGC~3423	& $	11	\%$  &  $	89	\%$ &	NGC~3423	& $	12	\%$  &  $	88	\%$ &	NGC~0428	& $	36	\%$  &  $	64	\%$ &	NGC~3686	& $	22	\%$  &  $	78	\%$ \\
NGC~4136	& $	10	\%$  &  $	90	\%$ &	NGC~3319	& $	11	\%$  &  $	89	\%$ &	NGC~3593	& $	36	\%$  &  $	64	\%$ &	NGC~4123	& $	22	\%$  &  $	78	\%$ \\
NGC~3319	& $	9	\%$  &  $	91	\%$ &	NGC~3893	& $	11	\%$  &  $	89	\%$ &	NGC~4051	& $	36	\%$  &  $	64	\%$ &	NGC~4388	& $	22	\%$  &  $	78	\%$ \\
NGC~3893	& $	9	\%$  &  $	91	\%$ &	NGC~4457	& $	11	\%$  &  $	89	\%$ &	NGC~4100	& $	36	\%$  &  $	64	\%$ &	NGC~0864	& $	21	\%$  &  $	79	\%$ \\
NGC~4457	& $	9	\%$  &  $	91	\%$ &	NGC~4136	& $	10	\%$  &  $	90	\%$ &	NGC~4123	& $	36	\%$  &  $	64	\%$ &	NGC~4212	& $	20	\%$  &  $	80	\%$ \\
NGC~3684	& $	8	\%$  &  $	92	\%$ &	NGC~3684	& $	9	\%$  &  $	91	\%$ &	NGC~4618	& $	36	\%$  &  $	64	\%$ &	NGC~0428	& $	18	\%$  &  $	82	\%$ \\
NGC~4051	& $	8	\%$  &  $	92	\%$ &	NGC~4051	& $	8	\%$  &  $	92	\%$ &	NGC~0864	& $	34	\%$  &  $	66	\%$ &	NGC~3319	& $	18	\%$  &  $	82	\%$ \\
NGC~5371	& $	8	\%$  &  $	92	\%$ &	NGC~4303	& $	8	\%$  &  $	92	\%$ &	NGC~4145	& $	32	\%$  &  $	68	\%$ &	NGC~3507	& $	18	\%$  &  $	82	\%$ \\
NGC~3596	& $	7	\%$  &  $	93	\%$ &	NGC~5371	& $	8	\%$  &  $	92	\%$ &	NGC~4571	& $	32	\%$  &  $	68	\%$ &	NGC~4698	& $	17	\%$  &  $	83	\%$ \\
NGC~3949	& $	7	\%$  &  $	93	\%$ &	NGC~7741	& $	8	\%$  &  $	92	\%$ &	NGC~7741	& $	32	\%$  &  $	68	\%$ &	NGC~4772	& $	17	\%$  &  $	83	\%$ \\
NGC~4303	& $	7	\%$  &  $	93	\%$ &	NGC~1084	& $	7	\%$  &  $	93	\%$ &	NGC~1073	& $	30	\%$  &  $	70	\%$ &	NGC~5701	& $	17	\%$  &  $	83	\%$ \\
NGC~4490	& $	7	\%$  &  $	93	\%$ &	NGC~3596	& $	7	\%$  &  $	93	\%$ &	NGC~5334	& $	29	\%$  &  $	71	\%$ &	NGC~0779	& $	16	\%$  &  $	84	\%$ \\
NGC~7741	& $	7	\%$  &  $	93	\%$ &	NGC~3949	& $	7	\%$  &  $	93	\%$ &	NGC~3596	& $	28	\%$  &  $	72	\%$ &	NGC~4580	& $	15	\%$  &  $	85	\%$ \\
NGC~1084	& $	6	\%$  &  $	94	\%$ &	NGC~4490	& $	7	\%$  &  $	93	\%$ &	NGC~4457	& $	21	\%$  &  $	79	\%$ &	NGC~4691	& $	11	\%$  &  $	89	\%$ \\
NGC~1087	& $	5	\%$  &  $	95	\%$ &	NGC~1087	& $	6	\%$  &  $	94	\%$ &	NGC~3319	& $	18	\%$  &  $	82	\%$ &	NGC~5005	& $	11	\%$  &  $	89	\%$ \\
NGC~3938	& $	5	\%$  &  $	95	\%$ &	NGC~4654	& $	6	\%$  &  $	94	\%$ &	NGC~3810	& $	17	\%$  &  $	83	\%$ &	NGC~5921	& $	10	\%$  &  $	90	\%$ \\
NGC~4654	& $	5	\%$  &  $	95	\%$ &	NGC~3938	& $	5	\%$  &  $	95	\%$ &	NGC~5371	& $	16	\%$  &  $	84	\%$ &	NGC~1073	& $	8	\%$  &  $	92	\%$ \\
NGC~3810	& $	3	\%$  &  $	97	\%$ &	NGC~3810	& $	2	\%$  &  $	98	\%$ &	NGC~3684	& $	12	\%$  &  $	88	\%$ &	NGC~4448	& $	8	\%$  &  $	92	\%$ \\

\enddata
\tablecomments{$\Gamma$ or $\Lambda$ columns indicate a ``heavy'' or ``light'' TP-AGB model, respectively.
For each library comparison [(m2005 vs.\ bc03), etc.], the objects in the columns are ranked 
from the highest to the lowest $\Gamma$.
}
\end{deluxetable}


\begin{deluxetable}{ccccc}
\tabletypesize{\scriptsize}
\tablecaption{Global pixel percentages with best model fits~\label{tbl-5}}
\tablewidth{0pt}
\tablehead{
\colhead{TP-AGB model} &
\colhead{(m2005 vs.\ bc03)} &
\colhead{(m2005 vs.\ bc03-2016)} &
\colhead{(cb07 vs.\ bc03)} &
\colhead{(cb07-2016 vs.\ bc03-2016)}\\
}
\startdata
\vspace{1 mm}
Two decaying exponentials SFH:  &           ~           &            ~            &          ~            &     ~    \\
\vspace{1 mm}
$\Gamma$                        &        $28\%$         &         $30\%$          &       $40\%$          &  $34\%$  \\
$\Lambda$                       &        $72\%$         &         $70\%$          &       $60\%$          &  $66\%$  \\

\cline{1-5}\\
\vspace{1 mm}
Salpeter IMF:                   &           ~           &            ~            &          ~            &    ~     \\
$\Gamma$                        &        $28\%$         &         $30\%$          &       $38\%$          &  $31\%$  \\
$\Lambda$                       &        $72\%$         &         $70\%$          &       $62\%$          &  $69\%$  \\

\cline{1-5}\\
\vspace{1 mm}
Solar metallicity models:       &           ~           &            ~            &          ~            &    ~     \\
$\Gamma$                        &        $29\%$         &         $29\%$          &       $~4\%$          &  $~4\%$  \\
$\Lambda$                       &        $71\%$         &         $71\%$          &       $96\%$          &  $96\%$  \\

\enddata
\tablecomments{
This table shows the resulting percentages when changing the SHF (see Section~\ref{var_SFH}),
the IMF (see Section~\ref{var_IMF}), and the metallicity (see Section~\ref{var_metallicity})
of the CSP libraries (compare with Table~\ref{tbl-3}).
$\Gamma$ represents the pixel percentage best fitted by ``heavy'' TP-AGB models,
and $\Lambda$ the pixel percentages with best fits by ``light'' TP-AGB models.
The percentages in this table result from the whole sample of pixels.
}

\end{deluxetable}



\section{Conclusions}~\label{sec_conclusions}

The determination of the stellar properties in galaxies,
via stellar population model fitting techniques, is important for our interpretation
of galaxy evolution. Among these properties, e.g., the stellar mass
is essential to constrain the main sequence of star formation,
i.e., the relationship between the SFR and the stellar mass~\citep[e.g,][]{noe07,rodi11};
the stellar mass function; the dark matter content of galaxies~\citep[e.g.,][]{rep13,rep15,rep17,rep18};
and the stellar mass density of the universe.
For these reasons, it is of utmost importance to discriminate between stellar population synthesis models
whose different NIR stellar mass-lo-light ratios, due to distinct
luminosity contributions from TP-AGB stars, result in unequal
recovered stellar masses.
We have fitted a sample of nearby galaxies, on a pixel by pixel basis, using
various libraries of stellar population synthesis models, with both ``heavy'' and ``light'' contributions from
TP-AGB stars. On average, the fits to the pixels in our sample
favor ``light'' models over ``heavy'' ones.
However, $\sim$30\%-40\% of the individual pixels are better fitted with ``heavy'' models.

Our results also indicate that for nearby disk galaxies, the luminosity contribution of TP-AGB stars
may depend on Hubble type, and therefore on stellar age and metallicity.
This may be explained if there is a dependence of the TP-AGB mass-loss
rate with metallicity, where a {\it pre-dust} wind precedes the {\it dust-driven}
wind and leads to shorter TP-AGB lifetimes in metal-poor galaxies~\citep{gir10,ros14,ros16}.
Since heavy elements and dust content in galaxies evolve with
redshift~\citep{mai19,tri20}, the possibility exists that the contribution of TP-AGB stars
to the light of galaxies also varies across cosmic time~\citep{kri10}.\footnote{The highest
luminosity contribution of the first TP-AGB stars may take place at redshifts ($z\sim5-6$), when
the age of the Universe was T$_{\rm form}\sim1$ Gyr, although this may depend on the metallicity of
the objects as suggested by the results of this paper.}

\acknowledgments

We acknowledge the referee for her/his comments and suggestions.
EMG acknowledges support through the C\'atedras CONACYT program,
as well as the remote use of the computer ``galaxias'' at IRyA, UNAM.
R.A.G.L. acknowledges the financial support of DGAPA, UNAM, project IN108518,
and of CONACYT, Mexico, project A1-S-8263.
GB acknowledges financial support from the National Autonomous University of Mexico (UNAM),
through grant DGAPA/PAPIIT IG100319, and from CONACYT, through grant CB2015-252364. 
We thank Manuel Zamora for his valuable help.
The authors thankfully acknowledge computer resources,
technical advise and support provided by
{\it Laboratorio Nacional de Superc\'omputo del Sureste de M\'exico} (LNS),
a member of the CONACYT network of national laboratories.


\end{document}